\documentclass{emulateapj}
\usepackage{times}

\newcommand{\ms}{preprint} 
\usepackage{ifthen}

\ifthenelse{\equal{\ms}{preprint}}
{
\newcounter{subsubsubsectioncounter}[subsubsection]
\newcommand{\subsubsubsection}[1]{
\begin{center}
\arabic{section}.\arabic{subsection}.\arabic{subsubsection}.\arabic{subsubsubsectioncounter}
{\it #1}\end{center} \addtocounter{subsubsubsectioncounter}{1}}
\newcommand{\na}{New Astron.}
\newcommand{\jcap}{JCAP}
\renewcommand{\altaffiltext}[2]{\affil{#1. #2}}
}
{}
\newcommand{\Fermi}{\emph{Fermi}}
\newcommand{\FGST}{\emph{Fermi Gamma-ray Space Telescope}}
\newcommand{\Swift}{\emph{Swift}}
\newcommand{\gray}{$\gamma$-ray}

\newcommand{\app}{APh}
\newcommand{\sci}{{Science}}

\newcommand{\nimA}{Nucl.\ Instr.\ Meth.\ Phys.\ Res.\ A}

\newcommand{\plb}{Phys.\ Lett.\ B}
 
\newcommand{\ieeenuc}{IEEE\ Trans.\ Nuc.\ Sci.}

\shorttitle{The Large Area Telescope on the \FGST{} Mission}
\shortauthors{Atwood et al.}

\begin{document}

\title{The Large Area Telescope on the \emph{Fermi Gamma-ray Space Telescope} Mission}

\author{
W.~B.~Atwood\altaffilmark{1}, 
A.~A.~Abdo\altaffilmark{2,3}, 
M.~Ackermann\altaffilmark{4}, 
B.~Anderson\altaffilmark{1}, 
M.~Axelsson\altaffilmark{5}, 
L.~Baldini\altaffilmark{6}, 
J.~Ballet\altaffilmark{7}, 
D.~L.~Band\altaffilmark{8,9}, 
G.~Barbiellini\altaffilmark{10,11}, 
J.~Bartelt\altaffilmark{4}
% ...
}
\altaffiltext{1}{Santa Cruz Institute for Particle Physics, Department of Physics and Department of Astronomy and Astrophysics, University of California at Santa Cruz, Santa Cruz, CA 95064}
\altaffiltext{2}{National Research Council Research Associate}
\altaffiltext{3}{Space Science Division, Naval Research Laboratory, Washington, DC 20375}
\altaffiltext{4}{W. W. Hansen Experimental Physics Laboratory, Kavli Institute for Particle Astrophysics and Cosmology, Department of Physics and Stanford Linear Accelerator Center, Stanford University, Stanford, CA 94305}
\altaffiltext{5}{Stockholm Observatory, Albanova, SE-106 91 Stockholm, Sweden}
\altaffiltext{6}{Istituto Nazionale di Fisica Nucleare, Sezione di Pisa, I-56127 Pisa, Italy}
\altaffiltext{7}{Laboratoire AIM, CEA-IRFU/CNRS/Universit\'e Paris Diderot, Service d'Astrophysique, CEA Saclay, 91191 Gif sur Yvette, France}
\altaffiltext{8}{Center for Research and Exploration in Space Science and Technology (CRESST), NASA Goddard Space Flight Center, Greenbelt, MD 20771}
\altaffiltext{9}{NASA Goddard Space Flight Center, Greenbelt, MD 20771}
\altaffiltext{10}{Istituto Nazionale di Fisica Nucleare, Sezione di Trieste, I-34127 Trieste, Italy}
\altaffiltext{11}{Dipartimento di Fisica, Universit\`a di Trieste, I-34127 Trieste, Italy}
% ...

\begin{abstract}

The Large Area Telescope (\Fermi/LAT, hereafter LAT), the primary instrument on the
\FGST{} (\Fermi) mission, is an imaging,
wide field-of-view, high-energy \gray{} telescope, covering the
energy range from below 20 MeV to more than 300 GeV.  The LAT was
built by an international collaboration with contributions from space
agencies, high-energy particle physics institutes, and universities in
France, Italy, Japan, Sweden, and the United States.  This paper
describes the LAT, its 
pre-flight expected 
performance, and summarizes the key science
objectives that will be addressed. 
On-orbit performance will be presented in detail in a subsequent paper.
The LAT is a pair-conversion
telescope with a precision tracker and calorimeter, each consisting of
a $4\times4$ array of 16 modules, a segmented anticoincidence detector that
covers the tracker array, and a programmable trigger and data
acquisition system.  Each tracker module has a vertical stack of 18
$x,y$ tracking planes, including two layers ($x$ and $y$) of single-sided
silicon strip detectors and high-$Z$ converter material (tungsten) per
tray.  Every calorimeter module has 96 CsI(Tl) crystals, arranged in
an 8 layer hodoscopic configuration with a total depth of 8.6
radiation lengths, giving both longitudinal and transverse information
about the energy deposition pattern.  The calorimeter's depth and
segmentation enable the high-energy reach of the LAT and contribute
significantly to background rejection.  The aspect ratio of the
tracker (height/width) is 0.4, allowing a large field-of-view (2.4 sr) and
ensuring that most pair-conversion showers initiated in the tracker
will pass into the calorimeter for energy measurement.  Data obtained
with the LAT are intended to (i) permit rapid notification of
high-energy \gray{} bursts (GRBs) and transients and facilitate monitoring of variable sources,
(ii) yield an extensive catalog of several thousand high-energy
sources obtained from an all-sky survey, (iii) measure spectra from 20
MeV to more than 50 GeV for several hundred sources, (iv) localize
point sources to 0.3 -- 2 arc minutes, (v) map and obtain spectra of
extended sources such as SNRs, molecular clouds, and nearby galaxies,
(vi) measure the diffuse isotropic \gray{} background up to TeV
energies, and (vii) explore the discovery space for dark matter.

\end{abstract}


\section{Introduction}\label{s1}

A revolution is underway in our understanding of the high-energy sky.
The early SAS-2 \citep{Fichtel1975} and COS-B \citep{Bignami1975}
missions led to the EGRET instrument \citep{Thompson1993} on the
\emph{Compton Gamma-Ray Observatory (CGRO)}.  EGRET performed the first all-sky
survey above 50 MeV and made breakthrough observations of high-energy
\gray{} blazars, pulsars, delayed emission from \gray{} bursts
(GRBs), high-energy solar flares, and diffuse radiation from our
Galaxy and beyond that have all changed our view of the high-energy
Universe.  

Many high-energy sources revealed by EGRET have not yet
been identified.  The Large Area Telescope (LAT) on the \FGST{} (\Fermi),
formerly the \emph{Gamma-ray Large Area Space Telescope (GLAST)}, 
launched by NASA on 2008, June 11 on a Delta II Heavy launch vehicle, 
offers enormous opportunities for determining the nature
of these sources and advancing knowledge in astronomy, astrophysics,
and particle physics.
In this paper a comprehensive overview of the LAT instrument design is
provided, the pre-flight expected performance based on detailed
simulations and ground calibration measurements is given, and the
science goals and expectations are summarized.  The \Fermi\ observatory
had been launched shortly before the submission of this paper so no
details of in-flight performance are provided at this time, although
the performance to date does not deviate significantly from that
estimated before launch.  The in-flight calibration of the LAT is
being refined during the first year of observations and therefore
details of in-flight performance will be the subject of a future paper.

\Fermi\ follows the successful launch of \emph{Agile} by the 
Italian Space Agency in April 2007 \citep{Tavani2008}.
The scientific objectives addressed by the LAT
include (i) determining the nature of the unidentified sources and the
origins of the diffuse emission revealed by EGRET, (ii) understanding
the mechanisms of particle acceleration operating in celestial
sources, particularly in active galactic nuclei, pulsars, supernovae
remnants, and the Sun, (iii) understanding the high-energy behavior of
GRBs and transients, (iv) using \gray{} observations as a probe of
dark matter, and (v) using high-energy \gray{s} to probe the early
universe and the cosmic evolution of high-energy sources to $z\ge
6$. These objectives are discussed in the context of the LAT's
measurement capabilities in \S\ref{s3}.  

\begin{figure}[!tbhp]
\centerline{
\includegraphics[width=3.5in]{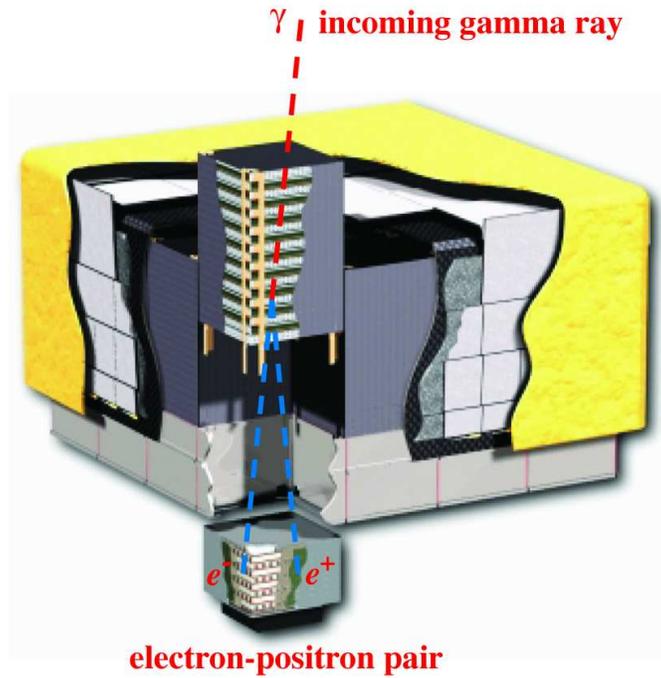}
}
\caption{Schematic diagram of the Large Area Telescope.  
The telescope's dimensions are $1.8\ {\rm m} \times 1.8\ {\rm m} \times 0.72$ m.  
The power required and the mass are 650 W and 2,789 kg, respectively.
}
\label{f1.1}
\end{figure}

\begin{figure}[!tbhp]
\centerline{
\includegraphics[width=5in]{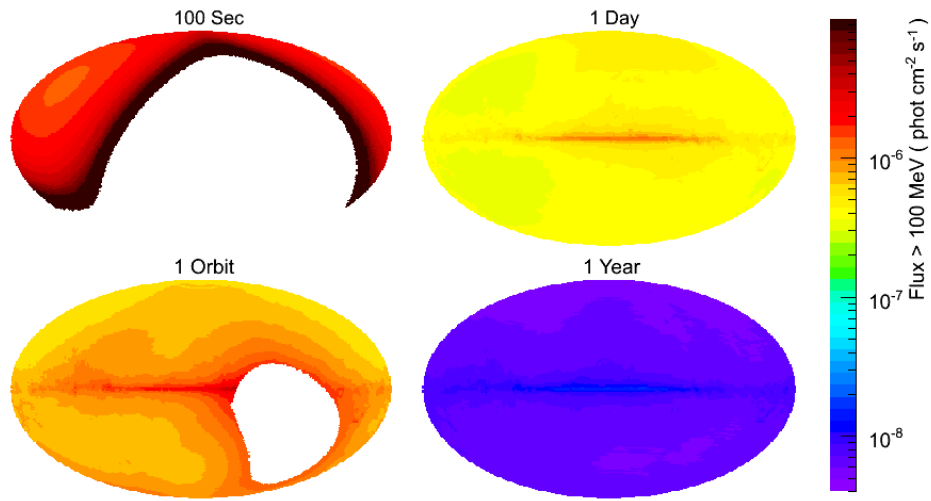}
}
\caption{LAT source sensitivity for exposures on various timescales.  
Each map is an Aitoff projection in galactic coordinates.  
In standard sky-survey mode, nearly uniform exposure is achieved every 2 orbits, 
with every region viewed for $\sim$30 min every 3 hours.
}
\label{f1.2}
\end{figure}

To make significant progress in
understanding the high-energy sky, the LAT, shown in Figure~\ref{f1.1}, has
good angular resolution for source localization and multi-wavelength
studies, high sensitivity over a broad field-of-view to monitor
variability and detect transients, good calorimetry over an extended
energy band to study spectral breaks and cut-offs, and good
calibration and stability for absolute, long term flux
measurement. The LAT measures the tracks of the electron ($e^-$) and
positron ($e^+$) that result when an incident \gray{} undergoes
pair-conversion, preferentially in a thin, high-$Z$ foil, and measures
the energy of the subsequent electromagnetic shower that develops in
the telescope's calorimeter. Table~\ref{t1.1} summarizes the scientific
performance capabilities of the LAT.  Figure~\ref{f1.2} illustrates the
sensitivity and field-of-view (FoV) achieved with the LAT for
exposures on various timescales.  To take full advantage of the
LAT's large FoV, the primary observing mode of \Fermi{}
is the so-called ``scanning'' mode in which the normal
to the front of the instrument ($z$ axis) on alternate orbits is pointed
to $+35^\circ$ from the zenith direction and towards the pole of the orbit
and to $-35^\circ$ from the zenith on the subsequent orbit.  In this way,
after 2 orbits, about 3 hours for \Fermi's orbit at $\sim$565 km and
$25.5^\circ$ inclination, the sky exposure is almost uniform.  For
particularly interesting targets of opportunity, the observatory can
be inertially pointed.  Details of the LAT design and performance are
presented in \S\ref{s2}.  

\placefigure{f1.1}
\placefigure{f1.2}
\placetable{t1.1}

The LAT was developed by an international
collaboration with primary hardware and software responsibilities at
Stanford University, Stanford Linear Accelerator Center, Agenzia
Spaziale Italiana, Commissariat \`a l'Energie Atomique, Goddard Space
Flight Center, Istituto Nazionale di Fisica Nucleare, 
Centre National de la Recherche Scientifique
/ Institut National de Physique Nucl\'eaire et de Physique des
Particules, Hiroshima University, Naval Research Laboratory, Ohio
State University, Royal Institute of Technology -- Stockholm,
University of California at Santa Cruz, and University of Washington.
Other institutions that have made significant contributions to the
instrument development include Institute of Space and Astronautical
Science, Stockholm University, University of Tokyo, and Tokyo
Institute of Science and Technology. 
All of these institutions as well as the Istituto Nazionale di 
Astrofisica in Italy are making significant contributions to LAT data analysis
during the science operations phase of the \Fermi{} mission.

\section{Large Area Telescope}\label{s2}
\subsection{Technical development path}\label{s2.1}

The LAT is designed to measure the directions, energies, and arrival
times of \gray{s} incident over a wide FoV, while
rejecting background from cosmic rays.  First, the design approach \citep{Atwood1994} 
that resulted in the instrument described in detail in
\S\ref{s2.2} made extensive use of detailed simulations of the
detector response to signal (celestial \gray{s}) and backgrounds
(cosmic rays, albedo \gray{s}, etc.).  Second, detector technologies
were chosen that have an extensive history of application in space
science and high-energy physics with demonstrated high reliability.
Third, relevant test models were built to demonstrate that critical
requirements, such as power, efficiency, and detector noise occupancy,
could be readily met.  Fourth, these detector-system models, including
all subsystems, were studied in accelerator test beams to validate both the
design and the Monte Carlo programs used in the simulations \citep{Atwood2000}.

The modular design of the LAT allowed the construction, at
reasonable incremental cost, of a full-scale, fully functional
engineering demonstration telescope module for validation of the
design concept.  This test engineering model was flown on a
high-altitude balloon to demonstrate system level performance in a
realistic, harsh background environment \citep{Thompson2002,Mizuno2004} and was
subjected to an accelerator beam test program \citep{Couto_e_Silva2001}.  
Particle beam tests were also done on spare flight tracker
and calorimeter modules (see (\S\ref{s2.5.1}).

\subsection{Technical description}\label{s2.2}

High-energy \gray{s} cannot be reflected or refracted; they interact
by the conversion of the \gray{} into an $e^+e^-$ pair.  The LAT is
therefore a pair-conversion telescope with a precision
converter-tracker (\S\ref{s2.2.1}) and calorimeter (\S\ref{s2.2.2}), each
consisting of a $4\times4$ array of 16 modules supported by a low-mass
aluminum grid structure.  A segmented anticoincidence detector (ACD,
\S\ref{s2.2.3}) covers the tracker array, and a programmable trigger and
data acquisition system (DAQ, \S\ref{s2.2.4}) utilizes prompt signals
available from the tracker, calorimeter, and anticoincidence detector
subsystems to form a trigger.  The self-triggering capability of the
LAT tracker in particular is an important new feature of the LAT
design that is possible because of the choice of silicon-strip
detectors, which do not require an external trigger, for the active
elements.  In addition, all of the LAT instrument subsystems utilize
technologies that do not use consumables such as gas.  Upon
triggering, the DAQ initiates the read out of these 3 subsystems and
utilizes on-board event processing to reduce the rate of events
transmitted to the ground to a rate compatible with the 1 Mbps average
downlink available to the LAT.  The on-board processing is optimized
for rejecting events triggered by cosmic-ray background particles
while maximizing the number of events triggered by \gray{s}, which are
transmitted to the ground.  Heat produced by the tracker, calorimeter,
and DAQ electronics is transferred to radiators through heat pipes in
the grid.

The overall aspect ratio of the LAT tracker (height/width) is 0.4,
allowing a large FoV\footnote{FoV $=\int A_{\rm eff}(\theta,\phi)
d\Omega/A_{\rm eff}(0,0)=2.4$ sr at 1 GeV, where $A_{\rm eff}$ is the effective
area of the LAT after all analysis cuts for background rejections have been made.}
and ensuring that nearly all
pair-conversion events initiated in the tracker will pass into the
calorimeter for energy measurement.

\subsubsection{Precision converter-tracker}\label{s2.2.1}

\begin{figure}[!tbhp]
\centerline{
\includegraphics[width=3.5in]{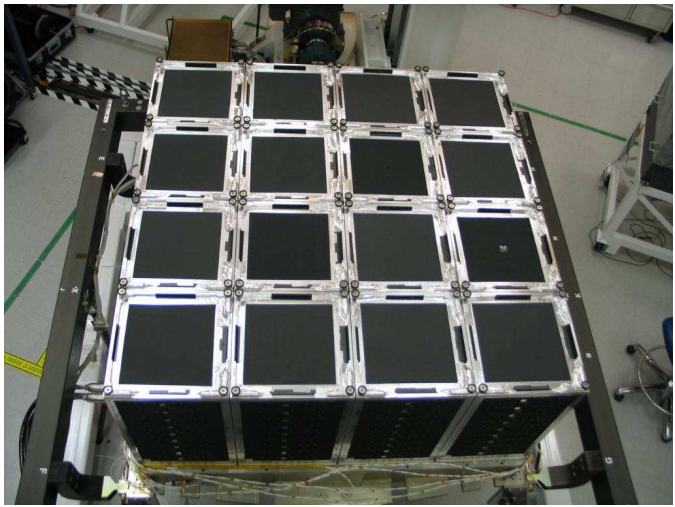}
}
\caption{Completed tracker array before integration with the ACD.
}
\label{f2.1}
\end{figure}

The converter-tracker has 16 planes of high-$Z$ material in which
\gray{s} incident on the LAT can convert to an $e^+e^-$ pair.  The
converter planes are interleaved with position-sensitive detectors
that record the passage of charged particles, thus measuring the
tracks of the particles resulting from pair conversion.  This
information is used to reconstruct the directions of the incident
\gray{s}.  Each tracker module has 18 $x,y$ tracking planes,
consisting of 2 layers ($x$ and $y$) of single-sided silicon strip
detectors.
The 16 planes at
the top of the tracker are interleaved with high-$Z$ converter material
(tungsten).  Figure~\ref{f2.1} shows the completed 16 module tracker array before
integration with the ACD.  Table~\ref{t2.1} is a summary of key parameters of
the LAT tracker.  See \citet{Atwood2007} for a more complete
discussion of the tracker design and performance.  We summarize here
the features most relevant to the instrument science performance.

\placefigure{f2.1}
\placetable{t2.1}

The single-sided SSDs are AC-coupled, with 384 56-$\mu$m wide aluminum
readout strips spaced at 228 $\mu$m pitch\footnote{
pitch = distance between centers of adjacent strips.
}.  They were produced on
$n$-intrinsic 15-cm wafers by Hamamatsu Photonics, and each has an area
of 8.95$\times$8.95 cm$^2$, with an inactive area 1 mm wide around the edges,
and a thickness of 400 $\mu$m.  Sets of 4 SSDs were bonded edge to edge
with epoxy and then wire bonded strip to strip to form ``ladders," such
that each amplifier channel sees signals from a 35 cm long strip.
Each detector layer in a tracker module consists of 4 such ladders
spaced apart by 0.2 mm gaps.  The delivered SSD quality was very high,
with a bad channel rate less than 0.01\% and an average total leakage
current of 110 nA.  The wafer dicing was accurate to better than
20 $\mu$m, to allow all of the assembly to be done rapidly with mechanical
jigs rather than with optical references.

The support structure for the detectors and converter foil planes is a stack
of 19 composite panels, or ``trays,'' supported by
carbon-composite sidewalls that also serve to conduct heat to the base
of the tracker array.  The tray structure is a low-mass,
carbon-composite assembly made of a carbon-carbon closeout,
carbon-composite face sheets, and a vented aluminum honeycomb core.
Carbon was chosen for its long radiation length, high
modulus (stiffness) to density ratio, good thermal conductivity, and thermal
stability.

\begin{figure}[!tbhp]
\centerline{\hfill
\includegraphics[width=3.5in]{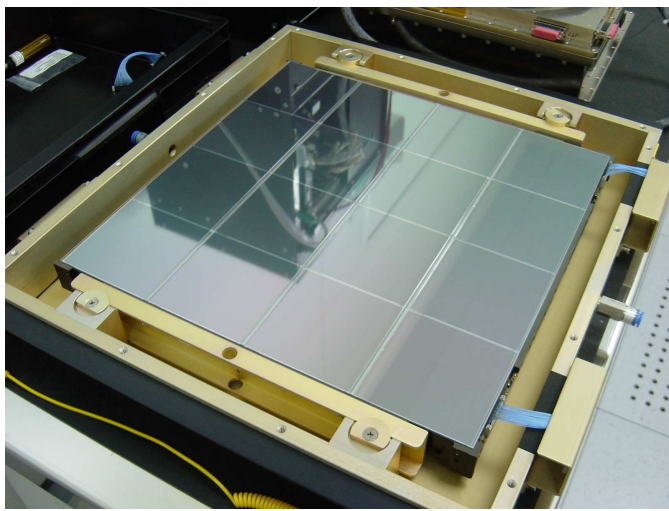}\hfill
\includegraphics[width=3.5in]{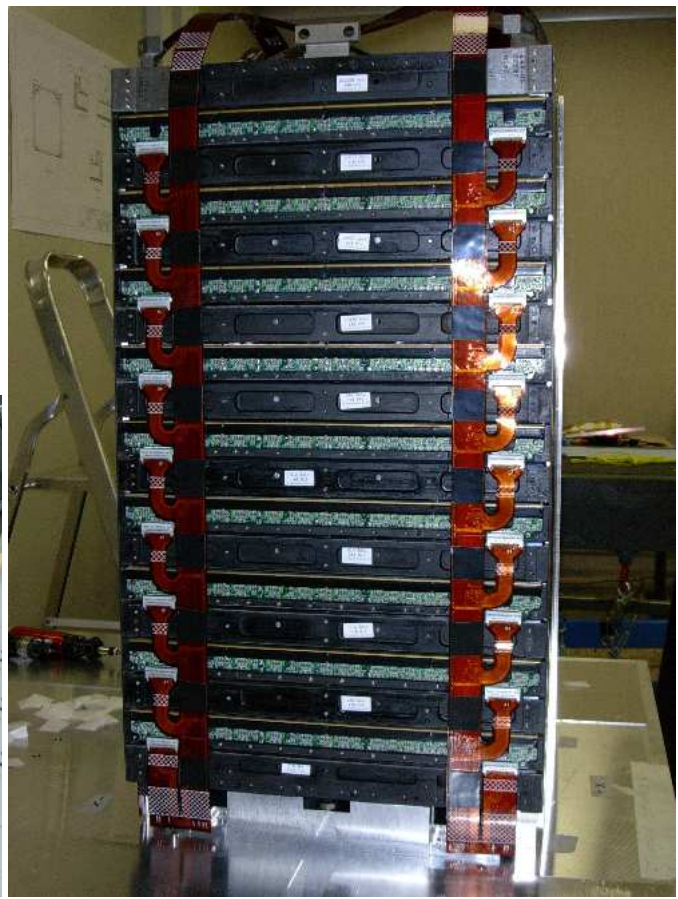}\hfill
}
\caption{(a) A flight tracker tray and (b) a completed tracker module with 
one sidewall removed.
}
\label{f2.2}
\end{figure}

The tray-panel structure is about 3 cm thick and is instrumented with
converter foils, detectors, and front-end electronics.  All trays are
of similar construction, but the top and bottom trays have detectors
on only a single face.  The bottom trays include the mechanical and
thermal interfaces to the grid, while the top trays support the
readout-cable terminations, mechanical lifting attachments, and
optical survey retro-reflectors.  Trays supporting thick converter
foils have stronger face sheets and heavier core material than those
supporting thin foils or no foils.  Figure~\ref{f2.2}a shows a flight tracker
tray and Figure~\ref{f2.2}b shows a completed tracker module with one sidewall removed.

\placefigure{f2.2}

The strips on the top and bottom of a given tray are parallel, while
alternate trays are rotated $90^\circ$ with respect to each other.  An $x,y$
measurement plane consists of a layer of detectors on the bottom of
one tray together with an orthogonal detector layer on the top of the
tray just below, with only a 2 mm separation. The tungsten
converter foils in the first 16 planes lie immediately above the upper
detector layer in each plane.  The lowest two $x,y$ planes have no
tungsten converter material.  The tracker mechanical design emphasizes
minimization of dead area within its aperture.  To that end, the
readout electronics are mounted on the sides of the trays and
interfaced to the detectors around the $90^\circ$ corner.  One fourth of the
readout electronics boards in a single tracker module can be seen in
Figure~\ref{f2.2}b.  The interface to the data acquisition and power
supplies is made entirely through flat cables constructed as long
4-layer flexible circuits, two of which are visible in Figure~\ref{f2.2}b.
As a result, the dead space between the active area of one tracker
module and that of its neighbor is only 18 mm.

Incident photons preferentially convert in one of the tungsten foils,
and the resulting $e^-$ and $e^+$ particles are tracked by the 
SSDs through successive planes.  The pair conversion
signature is also used to help reject the much larger background of
charged cosmic rays. The high intrinsic efficiency and reliability of
this technology enables straightforward event reconstruction and
determination of the direction of the incident photon.

\begin{figure}[!tbhp]
\centerline{
\includegraphics[width=3.5in]{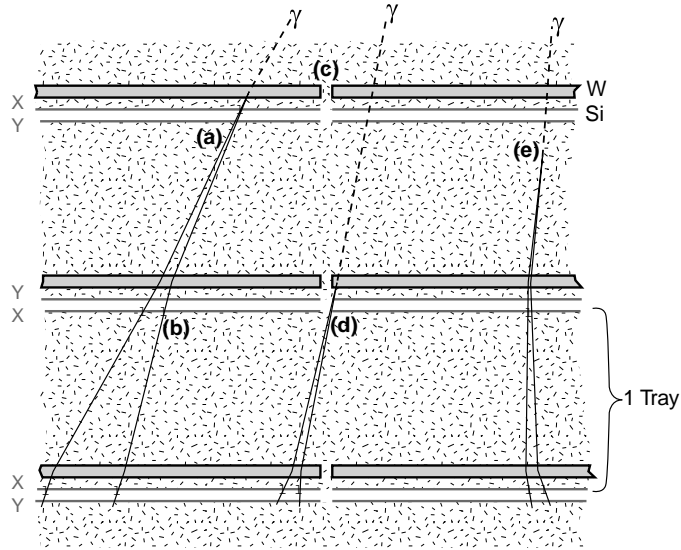}
}
\caption{Illustration of tracker design principles.  
The first two points dominate the measurement of the photon direction, especially at low energy.
(Note that in this projection only the $x$ hits can be displayed.)  
(a) Ideal conversion in W:  Si detectors are located as close as possible to the 
W foils, to minimize the lever arm for multiple scattering.  Therefore, scattering in the 
2nd W layer has very little impact on the measurement.  
(b)  Fine detector segmentation can separately detect the two particles in many cases, 
enhancing both the PSF and the background rejection. 
(c) Converter foils cover only the active area of the Si, to minimize conversions for 
which a close-by measurement is not possible.  
(d) A missed hit in the 1st or 2nd layer can degrade the PSF by up to a factor of two, 
so it is important to have such inefficiencies well localized and identifiable, rather 
than spread across the active area.  
(e) A conversion in the structural material or Si can give long lever arms for multiple 
scattering, so such material is minimized.  Good 2-hit resolution can help identify 
such conversions.
}
\label{f2.3}
\end{figure}

The probability distribution for the reconstructed direction of
incident \gray{s} from a point source is referred to as the Point
Spread Function (PSF). Multiple scattering of the $e^+$ and $e^-$ 
and bremsstrahlung
production limit the obtainable resolution.  To get optimal results
requires that the $e^-$ and $e^+$ directions be measured immediately
following the conversion.  At 100 MeV the penalty for missing one of
the first hits\footnote{
The term ``hit'' refers to the detection of the passage of a
charged particle through a silicon strip and the recording of the strip
address.} 
is about a factor of two in resolution, resulting in
large tails in the PSF.  Figure~\ref{f2.3} summarizes these and other
considerations in the tracker design that impact the PSF.  In
particular, it is important that the silicon-strip detector layers
have high efficiency and are held close to the converter foils, that
the inactive regions are localized and minimized, and that the passive
material is minimized.  To minimize missing hits in the first layer
following a conversion, the tungsten foils in each plane cover only
the active areas of the silicon-strip detectors.

\placefigure{f2.3}

One of the most complex LAT design trades was the balance between the
need for thin converters, to achieve a good PSF at low energy, where
the PSF is determined primarily by the $\sim$$1/E$ dependence of multiple
scattering, versus the need for converter material to maximize the
effective area, important at high energy.  The resolution was to
divide the tracker into 2 regions, ``front'' and ``back.''  The front
region (first 12 $x,y$ tracking planes) has thin converters, each 0.03
radiation lengths thick, to optimize the PSF at low energy, while the
converters in the back (4 $x,y$ planes after the front tracker section)
are $\sim$6 times thicker, to maximize the effective area at the expense of
less than a factor of two in angular resolution (at 1 GeV) for photons
converting in that region.  Instrument simulations show that the
sensitivity of the LAT to point-sources is approximately balanced
between the front and back tracker sections, although this depends on
the source spectral characteristics.

The tracker detector performance was achieved with readout electronics
designed specifically to meet the LAT requirements and implemented
with standard commercial technology \citep{Baldini2006}.  The system
is based on two Application Specific Integrated Circuits (ASICs).  The
first ASIC is a 64-channel mixed-mode amplifier-discriminator chip and
the second ASIC is a digital readout controller.  Each
amplifier-discriminator chip is programmed with a single threshold
level, and only a 0 or 1 (i.e., a ``hit'') is stored for each channel
when a trigger is generated.  Each channel can buffer up to 4 events,
and the system is able to trigger even during readout of the digital
data from previous events.  Thus the system achieves high throughput
and very low deadtime, and the output data stream is compact and
contains just the information needed for effective tracking, with 
$<$$10^{-6}$ noise occupancy, and with very little calibration required.  The
system also measures and records the time-over-threshold (TOT) of each
layer's trigger output signal, 
which
provides charge-deposition information that is useful for background
rejection.  In particular, isolated tracks that 
start from showers
in the calorimeter sometimes range out in the tracker, mimicking a
\gray{} conversion.  The TOT information is effective for detecting
and rejecting such background events 
because at the termination of such tracks the charge deposition is very 
large, often resulting in a large TOT in the last SSD traversed.

The tracker provides the principal trigger for the LAT.  Each detector
layer in each module outputs a logical OR of all of its 1536 channels, and a
first-level trigger is derived from coincidence of successive layers
(typically 3 $x,y$ planes).  There is no detectable coherent noise in
the system, such that the coincidence rate from electronics noise is
immeasurably small, while the trigger efficiency for charged particles
approaches 100\% when all layers are considered.

High reliability was a core requirement in the tracker design.  The 16
modules operate independently, providing much redundancy.  Similarly,
the multi-layer design of each module provides redundancy.  The
readout system is also designed to minimize or eliminate the impact of
single-point failures.  Each tracker layer has two separate readout
and control paths, and the 24 amplifier-discriminator chips in each
layer can be partitioned between the two paths by remote command.
Therefore, failure of a single chip or readout cable would result in
the loss of at most only 64 channels.

\subsubsection{Calorimeter}\label{s2.2.2}

\begin{figure}[!tbhp]
\centerline{
\includegraphics[width=3.5in,angle=0]{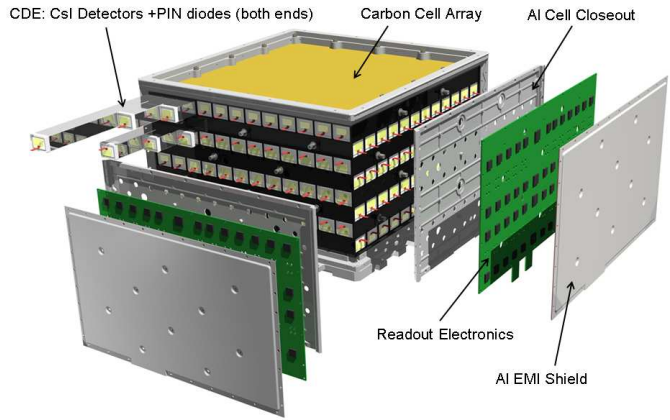}
}
\caption{LAT calorimeter module.  
The 96 CsI(Tl) scintillator crystal detector elements are arranged in 8 layers, 
with the orientation of the crystals in adjacent layers rotated by $90^\circ$.
The total calorimeter depth (at normal incidence) is 8.6 radiation lengths.
}
\label{f2.4}
\end{figure}

The primary purposes of the calorimeter are twofold: (i) to measure
the energy deposition 
due to the electromagnetic
particle shower
that results from the $e^+e^-$ pair produced by the incident photon;
and (ii) image the shower
development profile, thereby providing an important background
discriminator and an estimator of the shower energy leakage
fluctuations.  Each calorimeter module has 96 CsI(Tl) crystals, with
each crystal of size $2.7\ {\rm cm} \times 2.0\ {\rm cm} \times 32.6\ {\rm cm}$.
The crystals are
optically isolated from each other and are arranged horizontally in 8
layers of 12 crystals each.  The total vertical depth of the
calorimeter is 8.6 radiation lengths (for a total instrument depth of
10.1 radiation lengths).  Each calorimeter module layer is aligned $90^\circ$
with respect to its neighbors, forming an $x,y$ (hodoscopic) array
\citep{carson1996}.
Figure~\ref{f2.4} shows schematically the configuration of a calorimeter
module and Table~\ref{t2.2} is a summary of key parameters of the
calorimeter.

\placefigure{f2.4}
\placetable{t2.2}

The size of the CsI crystals is a compromise between electronic
channel count and desired segmentation within the calorimeter.  The
lateral dimensions of the crystals are comparable to the CsI radiation
length (1.86 cm) and Moli\`ere radius (3.8 cm) for electromagnetic
showers. 
Each CsI crystal provides 3 spatial coordinates for the energy
deposited within: two discrete coordinates from the physical location
of the crystal in the array and the third, more precise, coordinate
determined by measuring the light yield asymmetry at the ends of the
crystal along its long dimension.  This level of segmentation is sufficient to allow spatial
imaging of the shower and accurate reconstruction of its direction. The calorimeter's shower imaging
capability and depth enable the high-energy reach of the LAT and
contribute significantly to background rejection.  In particular, the
energy resolution at high energies is achieved through the application
of shower leakage corrections.

Each crystal element is read out by PIN photodiodes, mounted on both
ends of the crystal, which measure the scintillation light that is
transmitted to each end.  The difference in light levels provides a
determination of the position of the energy deposition along the CsI
crystal.  There are two photodiodes at each end of the crystal, a
large photodiode with area 147 mm$^2$ and a small photodiode with area 25
mm$^2$, providing two readout channels to cover the large dynamic range
of energy deposition in the crystal.  The large photodiodes cover the
range 2 MeV -- 1.6 GeV, while the small photodiodes cover the range
100 MeV -- 70 GeV.  Each crystal end has its own front-end electronics
and pre-amplifier electronics assembly.  Both low and high energy signals
go through a pre-amplifier and shaper and then a pair of Track and
Hold circuits with gains differing nominally by a factor of eight.  An
energy domain selection circuit routes the best energy measurement
through an analog multiplexer to an Analog to Digital Converter.  A
calibration charge injection signal can be fed directly to the front
end of the pre-amplifiers.

\begin{figure}[!tbhp]
\centerline{
\includegraphics[width=3.5in]{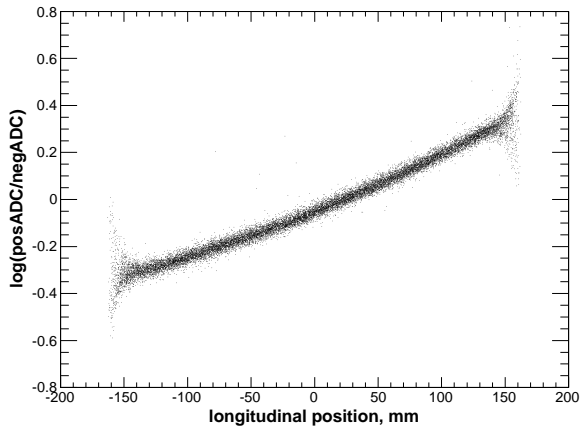}
}
\caption{Light asymmetry measured in a typical calorimeter 
crystal using sea level muons.  The light asymmetry is defined as the 
logarithm of the ratio of the outputs of the diodes at opposite ends 
of the crystal.
The width of the distribution at each position is attributable to the 
light collection statistics at each end of the crystal for the $\sim$11 MeV 
energy depositions of vertically incident muons used in the analysis.  
This width scales with energy deposition as $E^{-1/2}$.
%
}
\label{f2.5}
\end{figure}

The position resolution achieved by the ratio of light seen at each
end of a crystal scales with the deposited energy and ranges from a
few millimeters for low energy depositions ($\sim$10 MeV) to a fraction of
a millimeter for large energy depositions ($>$1 GeV). Simple analytic
forms are used to convert the light asymmetry into a position (see
Figure~\ref{f2.5}).

\begin{figure}[!tbhp]
\centerline{
\includegraphics[width=6.0in]{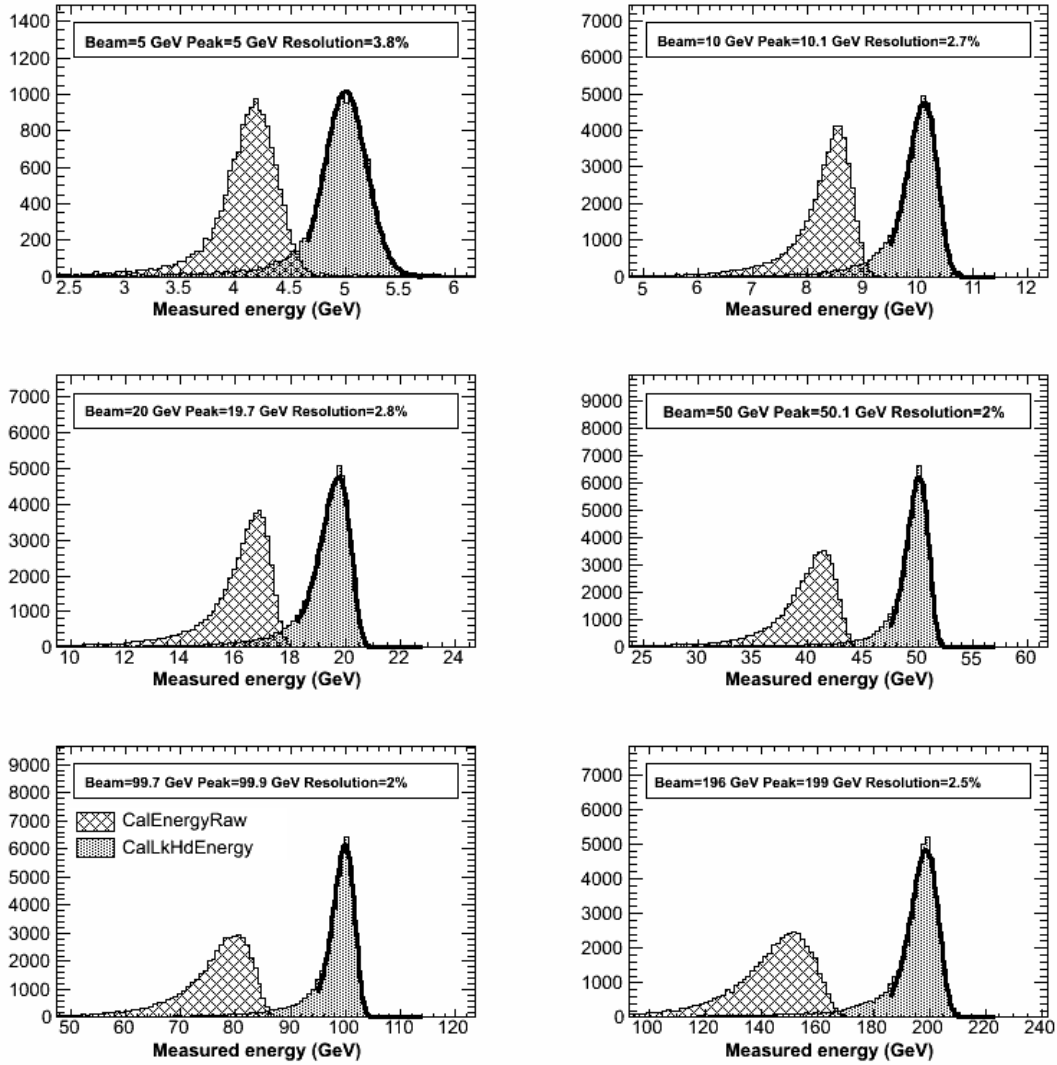}
}
\caption{Energy resolution as a function of electron energy as 
measured with the LAT calibration unit in CERN beam tests.  Each 
panel displays a histogram of the total measured energy (hatched 
peak) and the reconstructed energy (solid peak), 
using the LK method, 
at beam 
energies of 5, 10, 20, 50, 99.7 and 196 GeV, respectively.  The beams 
entered the calibration unit at an angle of $45^\circ$ to the detector 
vertical axis.  As long as shower maximum is within the calorimeter, the 
energy measurement and resolution are considerably improved
by the energy reconstruction algorithms.  The 
measured energy resolutions ($\Delta E/E$) are indicated in the figure.
}
\label{f2.6}
\end{figure}

Although the calorimeter is only 8.6 radiation lengths deep, the
longitudinal segmentation enables energy measurements up to a TeV.
From the longitudinal shower profile, an unbiased estimate of the
initial electron energy is derived by fitting the measurements to an
analytical description of the energy-dependent mean longitudinal
profile. Except at the low end of the energy range, the resulting
energy resolution is limited by fluctuations in the shower leakage.
The effectiveness of this procedure was evaluated in beam tests with
the flight-like calibration unit at CERN.
Figure~\ref{f2.6} shows the
measured energy loss and the leakage-corrected energy loss in the
calorimeter for electron beams of various energies.  Further details
of the calorimeter are in \citet{Grove2008}, \citet{Johnson2001}, 
and \citet{Ferreira2004}. Details of the
energy reconstruction are discussed in \S\ref{s2.4.2}.

\placefigure{f2.5}
\placefigure{f2.6}

\subsubsection{Anticoincidence detector}\label{s2.2.3}

The purpose of the ACD is to provide charged-particle background
rejection; therefore its main requirement is to have high detection
efficiency for charged particles.  The ACD is required to provide at
least 0.9997 efficiency (averaged over the ACD area) for detection of
singly charged particles entering the field-of-view of the LAT.

The LAT is designed to measure \gray{s} with energies up to at least
300 GeV.  The requirement to measure photon energies at this limit
leads to the presence of a heavy calorimeter ($\sim$1800 kg) to absorb
enough of the photon-induced shower energy to make this measurement.
The calorimeter mass itself, however, creates a problem we call the
backsplash effect: isotropically distributed secondary particles
(mostly 100--1000 keV photons) from the electromagnetic shower created
by the incident high-energy photon can Compton scatter in the ACD and
thereby create false veto signals from the recoil electrons.  This
effect was present in EGRET, where the instrument detection efficiency
above 10 GeV was a factor of at least two or more lower than at 1 GeV
due to false vetoes caused by backsplash.  
A design
requirement was established that vetoes created by backsplash
(self-veto) would reject not more than 20\% of otherwise accepted
photons at 300 GeV.  To suppress the backsplash effect, the ACD is
segmented so that only the ACD segment nearby the incident candidate
photon may be considered, thereby dramatically reducing the area of
ACD that can contribute to backsplash \citep{Moiseev2004}.  In
addition, the onboard use of the ACD veto signals is disengaged when
the energy deposition in the calorimeter is larger than an adjustable
preset energy (10 to 20 GeV). Such events are subsequently analyzed
using more complex software than can be implemented on board.

\begin{figure}[!tbhp]
\centerline{
\includegraphics[width=3.5in]{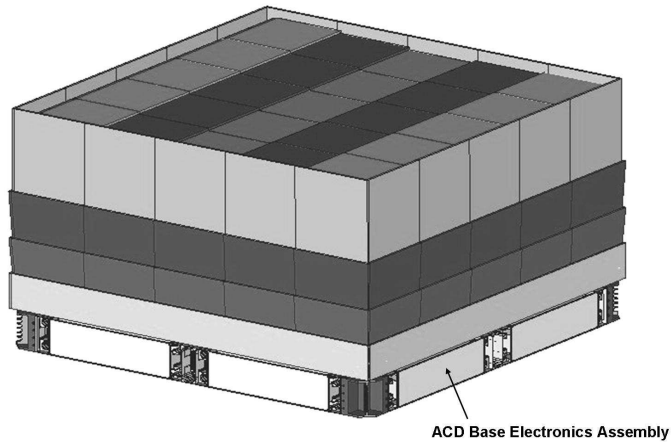}
}
\caption{LAT Anticoincidence Detector (ACD) design.  The ACD 
has a total of 89 plastic scintillator tiles with a $5\times5$ 
array on the top and 16 tiles on 
each of the 4 sides.  Each tile is readout by 2 photomultipliers 
coupled to wavelength shifting fibers embedded in the scintillator.
The tiles overlap in one dimension to minimize gaps between tiles.  
In addition, 2 sets of 4, 
scintillating fiber ribbons are used to cover the remaining gaps.
The ribbons, which are under the tiles, run up the side, across the top, and down
the other side. Each ribbon is
readout with photomultipliers at both ends.
%
}
\label{f2.7}
\end{figure}

Numerous trade studies and tests were performed in order to optimize
the ACD, resulting in the design shown schematically in Figure~\ref{f2.7}.
Plastic scintillator tiles were chosen as the most reliable,
efficient, well-understood, and inexpensive technology, with
much previous use in space applications.  Scintillation light from
each tile is collected by wavelength shifting fibers (WLS) that are
embedded in the scintillator and are coupled to two photomultiplier
tubes (PMTs) for redundancy.  This arrangement provides uniformity of
light collection that is typically better than 95\% over each detector
tile, only dropping to $>$75\% within 1--2 cm of the tile edges.  Overall
detection efficiency for incident charged particles is maintained by
overlapping scintillator tiles in one dimension.  In the other
dimension, gaps between tiles are covered by flexible scintillating
fiber ribbons with $>$90\% detection efficiency.

\placefigure{f2.7}

To minimize the chance of light leaks due to penetrations of the
light-tight wrapping by micrometeoroids and space debris, the ACD is
completely surrounded by a low-mass  micrometeoroid shield (0.39 g
cm$^{-2}$).

All ACD electronics and PMTs are positioned around the bottom
perimeter of the ACD, and light is delivered from the tiles and WLS
fibers by a combination of wavelength-shifting and clear fibers.  The
electronics are divided into 12 groups of 18 channels, with each group
on a single circuit board.  Each of the 12 circuit boards is
independent of the other 11, and has a separate interface to the LAT
central electronics.  The PMTs associated with a single board are
powered by a High Voltage Bias Supply (HVBS), with redundant HVBS for
each board. 
The tile readout has two thresholds: 
an onboard threshold of about 0.45 MIP for the initial rejection of charged 
particles, and a ground analysis threshold of about 0.30 MIP for the final analysis.

Further details of the ACD design, fabrication, testing, and
performance are given by \citet{Moiseev2007}.  Table~\ref{t2.3} is a
summary of key parameters of the LAT ACD.

\placetable{t2.3}

\subsubsection{Data acquisition system (DAQ) and trigger}\label{s2.2.4}

\begin{figure}[!tbhp]
\centerline{
\includegraphics[width=3.5in]{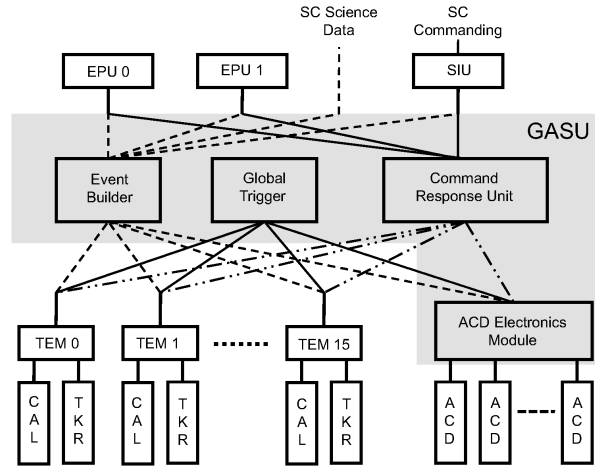}
}
\caption{LAT Data Acquisition System (DAQ) architecture.  
The Global-trigger/ACD-module/Signal distribution Unit (GASU) consists of the ACD 
Electronics Module, the Global Trigger Module (GTM), the Event Builder Module (EBM), 
and the Command Response Unit (CRU).  The trigger and data readout from each of the 
16 pairs of tracker and calorimeter modules is supported by a Tower Electronics 
Module (TEM).  There are two primary Event Processing Units (EPU) and one primary 
Spacecraft Interface Unit (SIU).  Not shown on the diagram are the redundant units 
(e.g.\ 1 SIU, 1 EPU, 1 GASU).
}
\label{f2.8}
\end{figure}

The Data Acquisition System (DAQ) collects the data from the other
subsystems, implements the multi-level event trigger, provides
on-board event processing to run filter algorithms to reduce the
number of downlinked events, and provides an on-board science analysis
platform to rapidly search for transients.  The DAQ architecture is
hierarchical as shown in Figure~\ref{f2.8}.  At the lowest level shown, each
of 16 Tower Electronics Modules (TEMs) provides the interface to the
tracker and calorimeter pair in one of the towers.  Each TEM generates
instrument trigger primitives from combinations of tower subsystem
(tracker and calorimeter) triggers, provides event buffering to
support event readout, and communicates with the instrument-level
Event Builder Module that is part of the
Global-trigger/ACD-module/Signal distribution Unit (GASU).

\placefigure{f2.8}

The GASU consists of (i) the Command Response Unit (CRU) that sends
and receives commands and distributes the DAQ clock signal, (ii) the
Global-Trigger Electronics Module (GEM) that generates LAT-wide
readout decision signals based on trigger primitives from the TEMs and
the ACD, (iii) the ACD Electronics Module (AEM) that performs tasks,
much like a TEM, for the ACD, and (iv) the Event Builder Module (EBM)
that builds complete LAT events out of the information provided by the
TEMs and the AEM, and sends them to dynamically selected target Event
Processor Units (EPUs).  

There are two operating EPUs to support
on-board processing of events with filter algorithms designed to
reduce the event rate from 2--4 kHz to $\sim$400 Hz that is then downlinked
for processing on the ground.  The on-board filters are optimized to
remove charged particle background events and maximize the rate of
\gray{} triggered events within the total rate that can be downlinked.
Finally, the Spacecraft Interface Unit (SIU) controls the LAT and
contains the command interface to the spacecraft.  Each EPU and SIU
utilizes a RAD750 Compact PCI Processor which, when operating at 115.5
MHz, provides 80 to 90 MIPS.  The instrument flight software runs only on
the EPUs and the SIU.  The TEMs and the GASU hardware have
software-controlled trigger configuration and mode registers.

Not shown in Figure~\ref{f2.8} is the redundancy of the DAQ system or the LAT's
Power Distribution Unit (PDU).  There are two primary EPUs and one
redundant EPU, one primary SIU and one redundant SIU, and one primary
GASU and one redundant GASU.   The PDU, which is also redundant,
controls spacecraft power to the TEMs, the GASU, and the EPUs.  The
feeds from the spacecraft to the PDU are fully cross-strapped.  In
turn, the TEMs control power to the tracker and the calorimeter
modules and the GASU controls power to the ACD.  Power to the SIUs is
directly provided by the spacecraft.

An instrument-level Trigger Accept Message (TAM) signal is issued by
the GEM only if the GEM logic is satisfied by the input trigger
primitives within the (adjustable) trigger window width.  The TAM
signal is sent to each TEM and to the AEM with no delays.  Upon
receipt of the TAM signal, a Trigger Acknowledge (TACK) signal with an
adjustable delay is sent by the TEM to the tracker front-ends and a
command, also with an adjustable delay, is sent to the calorimeter
front-ends.  The AEM sends a signal to the ACD front-ends.  The TACK
causes the entire instrument to be read out (e.g., addresses of hit
strips in the tracker and TOT for each layer in each tracker module,
and pulse heights for all 3,072 calorimeter channels and 216 ACD
channels).  Any of the TEMs or the AEM can issue a trigger request to
the GEM.  The time between a particle interaction in the LAT that
causes an event trigger and the latching of the tracker discriminators
is 2.3 to 2.4 $\mu$s, much of this delay due to the analog rise times in
the tracker front-end electronics.  Similarly, the latching of the
analog sample-and-holds for the calorimeter and the ACD are delayed
(programmable delay of $\sim$2.5 $\mu$s) until the shaped analog signals peak.

The minimum instrumental dead time per event readout is 26.50 $\mu$s and
is the time required to latch the trigger information in the GEM and
send it from the GEM to the EBM.  The calorimeter readout can
contribute to the dead time if the full four-range CAL readout is
requested.  During readout of any of the instrument, any TEM and the
AEM send a ``busy'' signal to the GEM.  From these signals, the GEM
then generates the overall dead time and the system records this
information and adds it to the data stream transmitted to the ground.

Any of the TEMs can generate a trigger request in several ways:  (i)
If any tracker channel in the tracker module is over threshold, a
trigger request is sent to the module's TEM which then checks if a
trigger condition is satisfied, typically requiring triggers from 3
$x,y$ planes in a row.  If this condition is satisfied, the TEM sends a
trigger request to the GEM.  (ii) If a predetermined low-energy
(CAL-LO) or high-energy (CAL-HI) threshold is exceeded for any crystal
in the calorimeter module, a trigger request is sent to the GEM.

The prompt ACD signals sent to the GEM are of two types: (i) a
discriminated signal (nominal 0.4 MIPs threshold) from each of the 97
scintillators (89 tiles and 8 ribbons) of the ACD, used to
(potentially) veto tracker triggers originating in any one of the
sixteen towers, and (ii) a high-level discriminated signal (nominal 20
MIPs threshold) generated by highly ionizing heavy nuclei cosmic-rays
(carbon-nitrogen-oxygen or CNO). The high-level CNO signal is used as
a trigger, mostly for energy calibration purposes.  During ground
testing the CNO signal is only tested through charge injection.  In
addition, the GEM can logically group tiles and ribbons to form
Regions Of Interest (ROIs) for trigger/veto purposes.  An ROI can be
defined as any combination of the ACD tiles and ribbons.  Up to 16
ROIs can be defined through a series of configuration registers.  The
ROI signal is simply whether any one of the tiles 
that define the ROI is asserted.

Finally, non-detector based trigger inputs to the GEM are used for
calibration and diagnostic purposes. The GEM can utilize (i) a
periodic signal derived from either the instrument system clock
(nominally running at 20 MHz) or the 1 pulse-per-second GPS spacecraft
clock (accurate to $\pm$1.5 $\mu$s), and (ii) a solicited trigger signal
input that allows the instrument to be triggered through operator
intervention.  The spacecraft clock is also used to strobe the
internal time base of the GEM, thus allowing an accurate measurement
of the time of an event relative to the spacecraft clock.

\subsection{Instrument modeling}\label{s2.3}

The development and validation of a detailed Monte Carlo simulation of
the LAT's response to signals (\gray{s}) and backgrounds (cosmic-rays,
albedo \gray{s}, etc.) has been central to the design and optimization
of the LAT.  This approach was particularly important for showing that
the LAT design could achieve the necessary rejection of backgrounds
expected in the observatory's orbit.  The instrument simulation was
also incorporated into an end-to-end simulation of data flow, starting
with an astrophysical model of the \gray{} sky, used to support the
pre-launch development of software tools to support scientific data
analysis.

\begin{figure}[!tbhp]
\centerline{
\includegraphics[width=3.5in,angle=270]{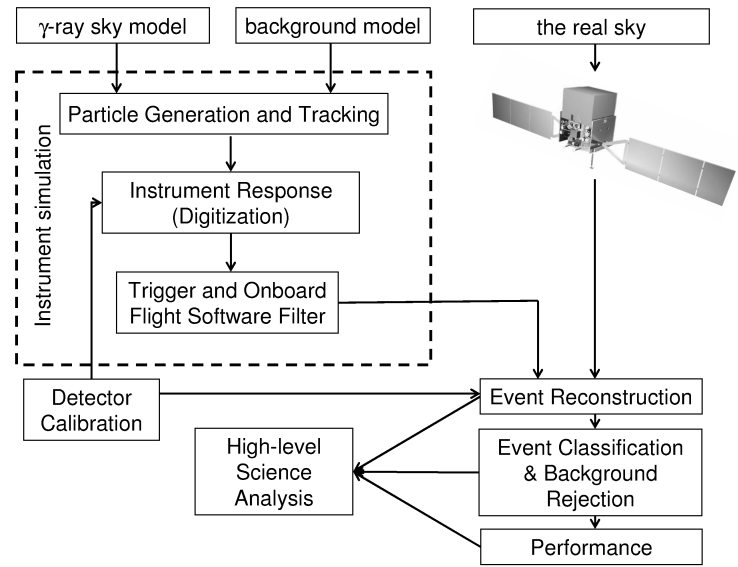}
}
\caption{Components of the instrument
simulation, calibration, and data analysis. 
}
\label{fX}
\end{figure}

Figure~\ref{fX} summarizes the various components of the instrument
simulation, calibration, and data analysis.  The instrument simulation
consists of 3 parts: (i) particle generation and tracking uses
standard particle physics simulators of particle interactions in
matter to model the physical interactions of \gray{s} and background
particle fluxes incident on the LAT.  In particular, the
simulation of events in the LAT is based on the Geant4 (G4) Monte
Carlo toolkit \citep{Agostinelli2003,Allison2006}, an object-oriented
simulator of the passage of particles through matter. G4 provides a
complete set of tools for detector modeling.  In the LAT application,
the simulation is managed by Gleam, our implementation of the Gaudi
software framework \citep{Barrand2001}, and so we use only a subset of
the G4 tools. 
(ii) For a given simulated event the instrument response
(digitization) is calculated parametrically based on the energy
deposition and location in active detector volumes in the
anticoincidence detector, tracker, and calorimeter.  (iii) From the
digitized instrument responses, a set of trigger primitives are
computed and a facsimile of the Trigger and On-board Flight Software
Filter (see \S\ref{s2.2.4}) is applied to the simulated data stream.  Events
that emerge from the instrument simulation (or real data) then undergo
event reconstruction and classification (\S\ref{s2.4}), followed by background
rejection analysis (\S\ref{s2.4.3}).  As discussed in \S\ref{s2.4.3}, the background
rejection can be tuned depending on the analysis objectives.

Information about the detector geometry and materials is stored in a
set of structured XML files.  These files are used by Gleam to build a
G4 representation of the detector (and also to provide information
about the detector to our event reconstruction packages).  The
geometry is quite detailed, particularly for the active elements,
namely, the tracker silicon strip detectors, CsI
crystals
and diodes of the calorimeter, and anticoincidence detector
scintillator tiles and ribbons.  The current implementation has about
54,000 volume elements, of which about 34,000 are active.

G4 contains a full suite of particle interactions with matter,
including multiple scattering and delta-ray production for charged
particles, pair production and Compton scattering for photons, and
bremsstrahlung for $e^-$ and $e^+$, and low-energy interaction
with atoms, as well as several models of hadronic interactions.  The
set of processes implemented is controlled by a ``physics list,''
which allows for considerable flexibility.  In fact, a special version
of the model of multiple scattering is used to provide better agreement
with our measured data.

Detector calibration data 
(thresholds, gains, non-uniformities,
etc.) are used to convert the energy deposited in the active elements
to instrument signals.  For the tracker, dead channels are removed
from the data at this stage, as well as any signals which would have
overflowed the electronic buffers. (These same effects are taken into
account again during event reconstruction, to aid the pattern
recognition.)

\subsection{Event reconstruction and classification}\label{s2.4}

The event reconstruction processes the raw data from the various
subsystems, correlating and unifying them under a unique event
hypothesis.  The development of the reconstruction relies heavily on
the Monte Carlo simulation of the events.  In the following
subsections, the basic blocks of the reconstruction are described.  We
start with track reconstruction, as it is key to developing the
subsequent analysis of the other systems: the found tracks serve as
guides as to what should be expected in both the calorimeter as well
as the ACD for various event types.
The analogous reconstruction processing for EGRET, a spark-chamber
pair conversion telescope, which did not benefit from a detailed Monte
Carlo model of the instrument, is described in \citet{Thompson1993}.

\subsubsection{Track reconstruction}\label{s2.4.1}

Spatially adjacent hit tracker strips are grouped together, forming
clusters, and the coordinates of these clusters are used in the track
finding and fitting.   Each cluster determines a precise location in $z$
as well as either $x$ or $y$.  Because the planes of silicon detectors are
arranged in closely spaced orthogonal pairs, both the $x$ and $y$
determinations can be made, albeit the choice of tracker technology
(single-sided silicon strip detectors) imposes the ambiguities
associated with projective coordinate readout on the initial pairing
of the $x$ and $y$ coordinates when 2 or more particles pass through a
detector plane.  This ambiguity is resolved for tracks associated with
particles that pass through more than one tracker module.  For events
with tracks confined to one module, the coordinate-pairing ambiguity
is resolved for $\sim$90\% of these events using calorimeter information.
Strictly, resolution of the coordinate-pairing ambiguity is only of
secondary importance, having primarily to do with background
rejection.

At the heart of track-finding algorithms is a mechanism to generate a
track hypothesis.  A track hypothesis is a trajectory (location and
direction) that can be rejected or accepted based on its consistency
with the sensor readouts.  The generation algorithm is combinatoric,
with a significant constraint imposed on the number of trial
trajectories considered because of the available computing power.  Two
algorithms, described below, are used.

\emph{Calorimeter-Seeded Pattern Recognition} (CSPR): For most of the LAT
science analysis, some energy deposition in the calorimeter is
required.  When present, both the centroid and shower axis of the
calorimeter energy deposition can be computed using a moments analysis
(see \S\ref{s2.4.2}) in most cases.   The first and most-often selected
algorithm is based on the assumption that the energy centroid lies on
the trajectory.  The first hit on the hypothesized track, composed of
an $x,y$ pair from the layer in the tracker furthest from the
calorimeter, is selected at random from the possible $x,y$ pairs.   If a
subsequent hit is found to be close to the line between the first hit
and the location of the energy centroid in the calorimeter, a track
hypothesis is generated.   The candidate track is then populated with
hits in the intervening layers using an adaptation of Kalman fitting
\citep[e.g.,][]{Fruhwirth2000}.  The process starts from the first hit.
A linear projection is made into the next layer.  The covariance
matrix is also propagated to the layer and provides an estimate of the
error ellipse that is searched for a hit to add to the track.  The
propagation of the covariance matrix includes the complete details of
the material crossed, thereby providing an accurate estimate of the
error caused by multiple scattering.  If a candidate hit exists in the
layer, it is incorporated into the trajectory weighted by the
covariance matrices.   The procedure is then iterated for subsequent
layers, allowing for missing hits in un-instrumented regions.  Adding
more hits to the track is terminated when more than a specified number
of gaps (planes without hits associated with the track) have
accumulated (nominally 2).  The whole process is repeated, starting
with each possible $x,y$ pair in the furthest plane from the calorimeter
and then continued using pairs from closer layers.   After a track of
sufficient quality is found and at least two layers have been looped
over, the process is terminated.

A byproduct of this process is the first Kalman fit to the track,
providing the $\chi^2$, the number of hits, the number of gaps, etc.  From
these quantities a track quality parameter is derived and used to
order the candidate tracks from ``best'' to ``worst''.

At high energies ($>$1 GeV)  the first-hit search is limited to a cone
around the direction provided by the calorimeter moments analysis in
order to minimize confusion with hits caused by secondary particles
generated by backsplash.  The cone angle is narrowed as the energy
increases, reflecting the improved directional information provided by
the calorimeter.

Following the completion of the CSPR, only the ``best'' track found is
retained.   The biasing caused by the track quality parameters makes
this ``the longest, straightest track'' and hence, for $\gamma$
conversions, preferentially the higher-energy track of the $e^+e^-$ pair.
The other tracks are discarded.  The hits belonging to the best track
are flagged as ``used'' and a second combinatoric algorithm is then
invoked.

\emph{Blind Search Pattern Recognition} (BSPR):  In this algorithm,
calorimeter information is not used for track finding.   Events having
essentially no energy deposition in the calorimeter are analyzed using
this algorithm as well as for subsequent track finding following the
stage detailed above.   The same procedure described for the CSPR is
used, but here the selection of the second hit used to make the
initial trajectory is now done at random from the next closest layer
to the calorimeter.  The trajectory formed by these two hits is
projected into the following layer and if a hit in that layer lies
sufficiently close to the projection a trial track is generated.
The mechanism of populating the track candidate with hits follows that
used in the CSPR, but without any estimation of the energy of the
track, the multiple scattering errors are set by assuming a minimum
energy (default: 30 MeV).   Hits are allowed to be shared between
tracks if the hit is the first hit on the best track (two tracks
forming a vertex) or if the cluster size (number of strips) is larger
than expected for the track already assigned to that hit.  The total
number of tracks allowed to be found is limited (default: 10).

The final track fits must await an improved energy estimate to be made
using the best track to aid in estimating the fraction of energy
deposited in the calorimeter (see \S\ref{s2.4.2}).   Once this is done, the
energy is apportioned between the first two tracks according to the
amount of multiple scattering observed on each.  A subsequent Kalman
fit is done but without re-populating the tracks with hits.

The final stage of track reconstruction combines tracks into vertices.
The process begins with the best track.   The second track is selected
by simply looping over the other tracks in the event.   The distance
of closest approach between the best track and the candidate second
track is computed and if within a specified distance (default: 6 mm)
a vertex solution is generated by covariantly combining the parameters
of the two tracks.  The $z$-axis location (coordinate along the
instrument axis) of the vertex candidate is selected using the
detailed topology of the first hits and is assigned either to be in
the center of the preceding tungsten foil radiator, in the silicon
detector itself, or within the core material of the tracker tray
directly above the first hit.  A quality parameter is created taking
into account the $\chi^2$ for the combination of tracks, the distance of
closest approach, etc.  The first track is paired with the second
track having the best quality parameter.  These tracks are marked as
``used'' and the next unused track is selected and the process
repeated.   If a track fails to make a satisfactory vertex it is
assigned to a vertex by itself.   Thus all tracks are represented by a
vertex.

In addition to the ``standard'' vertexing discussed above, an
additional improvement is possible if calorimeter information is
included.  In events where either during the conversion process or
immediately thereafter much of the energy is in \gray{s} (due to
Bremsstrahlung or radiative corrections), the charged tracks can point
well away from the incident \gray{} direction. However the location
of the conversion point is usually well determined and, when combined
with the energy centroid location in the calorimeter, can give a fair
estimate of the direction.  The ``best'' track as well as the first
vertex are combined covariantly with this direction using weights to
apportion the total energy between these directions.  These ``neutral
energy'' solutions result in significantly reducing the non-gaussian
tails of the PSF.

\subsubsection{Energy reconstruction}\label{s2.4.2}

Energy reconstruction begins by first applying the appropriate
pedestals and gains to the raw digitized signals.  Then, for each
calorimeter crystal, the signals from the two ends are combined to
provide the total energy in the crystal (independent of location) and
the position along the crystal where the energy was deposited.  The
result is an array of energies and locations.

The three-dimensional calorimeter energy centroid is computed along
with energy moments (similar to the moment of inertia, but with energy
in place of mass).  The shower direction is given by the eigenvector
with the smallest eigenvalue.   Initially, the overall energy is taken
to be the sum of the crystal energies (``CALEnergyRaw'' in Figure~\ref{f2.6}).  
Further improvements must await the completion of the fitted
tracks.

The trajectory provided by the best track (or best track vertex when
available) is used as input to estimate the energy correction
necessary to account for leakage out the sides and back of the
calorimeter and through the internal gaps between calorimeter modules.
Three different algorithms are applied to each event: a parametric
correction (PC)
based on the barycenter of the shower, 
a fit to the shower profile (SP)
taking into account the longitudinal and transverse development of the shower, and 
a maximum likelihood (LK) fit based on the correlations of the overall 
total energy deposited with the number of hits in the tracker and with 
the energy seen in the last layer.
%
%
Because the SP method starts to work beyond 1 GeV and the LK method
works below 300 GeV, only the PC method covers the entire phase space
of the LAT.  Figure~\ref{f2.6} shows the results of the LK method applied to
data obtained with electron beams at CERN entering the LAT calibration
unit at an angle of 45$^\circ$ to the detector vertical axis. The energy
resolutions obtained vary between 4\% at 5 GeV and 2\% at 196 GeV.

At low energy ($\sim$100 MeV), a significant fraction ($\sim$50\%) of
the energy in a $\gamma$ conversion event can be deposited in the
tracker and hence the determination of this contribution to the total
energy becomes important.  For this purpose the tracker is considered
to be a sampling calorimeter where the number of hit silicon strips in
a tracker layer provides the estimate of the energy deposition at that
depth.   The total number of hits in the thin radiator section, the
thick radiator section and the non-radiator last layers is computed
within a cone with an opening angle which decreases as $E^{-1/2}$,
where $E$ is the apparent energy in the calorimeter.   The ``tracker''
energy is added to the corrected calorimeter energy.

Because the PC method gives an energy estimate for all events, it is
used to iterate the Kalman track fits as mentioned in \S\ref{s2.4.1}.

\subsubsection{Background rejection}\label{s2.4.3}

The vast majority of instrument triggers and subsequently downlinked
data are background events caused by charged particles as well as
earth albedo \gray{s}.  The task of the hardware trigger is to
minimize their effects on the instrumental deadtime associated with
reading out the LAT.  Subsequently the task of the onboard filter is
to eliminate a sufficient number of background events without
sacrificing celestial \gray{} events such that the resulting data can
be transmitted to the ground within the available bandwidth.   The
final task is for the analysis on the ground to distinguish between
background events and \gray{} events and minimize the impact of
backgrounds on \gray{} science.  The combination of these 3 elements
reduces the background by a factor of almost $10^6$ while preserving
efficiency for \gray{s} exceeding 75\%.  For reference, the average
cosmic \gray{} event rate in the LAT is $\sim$2 Hz.

\ifthenelse{\equal{\ms}{preprint}}
{
\setcounter{subsubsubsectioncounter}{1}
}{}
\subsubsubsection{Background model}\label{s2.4.3.1}

In order to facilitate the development of the on-board triggering and
filtering and subsequent event reconstruction and classification
algorithms, a model of the background the LAT encounters in space
has been developed.

As shown in Table~\ref{t2.4}, the background model includes cosmic rays and
earth albedo \gray{s} within the energy range 10 MeV to $10^6$ MeV.  Any
particles that might either make non-astrophysical \gray{s} and/or
need to be rejected as background are included.  The model does not
include X-rays or soft \gray{s} that might cause individual
detectors within the LAT to be activated.  The model is meant to be
valid outside the radiation belts and the South Atlantic Anomaly (SAA); no
particle fluxes from inside the radiation belts are included. The
boundaries of the belts are defined to be where the flux of trapped
particles is 1 proton cm$^{-2}$ s$^{-1}$ ($E > 10$ MeV).
LAT does not take data inside the SAA. The fraction of time spent in the SAA is 14.6\%.

\placetable{t2.4}

The AMS \citep{Aguilar2002} and BESS \citep{Haino2004} experiments
provided important and accurate new measurements of the spectra of the
protons and alpha particles, the most abundant of the various galactic
cosmic-ray (GCR) components.  AMS made detailed latitude-dependent
measurements of the splash and reentrant albedo particles ($e^+, e^-$ and
protons) in the energy range from $\sim$150--200 MeV up to the cutoff
energies where the earth albedo components become lost in the much
greater GCR fluxes.  These fluxes will be updated with results from
the Pamela satellite \citep{Picozza2007}.

For albedo fluxes of particles with energies below $\sim$150 MeV,
inaccessible to the AMS and other large instruments, measurements made
by NINA and NINA-2 and a series of Russian satellite experiments with
an instrument known as Mariya are used.  The albedo \gray{} fluxes are
taken from a reanalysis of the data collected by EGRET when the CGRO
satellite was pointed at the Earth.

The model is based on empirical fits to the referenced data.  No time
variability is included.  The GCR fluxes are taken to be the same as
those observed near solar minimum (maximum GCR intensities).  The
albedo fluxes may vary with time and be correlated with the GCR
fluxes.  The fluxes as observed by the NINA and Mariya experiments are
used without correcting them for solar cycle variations.
While an East-West cutoff variation was included that affects galactic 
cosmic ray components, all fluxes except albedo protons are assumed to be isotropic.  
The measurements are not complete enough for us to be able to 
account for variation in parameters such as the zenith angle of the
particles 
or
their pitch angles with respect to the local field.  We
have attempted to model some 
the zenith angle dependence for albedo protons, based not on
measurements, but on modeling of the albedo \citep{Zuccon2003}.
Further verification and improvement to 
the model 
are being done on orbit.

\begin{figure}[!tbhp]
\centerline{
\includegraphics[width=3.5in,angle=270]{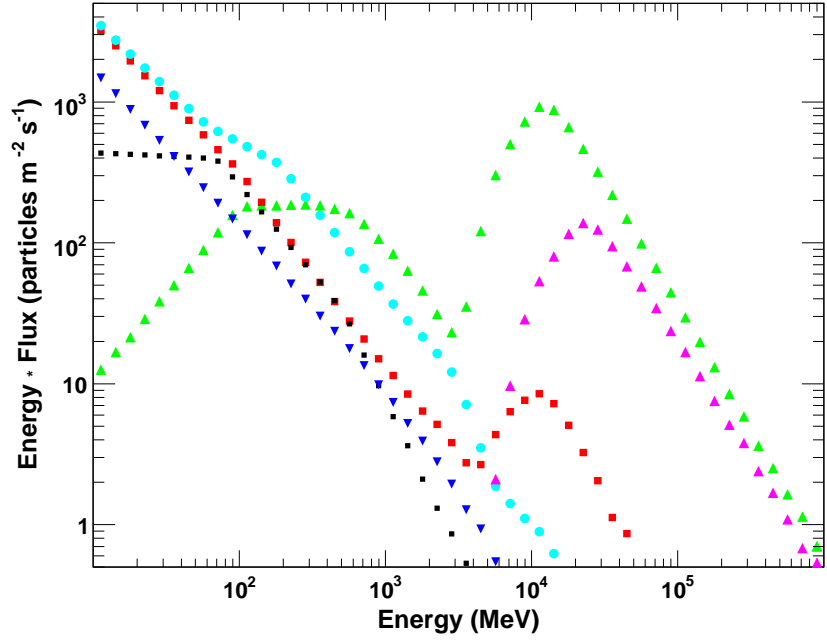}
}
\caption{Orbit averaged background fluxes of the various components
incident on the LAT
used in the background model.  The fluxes are shown as a function
of total kinetic energy of the particles:  protons (green filled
triangles up), He (purple filled triangles up), electrons (filled red
squares), positrons (light blue squares), Earth albedo neutrons (black
squares), and Earth albedo \gray{s} (dark blue filled triangles
down).  The effect of geomagnetic cutoff is seen at 3 GeV for protons
and electrons, and at higher energy for helium nuclei.  At low
energies the curves show the sum of re-entrant and splash albedo for
electrons and positrons.  
%
}
\label{f2.9}
\end{figure}

The orbit averaged background fluxes in the model are shown in
Figure~\ref{f2.9}.  For charged particles, these fluxes are integrated
over solid angle. It is straightforward to obtain fluxes per unit
solid angle. For galactic cosmic ray components, divide by 8.7 sr,
the solid angle of the visible sky that is not blocked by the Earth at
\Fermi{'s} orbital altitude.  For the albedo components we have taken the
reentrant and albedo fluxes to be the same.

\placefigure{f2.9}

\subsubsubsection{Event classification and background rejection}\label{s2.4.3.2}

After track reconstruction, vertexing, and energy reconstruction, the
events are analyzed to determine the accuracy of the energy
determinations, the directional accuracy, and whether they are \gray{s}.

All of the estimates are based on classification tree (CT) generated
probabilities.   This statistical tool was found to give the highest
efficiency with the greatest purity, exceeding that which we obtained
with either a more traditional cut-based analysis or with neural nets.
Our usage of classification trees involves training a modest number of
trees (a few to $\sim$10) and averaging over the results.  The trees are
``grown'' by minimizing ``entropy'' as defined in statistics \citep{Breiman1984}.

The final energy estimate for each event is made by first dividing the
sample up into subsets according to which energy methods were
reporting results (PC+LF+SP, PC+LF, PC+SP, and PC).   When more than
one energy method is available, the method selected is determined
using a CT.  The probability that the selected energy is better than
the 1$\sigma$ resolution limit is estimated using a second CT.   The subsets
are then merged, now with a single ``best'' energy and a probability
``knob'' that can be used to lessen the presence of tails (both high
and low) in the distribution of reconstructed energies at the expense
of effective area.

The analysis sorts the events according to where they occurred in the
LAT tracker.  (Events in the thick radiator portion have about a
factor of 2 worse angular resolution due to increased multiple
scattering.)   When there is sufficient energy in the calorimeter
(default: $>$10 MeV), the neutral energy solutions are used.  If a
2-track vertex is present, a CT determines whether the vertex derived
direction or the best track direction is used.  As such there are four
basic subsets:  thin and thick radiator events and vertexed and
1-track events.   For each of these subsets the probability that the
reconstructed direction is more accurate than the theoretical 68\%
containment PSF is determined using a CT.   The events are re-merged
now with a ``best'' direction solution and associated CT-based
probability.   This image ``knob'' can be used to limit the long tails
often associated with the PSFs of \gray{} instruments.

The background rejection is by far the most challenging of all the
reconstruction analysis tasks.   This is due to the large phase space
covered by the LAT and the very low signal-to-noise ratio in the
incoming data ($\sim$1:300 for down-linked data).  The first task is to
eliminate the vast majority of the charged particle flux that enters
within the FoV using the ACD in conjunction with the
found tracks.  One cannot simply demand that there are no triggers
from the ACD because high-energy \gray{s} generate a considerable
amount of back splash, from the shower that develops in the
calorimeter, in the form of hard X-rays that can trigger several ACD
tiles.  Consequently only the tiles pointed at by the reconstructed
tracks are used to establish a veto by the presence of a signal in
excess of $\sim$$1/4$ of a minimum ionization event.   Because the accuracy of
the pointing is energy dependent due to multiple scattering, at low
energy, only tiles within the vicinity of the track intersection with
the ACD are used, while at high energy the region is restricted to
essentially the one tile being pointed to.   In addition there are
several areas in the ACD where it is not possible to completely cover
the acceptance region (e.g.\ the four vertical edge corners, the screw
holes used to mount the tiles, etc.).   Since these are known
locations, tracks pointing at them must also be eliminated.   However,
these holes are small and account for a few percent of the surface
area and reduce the events sample by $<$2\%.

The considerations for rejecting backgrounds involve the detailed
topology of the events within the tracker and the overall match of the
shower profile in 3D in both the tracker and the calorimeter.    The
tracker provides a clear picture of the initial event topology.   For
example the identification of a 2-track vertex immediately reduces the
background contamination by about an order of magnitude.   However a
majority of events above 1 GeV don't contain such a recognizable
vertex due to the small opening angle of the $e^+ e^-$ pair along the
incoming \gray{} direction.   The observation of a significant
number of extra hits in close proximity to the track(s) indicates they
are electrons and hence from the conversion of a \gray{} while the
presence of unassociated hits or tracks are a strong indicator of
background.   These as well as other considerations are used for
training background rejection CTs.

The final discriminator of background is the identification of an
electromagnetic shower.   Considerations such as how well the tracker
solution points to the calorimeter centroid, how well the directional
information from the calorimeter matches that of the track found in
the tracker, as well as the width and longitudinal shower profile in
the various layers of the calorimeter, are important in discrimination
of backgrounds.  Again the information from the reconstruction is used
to train CTs and the resulting probability is used to eliminate
backgrounds.

The broad range of LAT observations and analysis, from GRBs to
extended diffuse radiation, leads to different optimizations of the
event selections and different rates of residual backgrounds.  For
example, in analysis of a GRB, the relatively small region of the sky
as well as the very short time window allow the background rejection
cuts to be relaxed relative to an analysis of a diffuse source
covering a large portion of the sky.   Furthermore a key science
attribute for GRB observations is the time evolution and the
sensitivity of a measurement to rapid time variation scales as the
square root of the number of detected burst photons.    The background
rejection analysis has been constructed to allow analysis classes to be
optimized for specific science topics.

Table~\ref{t2.5} lists 3 analysis classes that have been defined based on the
backgrounds expected in orbit, current knowledge of the \gray{} sky,
and the performance of the LAT.  Our 
estimates of LAT
performance are given in terms of these analysis classes.  Common to all
of these analysis classes is the rejection of the charged-particle
backgrounds entering within the field of view.  The classes are
differentiated by an increasingly tighter requirement that the
candidate photon events in both the tracker and the calorimeter behave
as expected for \gray{} induced electromagnetic showers.  The loosest
cuts apply to the Transient class, for which the background rejection
was set to allow a background rate of $<$2 Hz, estimated using the
background model described in \S\ref{s2.4.3.1}, which would result in no more
than one background event every 5 sec inside a $10^\circ$ radius about a
source.  The Source class was designed so that the residual background
contamination was similar to that expected from the extragalactic
\gray{} background flux over the entire field of view.  Finally, the
Diffuse class has the best background rejection and was designed such
that harsher cuts would not significantly improve the signal to noise.
The various analysis cuts and event selections will be
optimized for the conditions found on-orbit during the 1st year all-sky 
survey phase.  Note that these 3 analysis
classes are hierarchical; that is all events in the Diffuse class are
contained in the Source class and all events in the Source class are
in the Transient class.

\begin{figure}[!tbhp]
\centerline{
\includegraphics[width=3.5in]{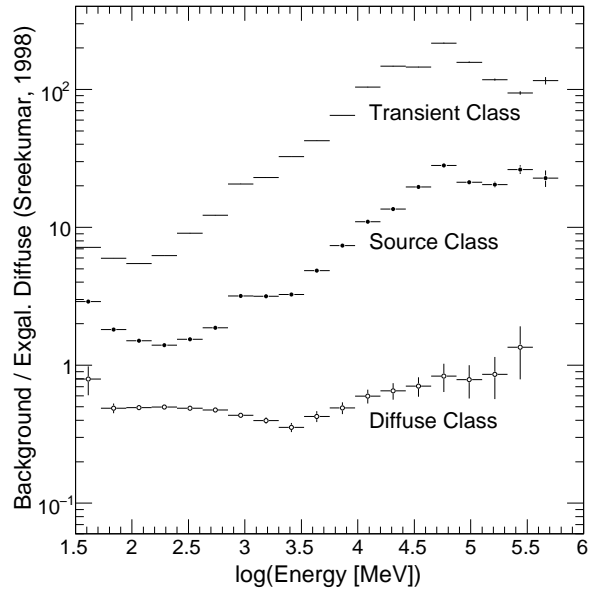}
}
\caption{Ratio of the residual background to the extragalactic diffuse background 
inferred from EGRET observations \citep{Sreekumar1998} for each of the three 
analysis classes.  The integral EGRET diffuse flux is $1.45\times10^{-7}$ ph cm$^{-2}$ 
s$^{-1}$ sr$^{-1}$ above 100 MeV.
}
\label{f2.10}
\end{figure}

The residuals of background events for the 3 analysis classes are shown
in Figure~\ref{f2.10}.  For the Diffuse class, the resulting rejection factor
is $\sim$1:$10^6$ at some energies (e.g., $\sim$10 GeV) while retaining $>$80\%
efficiency for retaining \gray{} events.  The residual background is
worse at low energy particularly for events originating in the thick
radiator portion of the tracker.   It is here that ``splash''
backgrounds, entering the backside of the calorimeter can undergo
interactions that result in low energy particles which range out in
the thick radiators, thus mimicking an event originating in the thick
tracker section.   In a sense the thick section shields the thin
section from this flux and hence the thin section is somewhat cleaner.

\placefigure{f2.10}
\placetable{t2.5}

The leaked background events generally fall into two categories:
irreducible events and reducible events.   The irreducible events are
events in which a background particle interacts in the passive
material outside the ACD or within the first $\sim$1 mm of the ACD
scintillator and the resulting secondaries contain \gray{s} which
enter inside the FoV.  This can happen in the case of entering $e^+$
which annihilate to two photons, entering $e^-$ or $e^+$ which
bremsstrahlung essentially all their energy to a single photon, and
proton interactions that make a $\pi^0$ which decays to 2 photons
with the rest of the secondaries either neutral or aimed away from the
LAT.  In these cases the ACD has no signals and a \gray{} is seen in
the LAT.  There is no way in principle to distinguish and eliminate
these events from the celestial \gray{} signal and this component is
the result of the reality of contemporary instrumentation and the
precautionary measures that must be taken to survive in low earth
orbit.  This irreducible component constitutes $\sim$60\% of the residual
background events with measured energies above 100 MeV.

The reducible background component comprises events that in principle
should be identifiable.   These events leak through the various
filters because they are in the far tails of their parent
distributions, overlapping the \gray{} (signal) distribution.  The
filter parameters are chosen to optimize efficiency for \gray{s}
versus background rejection.  Additional contributions to the
reducible background component come from the fact that any real
detector will have inefficiencies caused by real world design choices
such as gaps in the silicon detector planes of the tracker and in the
ACD.  This reducible component however is easily monitored by
comparing the apparent fluxes of events with and without vertices.
The difference is essentially the reducible component because the
vertexed event sample has 10 times fewer such reducible background
events.

\subsection{Performance of the LAT}\label{s2.5}

The performance of the LAT is basically determined by the design of
the LAT hardware, the event reconstruction algorithms (i.e., the
accuracy and efficiency with which the low-level event information is
used to determine energy and direction), and event selection
algorithms (i.e., the efficiency for identifying well reconstructed
\gray{} events).

\begin{figure}[!tbhp]
\centerline{
\includegraphics[width=3.5in]{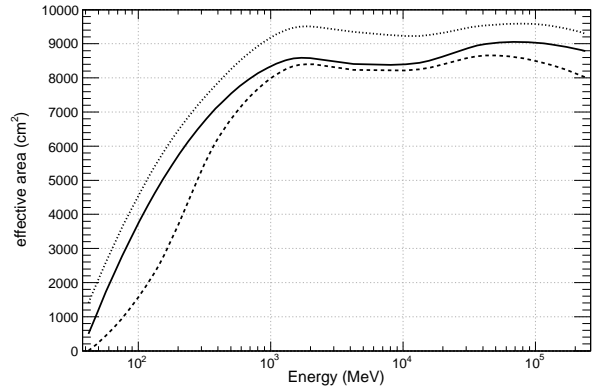}
}
\caption{
Effective area versus energy at normal incidence for Diffuse (dashed curve),
Source (solid curve), and Transient (dotted curve) analysis classes.
}
\label{f2.11}
\end{figure}

\begin{figure}[!tbhp]
\centerline{
\includegraphics[width=3.5in]{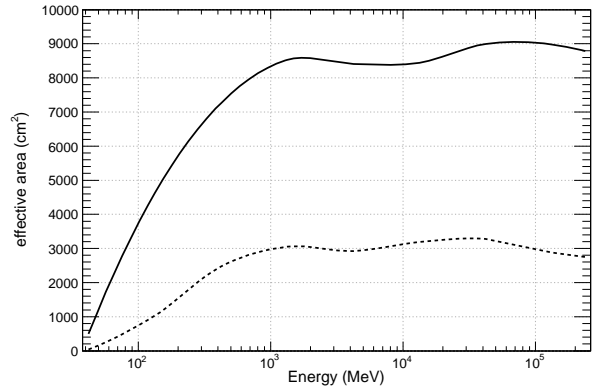}
}
\caption{
Effective area versus energy at normal incidence (solid curve) and at $60^\circ$ 
off-axis (dashed curve) for Source analysis class.
}
\label{f2.12}
\end{figure}

\begin{figure}[!tbhp]
\centerline{
\includegraphics[width=3.5in]{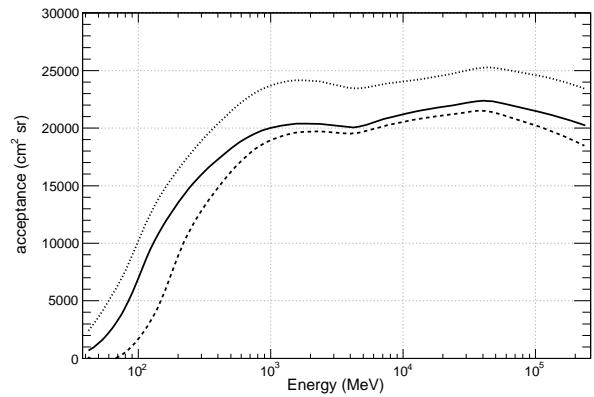}
}
\caption{
Acceptance versus energy for Diffuse (dashed curve), Source (solid curve), and
Transient (dotted curve) analysis classes.
}
\label{f2.13}
\end{figure}

\begin{figure}[!tbhp]
\centerline{
\includegraphics[width=3.5in]{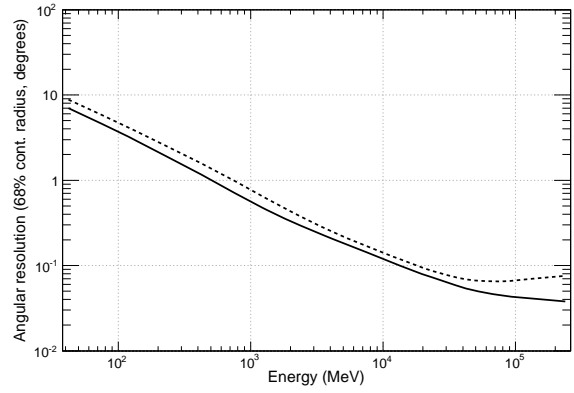}
}
\caption{
68\% containment radius versus energy at normal incidence (solid curve) and at $60^\circ$ 
off-axis (dashed curve) for conversions in the thin section of the tracker.
}
\label{f2.14}
\end{figure}

\begin{figure}[!tbhp]
\centerline{
\includegraphics[width=3.5in]{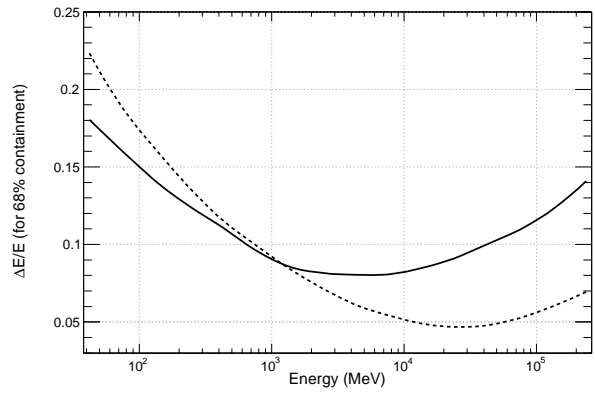}
}
\caption{Energy resolution versus energy for normal incidence (solid curve) and at
$60^\circ$ off-axis (dashed curve).
}
\label{f2.15}
\end{figure}

Figures~\ref{f2.11} -- \ref{f2.15} summarize the performance of the LAT.  The
performance parameters are subject to change as event selection
algorithms are further optimized, particularly during the early part
of on-orbit operations of \Fermi.  For the most up-to-date performance
parameters go to
http://www-glast.slac.stanford.edu/software/IS/glast\_lat\_performance.htm.

Figure~\ref{f2.11} shows the on-axis effective area versus energy for each of
the analysis classes defined in Table~\ref{t2.5}.  Contributions from
conversions in both the thin and thick sections of the tracker are
included, with each contributing about 50\% of the effective area.
Note that the peak effective area, near 3 GeV, is nearly the same for
all 3 analysis classes, while at energies below 300 MeV the effective
area for the transient class is a factor of $\sim$1.5 larger than the for
the diffuse class.  Figure~\ref{f2.12} shows the effective area for the
source class on-axis and at $60^\circ$ off-axis.  Figure~\ref{f2.13} shows the
telescope's acceptance, the average effective area times the
field-of-view.  Again, the differences between the analysis classes are
largest at low energies.

\placefigure{f2.11}
\placefigure{f2.12}
\placefigure{f2.13}
\placefigure{f2.14}
\placefigure{f2.15}

Figure~\ref{f2.14} shows the energy dependence of the 68\% containment radius
(space angle) for \gray{} conversions in the thin section of the
tracker that are incident either on-axis or at $60^\circ$ off-axis for the
source class.  The PSF for \gray{s} converting in the thick
section of the tracker is about twice as wide.

Figure~\ref{f2.15} shows the energy resolution of the LAT versus energy for
the source class.

\begin{figure}[!tbhp]
\centerline{
\includegraphics[width=3.5in]{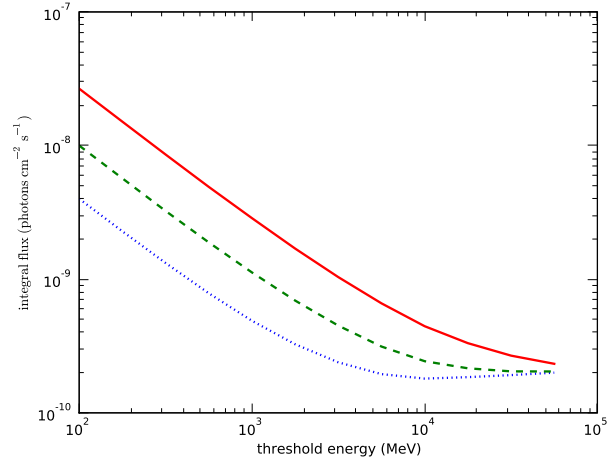}
}
\caption{Integral source sensitivity for $5\sigma$ detection for 1 year sky survey exposure. The
source is assumed to have a power law differential photon number
spectrum with index --2.0 and the background is assumed to be uniform with integral flux
(above 100 MeV) of $1.5 \times 10^{-5}$ ph cm$^{-2}$ s$^{-1}$ sr$^{-1}$ 
(dotted curve) and spectral index
--2.1, typical of the diffuse background at high galactic latitudes.  The background is 10 times higher and 100 times higher for the dashed and solid
curves, respectively, representative of the diffuse background near or on the galactic plane.
}
\label{f2.16}
\end{figure}

\begin{figure}[!tbhp]
\centerline{
\includegraphics[width=3.5in]{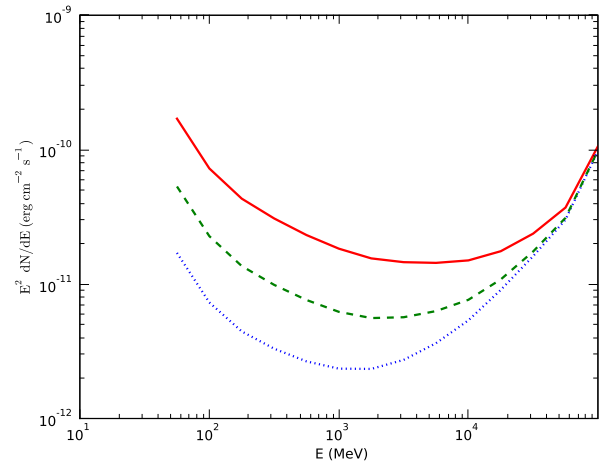}
}
\caption{Differential source sensitivity in 1/4 decade bins for $5\sigma$ detection for 1 year
sky survey exposure. The source is assumed to have a power law differential photon number
spectrum with index --2.0 and the background is assumed to be uniform with integral flux
(above 100 MeV) of $1.5 \times 10^{-5}$ ph cm$^{-2}$ s$^{-1}$ sr$^{-1}$ 
(dotted curve) and spectral index
--2.1, typical of the diffuse emission at high galactic latitudes.  The background is 10 times higher and 100 times higher for the dashed and solid
curves, respectively, representative of the diffuse background near or on the galactic plane.
}
\label{f2.17}
\end{figure}

With a diffuse \gray{} background model based on EGRET observations
and the instrument performance summarized above, the source
sensitivity of the LAT can be estimated.  The source sensitivity of
course depends not only on the flux of the source but it also depends
on the spectrum of the source.  Figure~\ref{f2.16} shows the integral source
flux above energy $E$ versus energy corresponding to a $5\sigma$ detection
after one year of scanning mode observations.  Figure~\ref{f2.17} shows the
differential source flux (in $1/4$ decade bins) corresponding to a
$5\sigma$ detection.

\placefigure{f2.16}
\placefigure{f2.17}

\subsubsection{LAT performance tests}\label{s2.5.1}

The design of the LAT was optimized using Monte Carlo simulations.
Verification of the design and simulations was done with a series of
beam tests at the SLAC, CERN and GSI heavy ion accelerator
laboratories.  In addition hardware prototypes as well as the flight
instrument have been tested using cosmic rays.  The early prototype
tests at SLAC have already been mentioned.  The most extensive beam
test was at CERN in 2006.  The CERN beams were selected because they
cover almost the entire energy range of the LAT for on-orbit
operations as well as provide large fluxes of hadrons to verify the
modeling of background interactions within the LAT.  Because schedule
prevented doing beam tests with the entire LAT, a Calibration Unit
(CU) consisting of two complete tracker and 3 calorimeter modules was
assembled.  The CU readout electronics is a copy of the flight
instrument data acquisition system.  The CU is also instrumented with
several ACD scintillator tiles to measure the backsplash response from
the calorimeter at high energies.  The overall agreement between the
Monte Carlo simulations of the CU and the beam test data are
excellent, including the overall tracker performance and the PSF, the
backsplash into the ACD, and the modeling of hadronic interactions.
The largest discrepancies involve the energy calibration in the
calorimeter which was found to be low by $\sim$7\%.  A much more complete
discussion of the preliminary
beam test results, comparing the CU
to the Monte Carlo simulations can be found in \citet{Baldini2007}.

In addition to the accelerator beam tests, several times during the
assembly of the LAT, cosmic ray triggers were recorded to verify the
proper functioning of the LAT modules as they were added to the
instrument array.  Collection of cosmic-ray data, recorded at a
trigger rate of $\sim$400 Hz, continued through the environmental testing
and pre-launch preparations of the \Fermi{} telescope.
While
terrestrial cosmic rays are quite messy (e.g., multiple particle types,
range of arrival directions) compared to a particle beam from an
accelerator, the LAT has sufficient power as a particle detector to
provide clean samples of sea-level muons, resulting in relatively
large samples of muon events that allow precision testing. With these
events, calibrations, efficiencies, and alignment issues were
successfully addressed.

The first 60 days after launch 
were
a commissioning period for the
\Fermi{} spacecraft and the LAT.  During part of this period the LAT 
was subjected to a 
relatively high rate albedo photon data by pointing at the
earth's limb, and 
directly observed the ``splash'' albedo background
component with nadir pointed runs, as well as run with modified
triggers to allow high-energy cosmic rays to be efficiently collected
to verify alignment and efficiencies that may have been affected
during the launch.  The early operations tests included checks of
internal timing and absolute timing, subsystem calibrations,
characterization of the perimeter of the South Atlantic Anomaly,
tuning the onboard event filters, and commissioning the on-board detection
of GRBs.

\subsection{Instrument operations}\label{s2.6}

\subsubsection{Onboard science processing}\label{s2.6.1}

A primary objective of onboard science processing is to provide rapid
detection and localization of GRBs.  The output of this processing can
trigger an autonomous re-pointing of the \Fermi{} 
to keep the
GRB within the LAT FoV for observation of high-energy afterglows and
is made available to support follow-up observations of afterglows by
other observatories.  
The \Fermi\ Gamma-ray Burst Monitor also produces onboard detections and
localizations, however for burst that trigger the LAT, the LAT's better
point spread function 
results in significantly improved localization. The
onboard estimates of the celestial coordinates of the GRB and the error
region are
distributed via the Gamma-ray burst Coordinate Network (GCN).

The onboard science processing consists of algorithms to (1) select
\gray{} candidate events, (2) reconstruct directions of \gray{}
candidate events and (3) search for and localize high energy
transients.

The information available to the onboard GRB search algorithm differs
substantially from that eventually available on the ground. The event
selection is based on parameters previously calculated for the onboard
filter (described in \S\ref{s2.2.4}).  The onboard software uses fairly
simple, computationally efficient algorithms to calculate the
directions of candidate \gray{} events.  The efficiency for
successfully reconstructing an event direction is within 25\%
of what can be achieved with subsequent ground processing however, the
reconstructed directions are about a factor of 2 to 5 worse.

The onboard GRB detection algorithm utilizes both the temporal and
spatial characteristics of GRBs.  It works by associating a
probability for a cluster of tracks to be located on a small part of
the sky during a short interval of time. \gray{} candidate events with
reconstructed directions are fed to the algorithm in time order. The
algorithm searches a list of the $n$ most recent events for the cluster
of events that has the smallest probability of occurring in time and
space. If the probabilities pass a pre-selected threshold, the time
and location of the cluster is passed to a second stage of processing
which considers events over a longer time interval. A GRB is declared
when cluster probabilities in the second stage exceed a pre-defined
threshold. The algorithm will then calculate refined localizations on
a configurable sequence of time intervals. The initial burst location
and each updated location is sent promptly to the ground via the
Tracking and Data Relay Satellite System (TDRSS) and then to the
GCN (http://gcn.nasa.gov/).  The
GCN provides locations of GRBs (the Notices) detected onboard by the
LAT or the GBM and reports on follow-up observations (the Circulars
and the Reports) made by ground-based and space-based optical, radio,
x-ray, TeV, and other observers.

Triggering on bursts depends on settable parameter choices.  We used a
phenomenological burst simulator and background model to guide the
initial choice of filter parameter values.  These parameters will be
optimized once a large enough sample of bursts has been identified via
ground reconstruction.

\subsubsection{Pipeline and data products}\label{s2.6.2}

LAT science data 
arrive at the LAT Instrument Science
Operations Center (ISOC) from the \Fermi{} Mission Operations Center at about
3 hour (2 orbits) intervals, 24 hours per day, in approximately 1.5 GB
downlinked data sets.  Automated processing of the data implements several
analysis functions. The primary function is to interpret the event
data, via pattern recognition and reconstruction, to indicate the
nature of the event as either celestial \gray{} photons or background,
and determine the direction, arrival time, and energy and provide
estimates of the associated errors.  During the process of correlating
data from all the subsystems, detailed information is available on the
operation of the LAT and is collected and trended for monitoring
purposes.  Once the photons have been isolated, the level-1 data are
immediately used to carry out several higher level science analysis
tasks that include searching for and refining GRB properties;
searching for flaring sources; and tracking the light curves of a
pre-selected list of sources.

The processing pipeline is designed to allow parallel processing of
events, with dependencies enabled so that processes can wait for the
parallel processing to finish before aggregating the results. It can
process an arbitrary graph of tasks.  The pipeline is run in a java
application server and interacts with farms of batch processors.
About
300 Hz of downlinked on-orbit data can be processed by 100
computing cores within 1-2 hours, allowing processing to finish before
the next downlink arrives.  Reconstruction inflates the raw science
data volume by approximately a factor of 20.  Keeping
all events
processed 
requires about 150 TB of disk per year.  The total LAT
pipeline processing compute facility is sized to accommodate prompt
processing, reprocessing and simulations.  300 computing cores for
reprocessing of data, allow one year of data to be reprocessed in
about one month.  This is an upper limit on the reprocessing time,
since use will be made of the much larger user batch processor pool in
the SLAC compute farm.

The reconstructed \gray{} photon events are then made available, along
with instrument response functions and high-level analysis tools, etc., 
to the \Fermi{} Science Support
Center (FSSC) for distribution to the community at the conclusion of
the first year on-orbit verification and sky-survey phase and during
subsequent mission phases.  After completion of the verification and
sky survey phase (year 1) of the mission, these data should arrive at
the FSSC within about 3-4 hours after arrival of unprocessed data at
the ISOC.  Automated science processing operates on 3 time scales: per
downlink, per week and per month.  During year 1 as well as beyond,
the ISOC will deliver high-level science data products, resulting from
Automated Science Processing (ASP), to the FSSC.  These include light
curves and GCN notices and circulars for GRBs and AGN flares as well
as fluxes, source locations and associated errors for transient or
flaring sources.

\subsubsection{Automated science processing}\label{s2.6.3}

Time critical analysis tasks related to detection and characterization
of transient sources, referred to as Automated Science Processing
(ASP), are performed on the reconstructed and classified events
from the level-1 pipeline to facilitate timely follow-up observations
by other observatories.

The ASP tasks relevant to GRBs (and other impulsive phenomena, such as
solar flares) are the refinement of information for GRBs that were
detected with onboard processing, the search for untriggered GRBs, and
the rapid search for and characterization of \gray{} afterglow
emission.  In this context, untriggered means not triggered by the
LAT; however, information about GRBs detected by the GBM and GCN
notices from other observatories will be used in conjunction with this
search.  The baseline for ASP processing uses an unbinned likelihood
analysis to determine the position and uncertainty of a GRB and
evaluates the spectral index and fluence by fitting a power-law
spectrum to the LAT events, also via an unbinned likelihood analysis. 
The refinement task uses a Bayesian Blocks temporal analysis \citep{Scargle1998,Jackson2005} 
to characterize the prompt burst light curve, and from
that analysis, it determines the burst start time and duration.  The
search for untriggered GRBs uses an algorithm similar to the one
developed for onboard detection; but since ASP analysis uses ground
processed events, it benefits from a substantially lower residual
background rate as well as from more accurate energy and directional
reconstructions.  Any independently-available information, such as
directions and times of GRBs seen by other instruments, is used to
increase the sensitivity of the search.  The afterglow search uses an
unbinned likelihood analysis to fit for a point source at the best-fit
GRB position using a likelihood analysis on all data available for $\sim$5
hours after the time of the GRB.  The principal products of ASP
processing for GRBs are Notices and Circulars released via GCN; for
GRB refinements the latency for release of these products will eventually be no
more than 15 minutes from the availability of the necessary level-1
data.  For GRB and afterglow searches the latency will be less than 1
hour.  Overall catalogs of LAT GRBs will be produced by the LAT
collaboration.

The ASP tasks relevant to blazars and other long-term transient
sources relate to monitoring for episodes of flaring.  For optimum
sensitivity this involves both routinely evaluating the fluxes for a
set of sources as well as searching for new transients not already on
the list of monitored sources.  The flux monitoring task uses an
unbinned likelihood analysis to evaluate the fluxes and upper limits
of a specified list of sources on daily and weekly bases.  The
ASP-monitored source list is not static -- bright transients will be
added as they are found, for example.  The general search for flaring
sources, to find transients that are not on the list of sources being
monitored, is run on daily and downlink ($\sim$3--4 hr) time scales. 
The baseline algorithm for this search monitors for changes in
exposure-corrected maps of counts. Newly-detected transients meeting
the detection criteria are released via GCN notices or
Astronomers Telegrams (ATELs).  The latency for updating daily light
curves of monitored sources will eventually be less than 6 hours after the
availability of the needed level-1 data.  The general search for
flaring sources is expected to take less than 1 hour per downlink.

The algorithms and event classification cuts used for the ASP analyses
are continuing to be refined during flight.
The ASP processing tasks are
built as part of the general pipeline system in the LAT ISOC, and are
extensible as needed.  A parallel set of tasks uses the LAT
science data for the bright pulsars to validate the instrument
response functions and to monitor the high-level performance of the
LAT.

\section{Key science objectives}\label{s3}


The LAT is designed to address a number of 
scientific objectives that 
include (i) resolving the \gray{} sky and determining the origins of diffuse 
emission and the nature of unidentified sources (\S\ref{s3.1}), (ii) 
understanding the mechanisms of particle acceleration in celestial sources (\S\ref{s3.2}),
(iii) studying the high-energy behavior of GRBs and transients (\S\ref{s3.3}), 
(iv) probing the nature of dark matter (\S\ref{s3.4}), and (v) using high-energy 
\gray{s} to probe the early universe (\S\ref{s3.5}).  The key objectives are 
largely motivated by the discoveries of EGRET ($\sim$30 MeV -- 10 GeV) and of 
ground-based atmospheric Cherenkov telescopes (ACT) above $\sim$100 GeV.  
Progress in several areas 
requires coordinated multi-wavelength observations 
with both ground and space-based telescopes.  The following sections describe 
how the LAT enables these scientific studies.

\subsection{Resolve the \gray{} sky: the origins of diffuse emission and the 
nature of unidentified sources}\label{s3.1}

High-energy \gray{} sources are seen against a diffuse background of
Galactic and extragalactic radiation.  Particularly at low Galactic
latitudes, the diffuse radiation is bright and highly structured.
About 80\% of the high-energy luminosity of the Milky Way comes from
processes in the interstellar medium (ISM).  Because these background
emissions are themselves not completely understood, analysis is an
iterative process.  As sources are discovered and distinguished from
the background, the diffuse background model can be improved, thus
allowing better analysis of the sources 
\citep[e.g.,][]{Hunter1997,Sreekumar1998,Hartman1999,Strong2004a}.

\subsubsection{Unidentified EGRET sources}\label{s3.1.1}

Although time signatures allowed identification of many EGRET sources
as pulsars or blazars, in the third EGRET catalog \citep{Hartman1999}
170 of the 271 sources had no firm identifications.
Progress towards identifications has been limited primarily by the
relatively large EGRET error boxes that often contain many potential
counterparts.

A wide variety of astrophysical objects have been suggested as
possible counterparts for some of these sources. Some examples are:
newly-found radio or X-ray pulsars 
\citep[e.g.,][]{Kramer2003,Halpern2001}, isolated neutron stars 
\citep[e.g., RX J1836.2+5925][]{Mirabal2001,Reimer2001,Halpern2002}, star forming
regions or association of hot and massive stars 
\citep[e.g.,][]{Kaaret1996,Romero1999}, supernova remnants 
\citep[e.g.,][]{Sturner1995,Esposito1996}, pulsar wind nebulae 
\citep[e.g..][]{Roberts2001}, and microquasars, such as LSI 61$^\circ$303 
\citep{Tavani1998,Paredes2000}.  Figure-of-Merit approaches have
increased the number of \gray{} sources at high Galactic latitudes
identified, with moderate confidence, with blazars 
\citep{Mattox2001,Sowards-Emmerd2003,Sowards-Emmerd2004,Sowards-Emmerd2005}.

Population studies of the unidentified EGRET sources have also
provided clues about their natures.  For example, spatial-statistical
considerations and variability studies provide evidence for a
population of Galactic and variable GeV \gray{} emitters among the
unidentified EGRET sources \citep{Nolan2003}.  Many sources may be
related to star-forming sites in the solar neighborhood or a few
kiloparsecs away along the Galactic plane \citep{Gehrels2000}.
These sites harbor compact stellar remnants, SNRs and massive stars,
i.e., many likely candidate \gray{} emitters.  Evidence exists for a
correlation with SNRs \citep{Sturner1995} as well as OB
associations \citep{Romero1999}, reviving the SNOB concept of
\citet{Montmerle1979} or making the pulsar option attractive.  Pulsar
populations may also explain a large fraction of unidentified sources
close to the Galactic plane \citep{Yadigaroglu1997} and possibly
in the nearby starburst Gould Belt \citep{Grenier2000}.  Other candidate
objects among the unidentified sources include radio-quiet neutron
star binary systems \citep{Caraveo1996} and systems
with advection-dominated accretion flows onto a black hole such as
Cygnus X-1, recently detected as a flaring source by MAGIC \citep{Albert2007}.

With regard to extragalactic sources, understanding the nature of the
unidentified sources is important because new \gray{} emitting source
classes (e.g., normal galaxies, clusters, etc.) 
are likely to be found in addition to the well-established
blazars.  A census of these sources is important for establishing
their contribution to the extragalactic \gray{} background (EGRB; see
\S\ref{s3.1.3}).  High-confidence detections and identifications of the first
representatives of other extragalactic \gray{} sources, such as galaxy
clusters \citep{Dar1995,Colafrancesco1998,Totani2000,Loeb2000,Gabici2003}, will enable
comparisons and normalization of theoretical predictions of their
contributions to the EGRB.

\begin{figure}[!tbhp]
\centerline{
\includegraphics[width=3.5in]{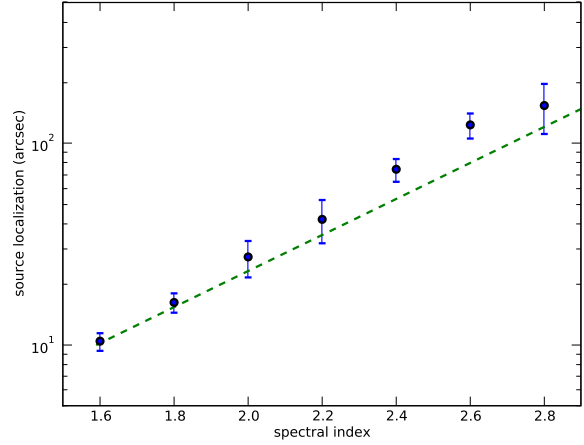}
}
\caption{LAT 68\% confidence radii localizations for a source with integral flux 
(above 100 MeV) of $10^{-7}$ ph cm$^{-2}$ s$^{-1}$ versus source spectral index 
for a source detected in the one-year sky survey. The variation of angular 
resolution with energy and viewing angle from the instrument axis is taken 
into account.  In effect, the source viewing angle is averaged over in 
sky-survey mode. The source is assumed to be 
located in a region with uniform background with integral diffuse flux (above 100 MeV) 
of $1.5 \times 10^{-5}$ ph cm$^{-2}$ s$^{-1}$ sr$^{-1}$ and spectral index --2.1.  
The source localization radius scales as (flux)$^{-1/2}$.
}
\label{f3.1}
\end{figure}

The LAT addresses these challenges with good source localization,
energy spectral measurement over a broader range, and nearly
continuous monitoring of sources for temporal variability.  These
capabilities greatly facilitate the source identification process in
the following ways:

(1) \emph{Provide good source localization for the majority of
    \gray{} sources, including all of the EGRET detected sources.}
    For $5\sigma$ one-year LAT survey sources and for EGRET sources
    (Figure~\ref{f3.1}), the typical error box sizes (68\% confidence
    radius) are $2.5\arcmin$ and $<$$0.4\arcmin$ respectively, for an
    $E^{-2}$ source and $12\arcmin$ and $2\arcmin$ respectively, for a
    source with a spectral cut-off at $\sim$3 GeV, as anticipated for
    pulsars.

\placefigure{f3.1}


More precise source locations and smaller positional uncertainties are
a prerequisite for more efficient and conclusive source
identifications, with the exception of \gray{} variability that is
tightly correlated with variability in another band.  Small error
boxes significantly reduce the number of potential counterparts at
other wavebands.  Better source localization will also improve
spatial-statistical correlation studies by reducing the number of
chance coincidences.  Finally, a number of unidentified EGRET sources
that are likely unresolved composite sources \citep[e.g.,][]{Sowards-Emmerd2003}, 
will be resolved into individual sources.

(2) \emph{Measure source spectra over a broad energy range.}   Determining
    \gray{} spectra with the LAT's resolution will allow investigation
    of features intrinsic to the sources such as absorption
    signatures, spectral breaks, transitions, and cutoffs
    (e.g., attenuation of blazar spectra at high-energy due to $\gamma+
    \gamma \to e^+ + e^-$ in the extragalactic background light).  The LAT's wide
    energy coverage will connect the GeV sky to ground-based very high
    energy \gray{} observations.  For example, LAT spans the energy
    range where the pulsed emission component in pulsars appears to
    fade out (a few GeV), to be dominated at higher energies by
    energetic synchrotron nebulae powered by the pulsar.

\begin{figure}[!tbhp]
\centerline{
\includegraphics[width=3.5in]{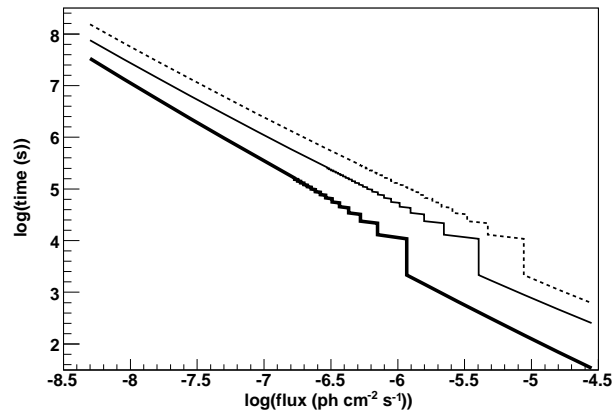}
}
\caption{Minimum time necessary to detect a source at high latitude with $5\sigma$ 
significance (thick solid curve), to measure its flux with an accuracy of 20\% 
(thin solid curve) and its spectral index with an uncertainty of 0.1 (dashed curve), 
as a function of source flux.  A photon spectral index of 2.0 is assumed.  
The steps at short times are due to the discontinuous source coverage due to the 
observatory survey mode.
}
\label{f3.2}
\end{figure}

(3) \emph{Measure \gray{} light curves over a broad range of timescales.}
    The large effective area, wide field of view, stability, and low
    readout deadtime of the LAT enable measurement of source flux
    variability over a wide range of timescales.  Figure~\ref{f3.2}
    illustrates this capability.  Coupled with the
    scanning mode of operation, this capability enables continual monitoring of source
    fluxes that will greatly increase the chances of detecting
    correlated flux variability with other wavelengths.  It allows
    periodicity and modulation searches, for example, for orbital
    modulation in close binaries.  LAT sources can be investigated for
    potential periodicities on time scales of milliseconds to years,
    encompassing millisecond pulsars, pulsars and binary systems
    hosting a neutron star. Extrapolating from EGRET analyses of
    Geminga \citep[e.g.,][]{Mattox1996,Chandler2001}, the LAT
    sensitivity allows searches in sources as faint as $\sim$$5\times10^{-8}$
    photons cm$^{-2}$ s$^{-1}$ ($E > 100$ MeV) without prior knowledge of the
    period and period derivative from radio, optical, or X-ray
    observations \citep{Atwood2006,Ziegler2008}.  Such a
    capability is crucial for revealing  radio-quiet, Geminga-like
    sources \citep{Bignami1996} which are expected to contribute
    significantly to the galactic unidentified source population
    \citep{Gonthier2007,Harding2007}.

\placefigure{f3.2}


In general, variability can be a discriminator for different source
populations, i.e., expected steadiness in the \gray{} emission in the
case of molecular-cloud-related CR interactions, \gray{s} from SNRs,
starburst galaxies, or in galaxy clusters versus modulated or
stochastic variable emission from Active Galactic Nuclei, Galactic
relativistic jet sources, black hole or neutron stars in binary
systems with massive stars, and pulsar wind nebulae.

Population studies for a prospective source class help to select the
most promising individual candidate sources for carrying out deep
multi-frequency identification campaigns based on their broadband
non-thermal properties and also help with investigating common
characteristics of the candidate population.  For example, galaxy
clusters, as a candidate population, can be characterized by mass as
deduced from optical richness, by temperature and mass functions, by
applying virial mass-over-distance constraints, and by observational
characteristics such as the presence or absence of merger activity,
the presence or absence of diffuse radio halos or indications of
nonthermal spectral components in the hard-X-rays.

LAT observations should allow at least several members among each new
candidate source populations to be individually discovered and
characterized.  In view of the large number of expected detections,
most probably representing different source classes, confirmation of a
given population as \gray{} emitters will require a common criteria
for statistical assessment \citep[e.g.,][]{Torres2005}, as well as
dedicated multiwavelength observing campaigns \citep[e.g.,][]{Caraveo2007}.

Given the advance for point-source detection provided by
the LAT, anticipating new observational features presently unknown in
GeV astrophysics is also important.  Although speculative at present,
GeV \gray{} phenomena might be found that initially, or ultimately,
have no detectable correspondence in other wavebands (e.g., GeV forming
galaxy clusters: \citealt{Totani2000}; dark matter clumps: 
\citealt{Lake1990,Calcaneo-Roldan2000}).

\subsubsection{Interstellar emission from the Milky Way, nearby galaxies, and 
galaxy clusters}\label{s3.1.2}

The diffuse emission of the Milky Way is an intense celestial
signal
that dominates the \gray{} sky.  The diffuse emission traces energetic
particle interactions in the ISM, primarily protons and electrons,
thus providing information about cosmic-ray spectra and intensities in
distant locations \citep[e.g.,][]{Hunter1997}.  This information is
important for studies of cosmic-ray acceleration and propagation in
the Galaxy \citep{Moskalenko2005}. 
\gray{s} can be used to trace the interstellar gas independently of
other astronomical methods, e.g., the relation of molecular H$_2$ gas to
CO molecule \citep{Strong2004c} and hydrogen overlooked by other methods \citep{Grenier2000a}.
The diffuse emission may also contain signatures of new physics,
such as dark matter, or may be used to put restrictions on 
the parameter space of supersymmetrical particle models 
and on cosmological models (see \S\ref{s3.4}).
The Galactic diffuse emission must also be
modeled in detail in order to determine the Galactic and extragalactic
\gray{} backgrounds and hence to build a reliable source catalog.

Accounting for the diffuse emission requires first a calculation of the cosmic-ray
(CR) spectra throughout the Galaxy \citep{Strong2000}.  
A realistic calculation that solves the transport equations
for CR species must include gas and source distributions, interstellar
radiation field (ISRF), nuclear and particle cross sections and
nuclear reaction network, \gray{} production processes, 
and energy losses. Finally, the spectrum and spatial
distribution of the diffuse \gray{s} are the products of CR particle
interactions with matter and the ISRF.

One of the critical issues for diffuse emission remaining from the 
EGRET era is the so-called ``GeV excess''.
This puzzling excess emission
above 1 GeV relative
to that expected \citep{Hunter1997,Strong2000} has shown up in all models
that are tuned to be consistent with directly measured cosmic-ray nucleon 
and electron spectra \citep{Strong2004a}. The excess has shown
up in all directions, not only in the Galactic plane. The origin of the excess
is intensively debated in the literature since its discovery by \citet{Hunter1997}.

The excess can be the result of an error in the determination of the EGRET 
effective area or energy response or could be the result of yet unknown physics
\citep[for a discussion of various hypotheses see][]{Moskalenko2005}.
Recent studies
of the EGRET data have concluded that the EGRET sensitivity above 1 GeV has been
overestimated \citep{Stecker2008} or underestimated \citep{Baughman2007}
or imply different cosmic-ray energy spectra in other
parts of the Galaxy compared to the local values \citep{Strong2004a,Porter2008}. 
If these possibilities are eliminated with high confidence then it may be
possible to attribute it to exotic
processes, e.g., dark matter annihilation products \citep{de_Boer2005}.
See, however, a discussion on limitations in the determination 
of the diffuse Galactic \gray{} emission using EGRET data
and a word of caution in \citet{Moskalenko2007}.


%
%


With its combination of good spatial and energy resolution over a
broad energy range, the LAT can test 
different hypotheses.
LAT measurements of the Galactic $\gamma$-radiation offer good
uniformity and high sensitivity as well.    As noted above,
understanding the Galactic diffuse emission is critical to analysis of LAT 
sources and important for cosmic ray and dark matter studies.
Optimizing this model over the entire sky will have a high priority in
the early phases of the mission.

The same basic considerations needed for the development of the model
of Galactic diffuse \gray{} emission also apply to other galaxies that
are candidates for study with the LAT.  For example, the LAT will
resolve the Large Magellanic Cloud in detail and, in particular, map
the massive star-forming region of 30 Doradus.  By detecting further
members among the normal galaxies in our Local Group, and galaxies
with enhanced star formation rates (e.g., Ultra Luminous Infrared
Galaxies and starburst galaxies), LAT observations can establish
independent measures of cosmic ray production and propagation.  Both
M31 and the Small Magellanic Cloud are predicted to be detectable with
LAT \citep{Pavlidou2001}, and the nearest starburst galaxies as
well \citep{Torres2004}.

Galaxy clusters emitting high-energy \gray{s} are, although well hypothesized, observationally not yet
established emitters in the GeV sky \citep{Reimer2003}. Predictions for galaxy clusters as a
candidate source class for detectable high-energy emission relate to observations of diffuse radio
signatures \citep[][and references therein]{Giovannini1999,Feretti2004}, revealing the existence of relativistic
electrons in a number of galaxy clusters. Further hints of the presence of nonthermal particles in
galaxy clusters arise from observations of hard emission components in case of a few nearby but X-ray
bright clusters \citep[][and references therein]{Rephaeli2008}. Similarly, large scale cosmological
structure formation scenarios predict high-energy \gray{} emission from galaxy clusters at a level
detectable for \Fermi/LAT \citep{Keshet2003}.

Both particle acceleration in merger processes as well as injection of relativistic particles through
feedback from AGN as cluster members can provide the mechanism to produce non-thermal particles
energized well into the energetic regime of LAT and perhaps beyond. Since galaxy clusters 
can store cosmic rays \citep{Berezinsky1997} injected either by AGNs or accelerated by primordial shocks,
\gray{s} can be produced in $pp$
interactions via production and decay of neutral pions and from annihilating DM or supersymmetrical
particles. However, weak constraints from measurements of the intercluster magnetic field ranging from
0.1 $\mu$G to 1 $\mu$G leave assessments of the total energy content, as well as the relative
fraction in relativistic electrons and protons still open to speculation. The first clear detection of
high-energy \gray{} emission from a galaxy cluster will undoubtedly constrain the baryonic particle
content as well as the uncertainly in the estimates of the magnetic field, and consequently enable
vastly improved modeling of galaxy clusters over the entire electromagnetic spectrum.

\subsubsection{Extragalactic diffuse emission}\label{s3.1.3}

An isotropic, apparently extragalactic component of the high-energy
\gray{} sky was studied by EGRET \citep{Sreekumar1998}.  This
extragalactic \gray{} background (EGRB) is a superposition of all
unresolved sources of high-energy \gray{s} in the universe plus any
truly diffuse component.  A list of the contributors to the EGRB
includes "guaranteed" sources such as blazars and normal galaxies
\citep{Bignami1979,Pavlidou2002}, and potential sources
such as galaxy clusters \citep{Ensslin1997}, shock waves associated
with large scale cosmological structure formation 
\citep{Loeb2000,Miniati2002}, distant \gray{} burst events 
\citep{Casanova2007}, pair cascades from TeV \gray{} sources and UHE cosmic rays at
high redshifts (so-called Greisen-Zatsepin-Kuzmin cut-off).  A
consensus exists that a population of unresolved AGN certainly
contribute to the EGRB inferred from EGRET observations; however
predictions range from 25\% up to 100\% of the EGRB 
\citep{Stecker1996,Mukherjee1999,Chiang1998,Mucke2000}.  
A number of exotic sources that may contribute to the EGRB
have also been proposed: baryon-antibaryon annihilation phase after
the Big Bang \citep{Stecker1971,Gao1990,Dolgov1993}, 
evaporation of primordial black holes \citep{Hawking1974,Page1976,Maki1996}, annihilation of
so-called weakly interacting massive particles (WIMPs) 
\citep{Silk1984,Rudaz1991,Jungman1996,Bergstrom2001,Ullio2002,Elsasser2005}, 
and strings \citep{Berezinsky2001}.

The EGRB is difficult to disentangle from the intense Galactic diffuse
foreground (see previous section) because it is relatively weak and
has a continuum spectrum with no strongly distinguishing features.
Indeed, determination of the EGRB spectrum depends on the adopted
model for the Galactic diffuse emission spectrum, which itself is not
yet firmly established.  Even at the Galactic poles, the EGRB does not
dominate over the Galactic component, with its flux comparable to the
Galactic contribution from inverse Compton scattering of the
interstellar radiation from stars and dust near the Galactic plane and
the cosmic microwave background \citep{Strong2000,Moskalenko2000}.  
The determination of the EGRB is thus model dependent
and influenced by the adopted size of the Galactic halo, the electron
spectrum there, and the spectrum of low-energy background photons
which must be determined independently.
Recent studies suggest that there are
two more diffuse emission components originating nearby in the solar system:
\gray{} emission due to inverse Compton scattering of solar photons on cosmic-ray
electrons \citep{Moskalenko2006,Orlando2007,Orlando2008} and a \gray{} glow
around the ecliptic due to the albedo of small solar system bodies (produced by
cosmic-ray interactions) in the
Main Asteroid Belt between the orbits of Mars and Jupiter and Kuiper Belt beyond
Neptune's orbit \citep{Moskalenko2008}, for more details see \S\ref{s3.2.3}.

Extensive work has been done \citep{Sreekumar1998} to derive the
spectrum of the EGRB from EGRET data.  Sreekumar et al. (1998) used
the relation of modeled Galactic diffuse emission to total measured
diffuse emission to determine the EGRB, as the extrapolation to zero
Galactic contribution of the total diffuse emission.  The derived
spectral index $-2.10\pm0.03$ appears to be close to that of \gray{}
blazars.  Using a different approach, \citet{Dixon1998} concluded
that the derived EGRB is affected by a significant contribution from a
Galactic halo component.  A new detailed model of the Galactic diffuse
emission \citep{Strong2004a} includes an anisotropic
Inverse Compton cross section, which brightens the high-latitude IC
intensity.  This re-analysis \citep{Strong2004b} gives a new estimate of the EGRB that is
lower in flux and steeper than found by \citet{Sreekumar1998} and is
not consistent with a power-law.

The sensitivity and resolution of the LAT allow it to resolve
many more individual sources, such as AGNs, not resolved by EGRET and
that contribute to current estimates of the EGRB.  Other components of
the remaining EGRB will therefore become more important.  Accurate
calculations of the "guaranteed background" from conventional sources
will make the limits and constraints imposed on exotic processes more
reliable.

Estimating point-source contributions to the EGRB requires statistical
information about the particular population under consideration, e.g.,
luminosity function, evolutionary properties, etc. This analysis has
been done for the \gray{} blazar population using a luminosity
function derived from EGRET observations to estimate the contribution
of unresolved point sources to the EGRB \citep[e.g.,][]{Chiang1998}
as $>$25\%.  The improved sensitivity of the LAT will reduce the
uncertainty of the LAT blazar luminosity function significantly, and
at the same time probe the blazar evolution to the redshifts of their
expected birth.

This approach will enable LAT observations to place interesting
constraints on the cosmological blazar formation rate.  So far
GeV-photon absorption in the cosmic background radiation field has not
been taken into account in any diffuse source background model.  With
LAT's sensitivity in a much broader energy range as compared to
previous pair conversion telescopes, the expected absorption imprints
on the diffuse spectrum may provide information on both the source
population as well as the background radiation field.  With the large
number of extragalactic sources resolved by the LAT, the extragalactic
component of the diffuse flux will be reduced accordingly; predictions
of the reduction due to radio-loud AGN are in the range $\sim$15\%--40\%.

Fluctuation analysis, where signatures of excess variance are searched
for in the surface brightness of the EGRB, is a very general approach
to estimating the contribution of any isotropically distributed source
population to the diffuse flux.  Application of this method to the
EGRET data set revealed a point source contribution to the EGRB of
5\%--100\% \citep{Willis1996} from analysis on an angular scale of $3.5^\circ
\times 3.5^\circ$, the scale of the \citet{Hunter1997} Galactic diffuse
emission model.  With LAT's sensitivity, point spread function and
more uniform exposure, smaller spatial scales can be probed, thereby
improving the detectability of a signal from contributing point
sources to the EGRB.

\subsection{Understand the mechanisms of particle acceleration in celestial sources}\label{s3.2}

\gray{} observations are a direct probe of particle acceleration
mechanisms operating in astrophysical systems.  Advances with LAT
observations in our understanding of these non-thermal processes can
be anticipated by reference to discoveries made with EGRET in several
important source categories: blazars, pulsars, supernovae remnants,
and the Sun.

\subsubsection{Blazar AGN jets}\label{s3.2.1}

\begin{figure}[!tbhp]
\centerline{
\includegraphics[width=3.5in]{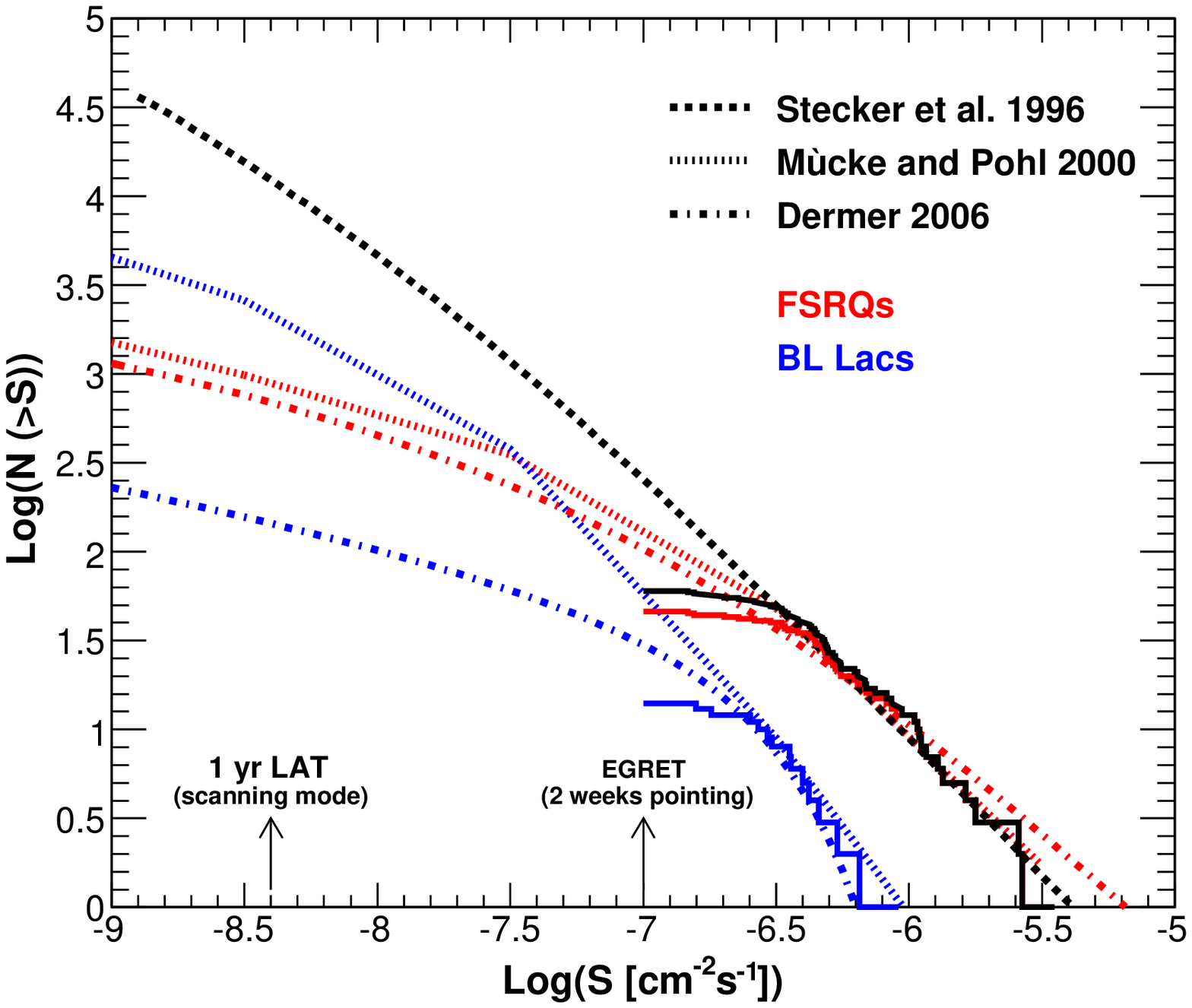}
\includegraphics[width=3.5in]{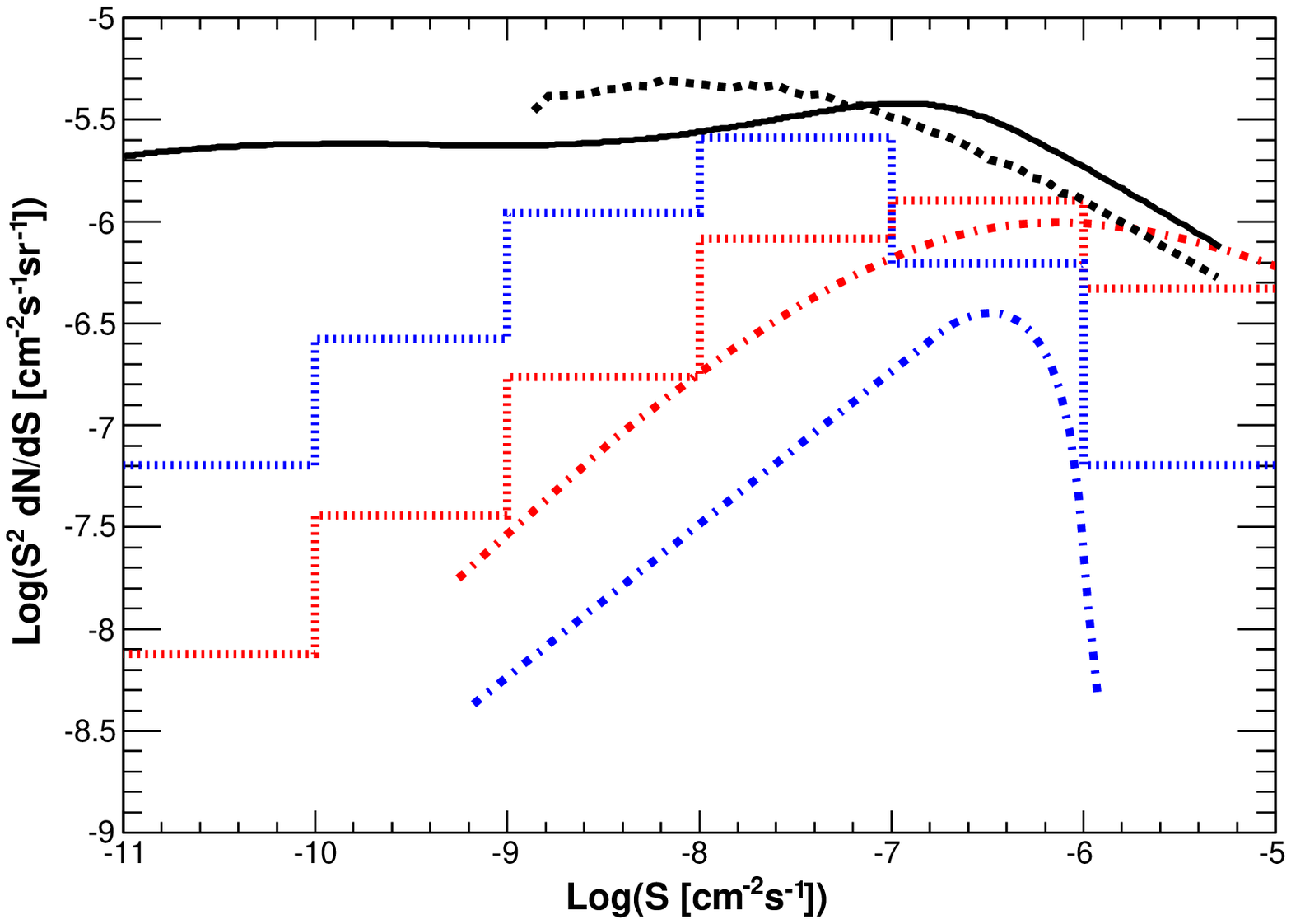}
}
\caption{(a) Cumulative number distribution of EGRET detected blazars measured over 
two-week intervals (FSRQs: blue curves, BL Lac objects: red curves) and various model 
predictions (\citealt{Stecker1996}: long-dashed line; \citealt{Mucke2000}: 
dashed-dotted lines; \citealt{Dermer2007}: dashed lines). 
The predicted number of radio-loud AGN ranges from $\sim$$10^3$ up to $\sim$$10^4$ sources.  
(b) The predicted power distribution of radio-loud AGN for the respective models, 
with the solid black line representing the model of \citet{Narumoto2006}, 
the black dotted line corresponds to the predicted power distribution of 
\citet{Stecker1996}.  The colored histograms correspond to the predicted 
power distributions for BL Lacs (blue) and FSRQs (red) of 
\citet{Mucke2000}, and the colored curves correspond to the power 
distributions of BL Lacs (blue) and FSRQs (red) of 
\citet{Dermer2007}.
The main contribution to the extragalactic diffuse \gray{} background is predicted 
to come from sources at the peak of the respective model distribution.
}
\label{f3.3}
\end{figure}

With high-confidence detections of more than 60 AGN, almost all of
them identified with BL Lacs or Flat Spectrum Radio Quasars (FSRQs)
\citep{Hartman1999}, EGRET established blazars as a class of
powerful but highly variable \gray{} emitters, in accord with the
unified model of AGN as supermassive black holes with accretion disks
and jets.  Although blazars comprise only several per cent of the
overall AGN population, they largely dominate the high-energy
extragalactic sky.  This is because most of the non-thermal power,
which arises from relativistic jets that are narrowly beamed and
boosted in the forward direction, is emitted in the \gray{} band
(Figure~\ref{f3.3}), whereas the presumably nearly-isotropic emission from
the accretion disk is most luminous at optical, UV, and X-ray
energies. Â  Most extragalactic sources detected by the LAT are
therefore expected to be blazar AGNs, in contrast with the situation
at X-ray frequencies, where most of the detected extragalactic sources
are radio-quiet AGN.

\begin{figure}[!tbhp]
\centerline{
\includegraphics[width=3.0in]{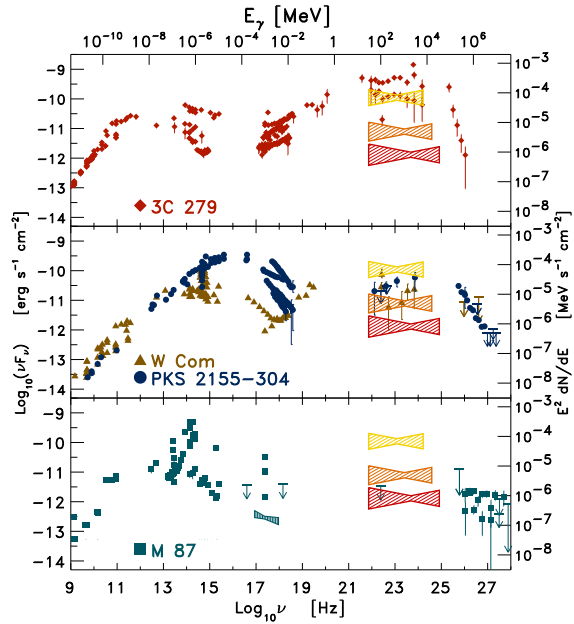}
}
\caption{Spectral energy distributions (SEDs) of four \gray{}
blazars: 3C 279  (a typical FSRQ, $z = 0.5362$, top); W Com (a low
energy peaked BL Lac object, LBL, $z = 0.102$)  and PKS 2155-304 (a
high energy peaked BL Lac object, HBL, $z = 0.116$) middle; M 87  (a
FR-I radio galaxy, $z = 0.00436$, bottom).  Included in the SEDs are
multiwavelength  data points collected in different epochs (different
brightness states) for each source (errors bars not represented for
clarity).  A qualitative representation of the average  expected LAT
pass band and sensitivity for 1 year of observations is shown.   The
LAT integral sensitivity shows the minimum needed for a 20\%
determination of the  flux after a one-day (yellow/upper bowties),
one-month (orange/middle bowties), and  one-year (red/bottom bowties)
exposure of in all-sky survey mode for a blazar with a $E^{-2}$ \gray{}
spectrum.  The resulting significance at each of these levels is about
$8\sigma$,  the spectral index is determined to about 6\%, and the
bowtie shape indicates the energy  range that contributes the most to
the sensitivity.  To make a measurement at that level  or better, a
flat spectral energy density curve must lie above the axis of the
bowtie.  }
\label{f3.4}
\end{figure}

The estimated number of blazars that \Fermi/LAT will detect ranges from a
thousand \citep{Dermer2007} to several thousand 
(\citealt{Stecker1996,Chiang1998,Mucke2000}: see Figure~\ref{f3.4}).  Such a
large and homogeneous sample will greatly improve our understanding of
blazars and will be used to perform detailed population studies and to
carry out spectral and temporal analyses on a large number of bright
objects. In particular, the very good statistics will allow us to a)
extend the $\log N-\log S$ curve to fluxes about 25 times fainter than
EGRET, b) estimate the luminosity function and its cosmological
evolution, and c) calculate the contribution of blazars and radio
galaxies to the extragalactic \gray{} background (see previous
section).  These observations will chart the evolution and growth of
supermassive black holes from high-redshifts to the present epoch,
probe a possibly evolutionary connection between BL Lacs and FSRQs,
verify the unified model for radio galaxies and blazars \citep{Urry1995}, 
and test the ``blazar sequence'' \citep{Fossati1998}. 
The redshift dependence of spectral parameters of blazars in the LAT
energy band, together with the measurements or limits from
ground-based TeV instruments, will be used to measure the evolution of
the Extragalactic Background Light (see \S\ref{s3.5}).  Finally, LAT blazar
detections will be essential in determining if a truly diffuse
component of extragalactic \gray{} emission is required, or if such
background can be accounted for by a superposition of various classes
of discrete objects.

\placefigure{f3.3}
\placefigure{f3.4}

The LAT's wide field of view will allow AGN variability to be
monitored on a wide range of time scales. Rapid flares as bright as
those observed by EGRET from 3C 279 \citep{Kniffen1993} and by \Swift{}
and \emph{Agile} from 3C454.3 \citep{Giommi2006,Vercellone2008} 
will be measurable with \Fermi{} at
\gray{} energies on time scales of hours (e.g., see Figure~\ref{f3.2}).  In
addition, the duty cycle of flaring of a large number of blazars will
be determined with good accuracy.  Measurements of the short
variability time scale for luminous \gray{} emission will place lower
limits on the Doppler factor of the jet plasma.  The values of the
Doppler factor can be correlated with \gray{} intensity states for a
specific blazar and correlated with membership in different subclasses
for many blazars.  The Doppler factors can also be compared with
values obtained from superluminal motion radio observations in order
to infer the location of the \gray{} emission site, with the goal to
study the evolution of jet Lorentz factor with distance from the black
hole.

Most viable current models of formation and structure of relativistic
jets involve conversion of the gravitational energy of matter flowing
onto a central supermassive black hole. \gray{} flares are most likely
related to the dissipation of magnetic accretion energy or extraction
of energy from rotating black holes \citep[e.g.,][]{Blandford1977}. 
However, the conversion process itself is not well understood, and
many questions remain about the jets, such as: How are they collimated
and confined?  What is the composition of the jet, both in the initial
and in the radiative phase?  Where does the conversion between the
kinetic power of the jet into radiation take place, and how?  What
role is played by relativistic hadrons.  There are also questions
about the role of the magnetic field, such as whether the total
kinetic energy of the jet is, at least initially, dominated by
Poynting flux.

The first step in answering these questions is to determine the
emission mechanisms in order to infer the content of the luminous
portions of jets.  This understanding should, in turn, shed light on
the jet formation process and its connection to the accreting black
hole.  Determining the emission mechanisms, whether dominated by
synchrotron self-Compton, external Compton, or hadronic processes,
requires sensitive, simultaneous multiwavelength observations. Such
observations can uncover the causal relationships between the variable
emissions in different spectral bands and provide detailed modeling of
the time-resolved, broadband spectra. The sensitivity and wide
bandpass of the LAT, coupled with well-coordinated multiwavelength
campaigns, are essential.  Figure~\ref{f3.4} shows representative spectral
energy distributions of \gray{} blazars and the detection pass-band
and sensitivity of the LAT.

Broadband campaigns have been organized to measure the total jet power
as compared with accretion power, and the spectra from these
observations should reveal whether a single zone structure is
sufficient or whether multiple zones are required.  Furthermore, the
content of the inner part of the jet will be tightly constrained by
broadband X-ray spectra and by temporal correlations between the X-ray
and \gray{} variability; this is because the radiative energy
density in the vicinity of black holes in AGN can be reliably
estimated from contemporaneous broadband data, and this circumnuclear
radiation must Compton-scatter with all "cold" charged particles
contained in the jet \citep[e.g.,][]{Sikora2000,Moderski2004}.  
Finally, the detection of anomalous \gray{}
spectral features will indicate the importance of hadronic processes,
with significant implications for the origin of ultra-high-energy
cosmic rays.

\subsubsection{Pulsars, pulsar wind nebulae and supernova remnants}\label{s3.2.2}

Pulsars, with their unique temporal signature, were the only
definitively identified EGRET population of Galactic point sources.
There were five young radio pulsars detected with high significance,
along with the radio-quiet pulsar Geminga and one likely millisecond
pulsar \citep[for a summary, see][]{Thompson2001}.  A number of other pulsars
had lower significance pulse detections and many of the bright,
unidentified \gray{} sources are coincident with known radio pulsars.
Surrounding young pulsars are bright non-thermal pulsar wind nebulae
(PWNe).  In the case of the Crab pulsar, EGRET detected a clear
signature of PWN emission on off-pulse phases; several other EGRET
sources near young pulsars/PWNe show strong variability, possibly
connected with variations in the wind shock termination.  Even more
encouraging has been the success in detecting PWN Compton emission in
the TeV band \citep{Aharonian2005b} from a number of PWNe.  Finally,
it has long been noticed \citep{Montmerle1979,Kaaret1996,Yadigaroglu1997} 
that \gray{} sources are spatially
correlated with massive star sites, including supernova remnants
(SNRs).  While EGRET was not able to make definitive associations with
SNRs, the LAT has the spatial and spectral resolution to do so.

\ifthenelse{\equal{\ms}{preprint}}
{
\setcounter{subsubsubsectioncounter}{1}
}{}
\subsubsubsection{Pulsar magnetospheric emission}\label{s3.2.2.1}

\begin{figure}[!tbhp]
\centerline{
\includegraphics[width=3.5in]{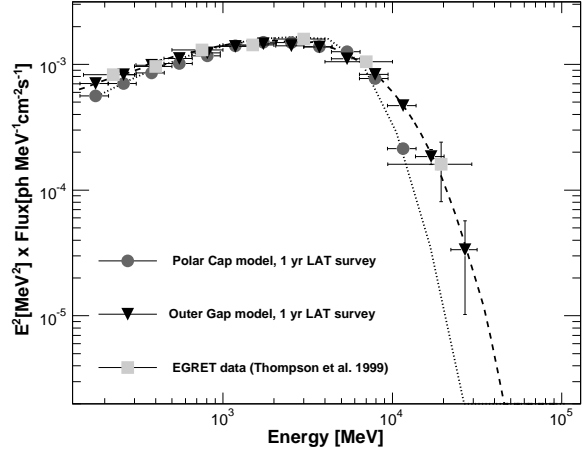}
}
\caption{The observed EGRET Vela pulsar spectrum, along with realizations of the 
expected spectrum after one year of GLAST LAT sky survey observations for two pulsar 
models.  The expected sensitivity allows discrimination between the two models and 
allows tests of the emission zone structure through phase resolved spectra.
}
\label{f3.5}
\end{figure}

Rotation-induced electric fields in charge-depleted regions of pulsar
magnetospheres (``gaps'') accelerate charges to ten's of TeV and produce
non-thermal emission across the electromagnetic spectrum. The coherent
radio emission, through which most pulsars are discovered, is however
a side-show, representing a tiny fraction of the spin-down power. In
contrast $\sim$GeV peak in the pulsed power can represent as much as
20-30\% of the total spin-down. This emission, with its complex pulse
profile and phase-varying spectrum, thus gives the key to
understanding these important astrophysical accelerators.  And,
despite 40 years of pulsar studies, many central questions remain
unanswered.  A basic issue is whether the high energy emission arises
near the surface, close to the classical radio emission \citep[``the polar
cap'' model,][]{Daugherty1996} or at a significant fraction
of the light cylinder distance \citep[``outer gap'' models,][]{Cheng1986,Romani1996}.  
In addition to geometrical
(beam-shape) differences, the two scenarios predict that different
physics dominates the pair production.  Near the surface $\gamma + B \to e^+ +
e^-$ is important, while in the outer magnetosphere $\gamma + \gamma \to e^+ + e^-$
dominates; these result in substantially different predictions for the
high energy pulsar spectrum (see Figure~\ref{f3.5}).

\placefigure{f3.5}

There are a number of pulsar models estimating detailed pulse profiles
and spectral variation with pulse phase \citep[for a recent summary, see]
[]{Harding2007,Takata2006}.  Some also predict emission
between the polar cap and outer magnetosphere extremes 
\citep{Muslimov2003,Dyks2003}.  The improved statistics, energy
resolution and high energy sensitivity provided by the LAT enable
serious tests of these models for individual bright pulsars.  Also,
with predicted numbers ranging from dozens to hundreds, the LAT survey
of the Galactic pulsar population will provide additional key tests of
massive star populations and pulsar evolution.

An extensive campaign of pulsar timing using radio telescopes at
Parkes, Jodrell Bank, Nancay, Green Bank, and Arecibo, plus X-ray
timing with the Rossi X-ray Timing Explorer has been started in order
to provide contemporaneous ephemerides with the \gray{} observations
\citep{Smith2008}.  As discussed in \S\ref{s3.1.1}, LAT's high sensitivity
also allows searches for pulsations in many sources independent of
external timing information.  Finding a larger population of
radio-quiet pulsars is another test of pulsar models \citep[e.g.,][]{Gonthier2007} 
as well as a new window on the neutron star population of
the Galaxy.
Indeed, shortly after in-orbit activation, \Fermi/LAT detected
a radio-quiet pulsar in the supernova remnant CTA 1 \citep{Abdo2008}.

\subsubsubsection{Plerions and unidentified sources}\label{s3.2.2.2}

For the Crab pulsar, EGRET detected unpulsed, possibly variable,
emission below $\sim$150 MeV (likely synchrotron) and Compton-scattered PWN
emission at higher energies \citep{de_Jager2006}.  In this and other
pulsars the connection with the IC flux observed in the TeV band is
particularly valuable in constraining the PWN B field and the injected
particle spectrum.  Recent successes with detecting PWN at TeV
energies show that the Galactic plane contains an abundance of such
sources.

\begin{figure}[!tbhp]
\centerline{\hfill
\includegraphics[width=3.5in]{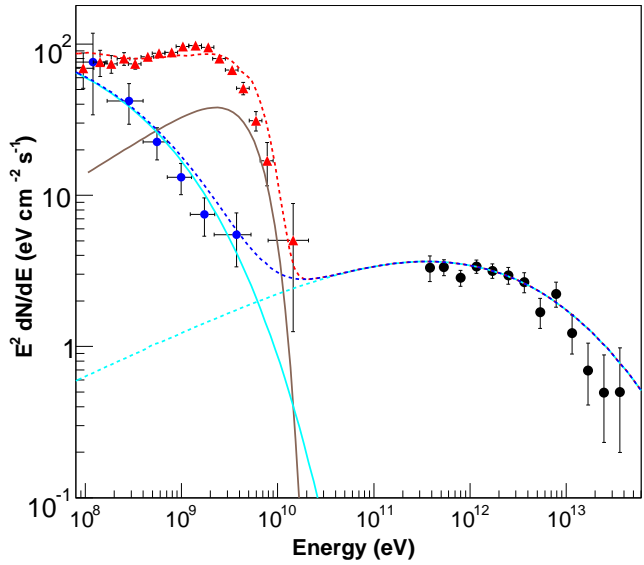}\hfill
\includegraphics[width=1.5in]{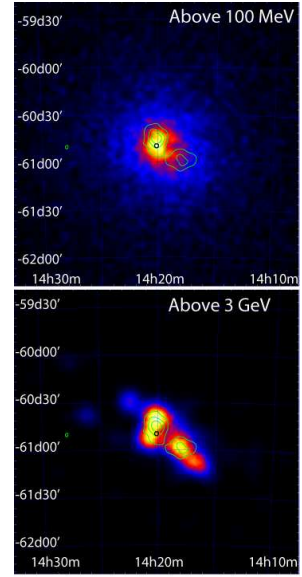}\hfill
}
\caption{Simulation of the K3/Rabbit PWN complex.  
(a): simulated smoothed LAT count maps from 5 years of sky survey-mode observation 
-– the K3 region is at upper left and the Rabbit nebula is at lower right.  
Shown are the full $E > 100$ MeV emission and the $E > 3$ GeV non-pulsed emission 
(obtainable by gating off of the PSR J1420-6048 pulse).  The green contours show the 
HESS TeV emission.  At high energies the two PWNe are clearly resolved.  
(b): the HESS spectrum of Compton emission from the PWNe along with simulated LAT 
spectra from five years of sky-survey type observations (red points).  The blue points 
show the simulated off-pulse spectrum measured with the LAT, indicating a clear 
detection of the synchrotron component of the PWNe.  Also shown are two pulsar-like spectra.  
The brighter pulsar model is for Vela-like emission from PSR J1420-60438; the 
fainter (dashed line) is for an unknown Geminga-like pulsar in the Rabbit.  
}
\label{f3.6}
\end{figure}

To illustrate the capability of the LAT for advancing PSR/PWN physics,
we have simulated one particularly interesting source, the
``Kookaburra/Rabbit'' complex \citep{Ng2005}. EGRET data
suggested that the source was composite and now X-ray \citep{Ng2005} 
and TeV \citep{Aharonian2006a} studies show that the
source contains two PWNe.  One contains the young energetic radio
pulsar PSR J1420-6048, for the other radio pulsations are not known
and the source may be Geminga-like.  We have simulated, see Figure~\ref{f3.6}, 
a plausible PWN spectrum for the two sources (following the HESS
morphology) along with a Vela-like pulsed emission for PSR J1420-6048,
in the K3 region, and Geminga-like emission for a pulsar in the
Rabbit.  At high energies the simulation indicates that the two PWNe
can be resolved.

\placefigure{f3.6}

\subsubsubsection{Host supernova remnants}\label{s3.2.2.3}

\begin{figure}[!tbhp]
\centerline{\hfill
\includegraphics[width=3.0in]{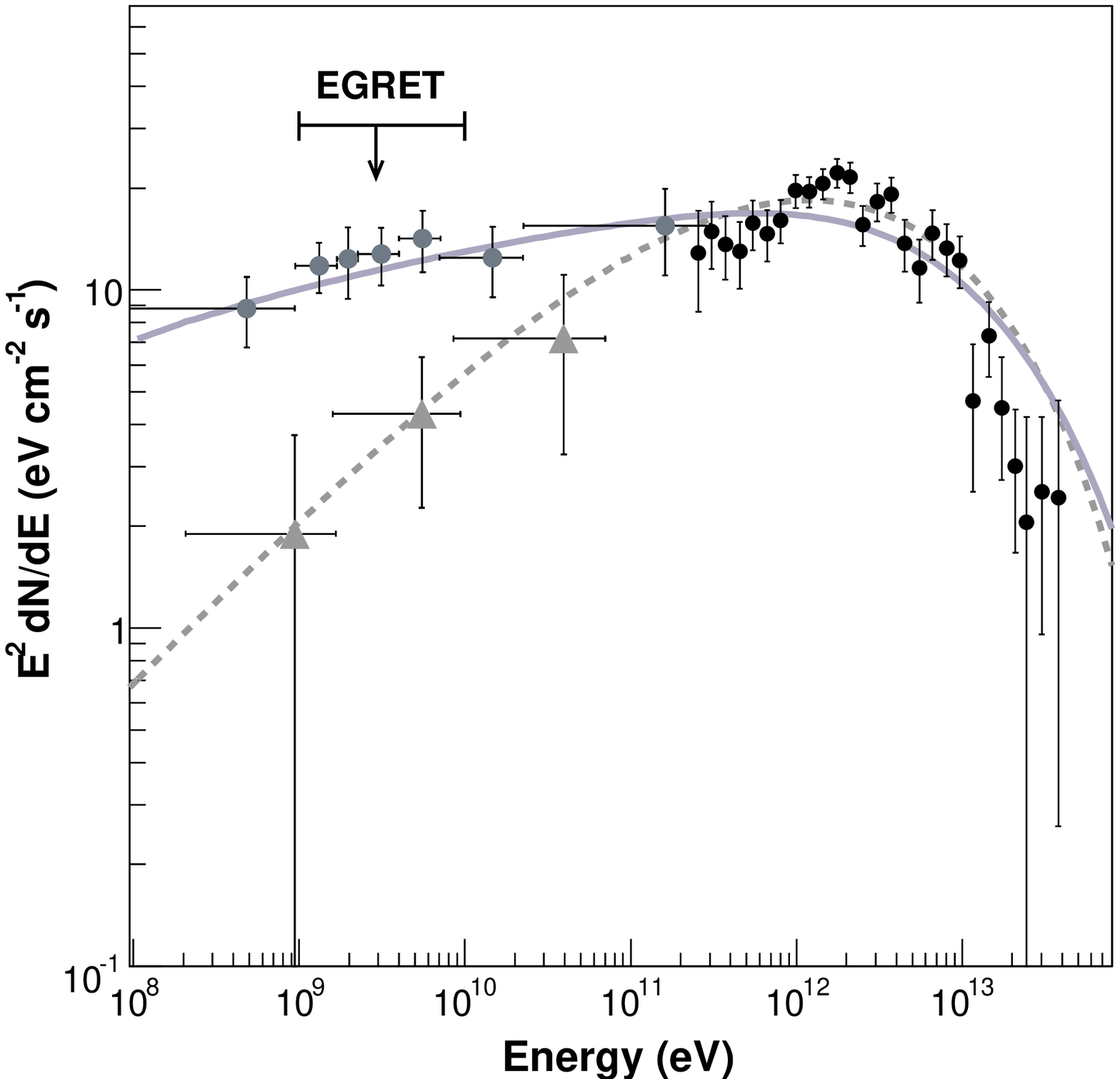}\hfill
\includegraphics[width=3.0in]{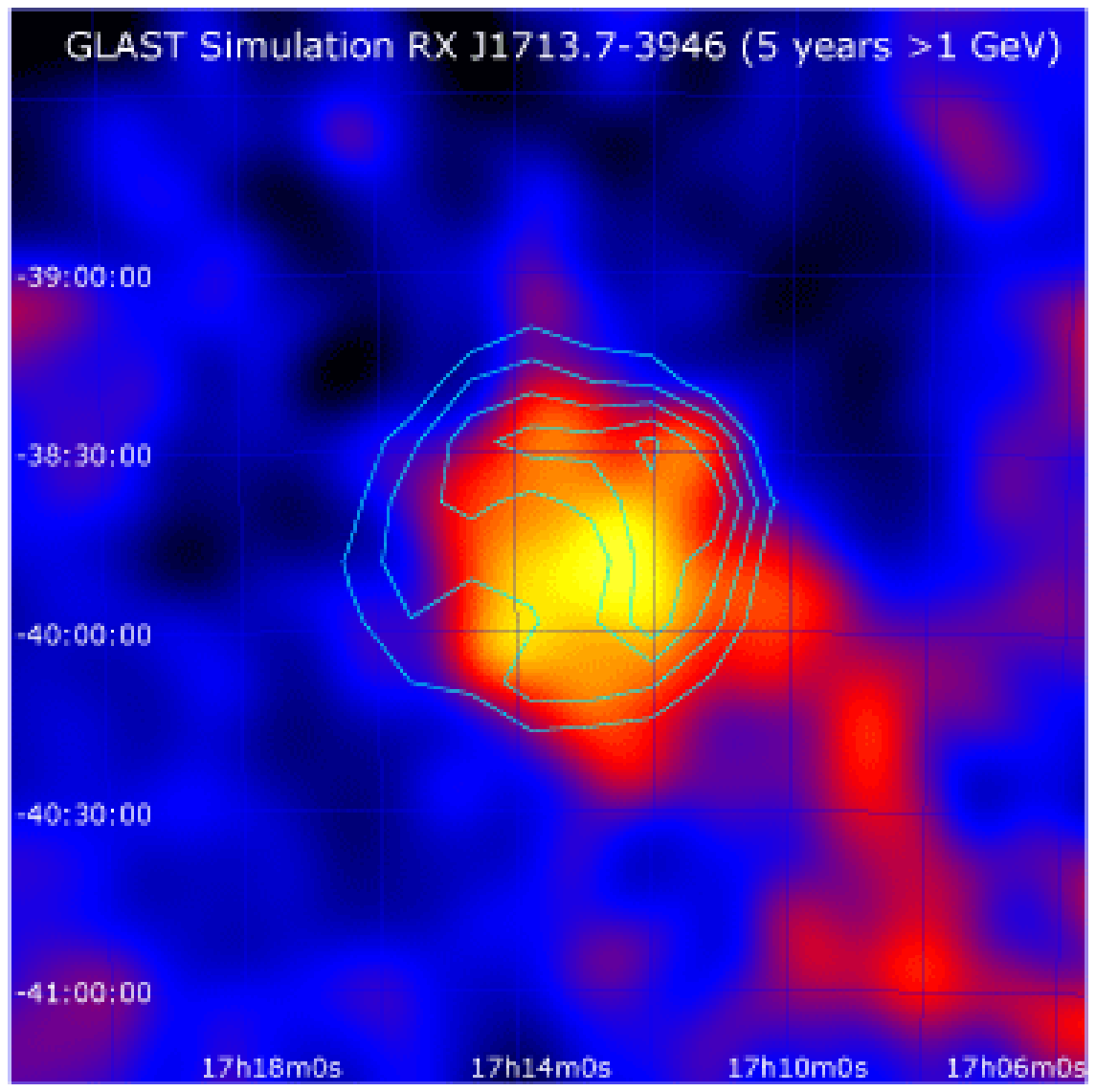}\hfill
}
\caption{The HESS spectrum of the shell SNR RX J1713.7-3946, with plausible leptonic 
and hadronic models.  Note that EGRET was not able to distinguish the SNR from the 
relatively bright nearby point source 3EG J1714-3857 and upper limits to the SNR 
flux from this source's spectrum did not allow EGRET to distinguish these possibilities. 
However, 5y of LAT observations at typical sky survey duty cycle can.  
(a) Simulated LAT spectra for the two cases \citep{Funk2008}.   
A differential spectral index of $\gamma=2$ 
has been assumed for both the parent proton and electron energy distributions.  
The observed \gray{} spectrum is sensitive to that assumption, which limits the ability 
to differentiate between parent species. 
(b) Result of a Lucy-Richardson deconvolution of the simulated LAT counts map, 
after cleaning of the point source.  The SNR is clearly resolved, although the bright 
background of the Galactic plane limits the S/N of the detection.
}
\label{f3.7}
\end{figure}

Cosmic rays with energy $\le10^{15}$ eV have long been thought to be
shock-accelerated in supernova remnants.  For some time, non-thermal
X-ray emission has implied a significant population of electrons
accelerated to TeV energies \citep{Allen1997}.
Moreover,
recently the HESS experiment has had great success in detecting TeV
emission from Galactic SNR \citep{Aharonian2005b}.  However the
origin of this emission -- inverse Compton scattering from a leptonic
component or $\pi^0$ decay from a hadronic component -- is still uncertain.
The \Fermi/LAT has the spatial and spectral sensitivity to resolve this
question and thus constrain the origin of cosmic rays.  
Particularly interesting
sources are G0.9+0.1 \citep{Aharonian2005a} and RX J1713.7-3946
\citep{Aharonian2006b}. In the case of G0.9+0.1, LAT observations will
probe the inverse Compton emission mechanism and the interstellar radiation field at
the Galactic center \citep{Porter2006}. In case of RX J1713.7-3946, 
extended TeV emission matches well spatially the lower energy
X-ray emission.  
This match might implicate inverse Compton emission from $e^+e^-$ populations
\citep{Porter2006} or can be easily accommodated by a $\pi^0$ model. In
the GeV band, well covered by the LAT, the spectra differ (see Figure~\ref{f3.7}), 
and can be distinguished.  In the particular case of RX
J1713.7-3946 and for perhaps a dozen additional objects, careful
analysis of LAT observations should be able to resolve the emission at
$E > 10$ GeV -- such spatial-spectral studies can further constrain the
particle acceleration physics and may isolate shell SNR emission from
core PWN emission in composite sources.

\placefigure{f3.7}

\subsubsection{
\gray{} emission from the Sun and solar system bodies}\label{s3.2.3}

The 2005 January 20 solar flare produced one of the most intense,
fastest rising, and hardest solar energetic particle events ever
observed in space or on the ground.  \gray{} measurements of the flare
\citep{Share2006,Grechnev2008} 
revealed what appear to be
two separate components of particle acceleration at the Sun: i) an
impulsive release lasting $\sim$10 min with a power-law index of $\sim$3
observed in a compact region on the Sun and, ii) an associated release
of much higher energy particles having an spectral index $\le$2.3
interacting at the Sun for about two hours.   Pion-decay \gray{s}
appear to dominate the latter component.  Such long-duration
high-energy events have been observed before, most notably on 1991
June 11 when the EGRET instrument on CGRO observed $>$50 MeV emission
for over 8 hours \citep{Kanbach1993}.  It is possible that these
high-energy components are directly related to the particle events
observed in space and at Earth.

Solar activity is expected to rise in 2008 with a peak occurring as
early as 2011.  During normal operations \Fermi{} will be able to observe
the Sun about 20\% of the time with the possibility of increasing that
to about 60\% during heightened solar activity.  With LAT's large
effective area and field-of-view, and its low deadtime it is expected
to observe tens of these high-energy events from the Sun.  For intense
events LAT may be able to localize the source to about $30\arcsec$,
sufficient to determine if it originates from the flare's X-ray
footpoints or from a different location that might be expected if the
high-energy particles were accelerated in a shock associated with a
coronal mass ejection.


The quiet Sun is also a source of \gray{s} which will be detectable by LAT. 
Estimates of the cosmic-ray proton interactions with the
solar atmosphere (solar albedo) were made by \citet{Seckel1991}, it is
expected that LAT will observe a flux of $\sim$$10^{-7}$ cm$^{-2}$ s$^{-1}$ 
above 100 MeV from pion decays that is 
at the limit of  
EGRET sensitivity \citep{Thompson1997b}.
In addition, a diffuse emission component with maximum in the direction of the Sun
due to the inverse Compton scattering of solar photons on cosmic-ray electrons was
predicted to be detected by LAT \citep{Moskalenko2006,Orlando2007,Orlando2008}.
A detailed analysis of the EGRET data \citep{Orlando2008} yielded the flux of
these two solar components at 4$\sigma$, consistent with the predicted level.
Observations of the
inverse Compton scattering of solar photons will allow for 
continuous monitoring of the cosmic-ray electron spectrum 
from the close proximity of the solar surface to Saturn's orbit at 10 AU,
important for heliospheric cosmic-ray modulation studies.
The fluxes of these components will vary over the solar cycle as solar
modulation increases, thus we can expect the highest fluxes to be
observed early in the \Fermi{} mission.


Recent studies suggest that LAT will be able to see
another diffuse emission component originating nearby in the solar system:
a \gray{} glow
around the ecliptic due to the albedo of small solar system bodies (produced by
cosmic-ray interactions) in the
Main Asteroid Belt between the orbits of Mars and Jupiter and Kuiper Belt beyond
Neptune's orbit \citep{Moskalenko2008}.
Observations of the albedo of small bodies can be used to 
derive their size distribution. Additionally 
\gray{} albedo of Kuiper Belt objects could be used to 
probe the cosmic-ray spectrum in the far outer solar system close to the heliospheric boundary.

Since the ecliptic is projected across the Galactic center, and passes through
high Galactic latitudes, both diffuse emission components (inverse Compton scattering of
solar photons and the albedo of small solar system bodies) 
are important to take into account when 
studying the sources in the direction of the Galactic center and extragalactic diffuse emission
(see also \S\S\ref{s3.1.2}, \ref{s3.4}).

The Moon is also a source of \gray{s} due to CR interactions
with its surface and has been detected by EGRET \citep{Thompson1997b}.
However, contrary to the CR interaction with the gaseous
atmospheres of the Earth and the Sun, the Moon surface is solid, 
consisting of rock, making its albedo spectrum unique.
The spectrum of \gray{s} from the Moon 
is very steep with an effective cutoff around 3--4 GeV (600 MeV
for the inner part of the Moon disk) and exhibits a narrow
pion-decay line at 67.5 MeV, perhaps unique in astrophysics \citep{Moskalenko2007a}. 
Apart from other astrophysical sources,
the albedo spectrum of the Moon is well understood, including
its absolute normalization; this makes it a useful
``standard candle'' for \gray{} telescopes.
The steep albedo spectrum also provides a unique opportunity for energy 
calibration of \gray{} telescopes such as LAT.

Finally, the brightest \gray{} source on the sky is the Earth's atmosphere due to
its proximity to the spacecraft. The Earth's albedo due to the cosmic-ray
interactions with the atmosphere has been observed by EGRET \citep{Petry2005}.
Its observations can provide important information
about interactions of cosmic rays and solar wind particles with Earth's magnetic field 
and the atmosphere.

\subsection{Study the high energy behavior of GRBs and transients}\label{s3.3}

Over the last decade the study of X-ray, optical, and radio afterglows
of \gray{} bursts (GRBs) has revealed their distance scale, helping to
transform the subject from phenomenological speculation to
quantitative astrophysical interpretation.  We now know that
long-duration GRBs ($\tau > 2$ s) and at least some short-duration GRBs lie
at cosmological distances and that both classes involve extremely
powerful, relativistic explosions.  Long GRBs are associated with low
metallicity hosts with high star formation rates, and have nuclear
offsets of $\sim$10 kpc \citep{Bloom2002}.  Long-duration bursts are
typically found in star-forming regions of galaxies and are sometimes
associated with supernovae, indicating that the burst mechanism is
associated with the collapse of very massive stars \citep{Zhang2004a}.  
Short-duration bursts are often located in much lower
star-formation rate regions of the host galaxy, suggesting that in
some cases these bursts arise from the coalescence of compact objects
\citep{Bloom2006,Nakar2007}.  For the $\sim$30\% of long-duration
bursts seen by \Swift{} that have measured redshifts, the redshift
distribution peaks near $z \sim 2.8$ \citep{Jakobsson2006}, comparable to
Type 2 AGN.  The sparse distribution for short bursts with
spectroscopic redshifts spans a much lower range, $z \sim 0.1 - 1.1$.
However, a photometric study of the host galaxies of short bursts
without spectroscopically determined redshifts indicates that the
fainter hosts tend to lie at redshifts $z > 1$ \citep{Berger2007}.

The standard picture that has emerged of GRB physics is that an
initial fireball powers a collimated, super-relativistic blast wave
with initial Lorentz factor $\sim 10^2 - 10^3$.  Prompt \gray{} and X-ray
emission from this ``central engine'' may continue for few $\times 10^3$ s.
Then external shocks arising from interaction of the ejecta with the
circumstellar environment at lower Lorentz factors give rise to
afterglows in the X-ray and lower-energy bands that are detected for
hours to months.  The physical details -- primal energy source and
energy transport, degree of blast wave collimation, and emission
mechanisms -- remain for debate \citep{Zhang2004}.  The LAT will
help constrain many uncertainties in these areas.

EGRET detected two components of high-energy \gray{} emission from
GRBs:  $>$100 MeV emission contemporaneous with the prompt pulsed
emission detected in the 10--1000 keV band, and a delayed component
extending to GeV energies that lasted more than an hour in the case of
GRB 940217 \citep{Hurley1994}.  Analogous components were detected
in the short burst GRB 930131 \citep{Sommer1994}.  Most importantly,
EGRET detected one burst (GRB 941017) in which a third power-law
component was evident above the usual Band function spectrum \citep{Band1993},
with an inferred peak in $\nu F(\nu)$ above 300 MeV during most of the prompt
emission phase \citep{Gonzalez2003}.  This indicates that some
bursts occur for which the bulk of the energy release falls in the LAT
energy band.  The prompt pulsed component in these bursts was poorly
measured by EGRET since the severe spark chamber deadtime ($\sim$100
ms/event) was comparable to or longer than pulse timescales.  The LAT
is designed with low deadtime ($\sim$26 $\mu$s/event) so that even very intense
portions of bursts will be detected with very little ($<$ few \%)
deadtime.

The delayed-emission component will also be much better measured
because of LAT's increased effective area, larger FoV, and low
self-veto at supra-GeV energies. These observations will test models
of delayed GeV emission, for example, those involving production of
\gray{s} from ultra-high-energy cosmic rays \citep{Bottcher1998},
impact of a relativistic wind from the GRB on external matter
\citep{Meszaros1994}, and synchrotron self-Compton radiation
\citep{Dermer2000}.

Internal and external shock models \citep{Zhang2004} are
currently constrained primarily by spectral and temporal behavior at
sub-MeV energies \citep{Fenimore1999}, where the most detailed
observations have been made.  But these observations span only a
relatively narrow energy range.  The LAT's sensitivity will force
comparison of models with observations over a dynamic range in energy
of $\sim 10^3-10^4$, and a factor of $\sim$$10^6$ including joint GBM observations.

The LAT can provide time-dependent spectral diagnostics of bright
bursts and will be able to measure high-energy exponential spectral
cutoffs expected for moderately high redshift GRBs caused by $\gamma\gamma$
absorption in the cosmic UV-optical background (complementing AGN
probes).  The LAT will distinguish such attenuation from $\gamma\gamma$ absorption
internal to the sources.  Internal absorption is expected to produce
time-variable breaks in power-law energy spectra.  Signatures of
internal absorption will constrain the bulk Lorentz factor and
adiabatic/radiative behavior of the GRB blast wave as a function of
time for sufficiently bright bursts \citep{Baring1997,Lithwick2001,Baring2006}.

\begin{figure}[!tbhp]
\centerline{
\includegraphics[width=3.5in]{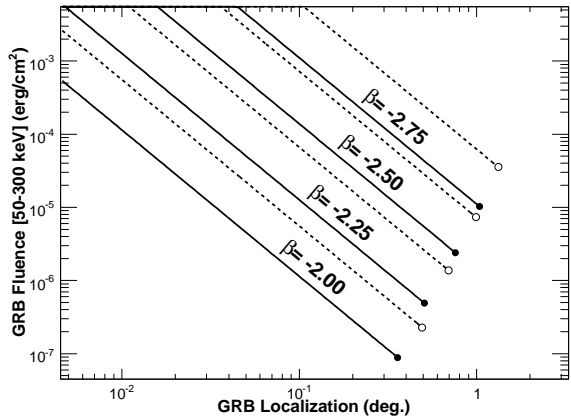}
}
\caption{\gray{} burst localization with the LAT.  The lines correspond
to the scaling law between the location accuracy (at $1\sigma$) and the
intensity of the burst, expressed as fluence in the 50--300 keV band. Solid
lines correspond to GRB at normal incidence, and dashed lines to $60^\circ$
off-axis. Different sets of lines are for different high-energy spectral
indexes (assuming the Band function describes the GRB spectral energy
distribution). The starting points of the lines, (filled circles for
on-axis, and empty for off-axis) correspond to the minimum fluence required
to detect a burst (at least 10 counts in the LAT detector).
}
\label{f3.8}
\end{figure}

To estimate the LAT sensitivity to GRB, a phenomenological GRB model
is adopted that assumes the spectrum of the GRB is described by the
Band function, and the high-energy power law
extends up to LAT energies. In order to compare the LAT sensitivity to
GRB with the BATSE catalog of GRB, we compute the fluence of GRBs in
the 50--300 keV energy band. Figure~\ref{f3.8} shows the minimum detectable
fluence as a function of the localization accuracy, for different
viewing angles and for different high-energy spectral indexes
keeping the peak energy and the low energy spectral index of the 
Band model fixed (to 500 keV and --1, respectively). The plot
showns the expected relation between the fluence and the
localization accuracy, which scales as the inverse of the square root
of the burst fluence. 

Detailed simulations, based on extrapolations from the BATSE-detected
GRBs, and adopting the distribution of Band parameters of the catalog
of bright BATSE bursts \citep{Kaneko2006}, suggest that the LAT may
detect one burst per month, depending on the GRB model for high
energy emission.  These estimations are in good agreement with the
observed number of GRBs.  In the first few months of operations LAT
has already detected high-energy emission from four GRBs: GRB 080825C
\citep[][GCN: 8183]{Bouvier2008}, the bright GRB 080916C \citep[][GCN: 8246]
{Tajima2008}, GRB 081024B \citep[][GCN 8407]{Omodei2008} and GRB 081215A,
\citep[][GCN 8684]{McEnery2008}.

For more than one-third of LAT-detected bursts, LAT localizations should be
sufficiently accurate for direct X-ray and optical counterpart
searches.  For instance, $\sim$50\% of the LAT bursts are projected to have
localization errors commensurate with the field of view of \Swift's XRT
($23\arcmin$), which very efficiently detects afterglows with few
arc-second error radii.  Burst positions are also calculated rapidly onboard,
albeit with less initial accuracy, by the LAT flight software, as well as on the ground by the science
analysis software pipeline, and distributed via the GCN network.
Searches are conducted during ground analysis for fainter bursts
not detected by the on-board trigger of the LAT.

\placefigure{f3.8}
\placefigure{f3.10}

Simulations show that LAT observations may constrain quantum gravity
scenarios that give rise to an energy-dependent speed of light and
consequent energy-dependent shifts of GRB photon arrival times
\citep{Amelino-Camelia1998,Alfaro2002}.  Short-duration
GRBs, which exhibit negligible pulse spectral evolution above $\sim$10 keV
may represent the ideal tool for this purpose \citep{Scargle2008}.  
The LAT properties important for such measurements are
its broad energy range, sensitivity at high energies, and $<$10 $\mu$s
event timing.  The LAT's low deadtime and simple event reconstruction,
even for multi-photon events, enable searches for
evaporation of primordial black holes with masses of $\sim$$10^{17}$ gm
\citep{Fichtel1994}.

\subsection{Probe the nature of dark matter}\label{s3.4}

Compelling evidence for large amounts of nonbaryonic matter in the
Universe is provided by the rotation curves of galaxies,
structure-formation arguments, the dynamics and weak lensing of
clusters of galaxies, and, most recently, WMAP measurements of the CMB
(\citealt{Spergel2007}, for review see e.g., \citealt{Bergstrom2000}).  One of
the most attractive candidates for Dark Matter is the Weakly
Interacting Massive Particle (WIMP).  Several theoretical candidates
for WIMPs are provided in extensions of the Standard Model of Particle
Physics such as Super-Symmetry. Searches for predicted particle states
of these theories are one of the prime goals of accelerator-based
particle physics, in particular the experiments at the Large Hadron
Collider (LHC), which is planned to be operational in 2008.

Annihilations of WIMPs can lead to signals in radio waves, neutrinos,
antiprotons and positrons and \gray{s}.  \gray{} observations have the
advantage over charged particles that the direction of the \gray{s}
points back to the source, and they are not subject to additional flux
uncertainties such as unknown trapping times \citep{Bergstrom2001,Ullio2002}.  
However, predicted rates are subject
to significant astrophysical uncertainties.  Substructure in Dark
Matter Halos is especially uncertain, with the predicted flux, for a
given annihilation cross section, varying by several orders of
magnitude.

Observations of the \gray{} signal of WIMPs may not only constrain the
particle nature of these particles but also, in the case that the LHC
experiments discover a WIMP candidate, establish the connection
between those particles and the Dark Matter. If the Dark Matter is
identified, the LAT may be able to image the distribution of Dark
Matter in the Galaxy which would constrain scenarios for structure
formation.

Two types of WIMP annihilation signals into \gray{s} are possible: a
spectrally \emph{continuous} flux below $m_\chi$ the mass of the annihilating
particle, resulting mainly from the decay of $\pi^0$ mesons produced in the
fragmentation of annihilation final states, and \emph{monoenergetic} \gray{}
lines resulting from WIMP annihilations into two-body final states
containing two photons or a Z boson and a photon.  Generally, the
continuous signal has a much larger rate, but with a signature that is
difficult to separate from the other Galactic diffuse foreground
contributions, while the monoenergetic line is a much smaller signal,
but, if detected, is more easily distinguished.  The basic quantities
that LAT observations can constrain are the total velocity-averaged
annihilation cross section, the branching fraction in different final
states, and the mass of the WIMPs.

Different astrophysical sources can be used to search for a signal
from WIMP annihilations, each with advantages and challenges. Table~\ref{t3.1} 
summarizes the different search strategies that we have studied.

\placetable{t3.1}

\begin{figure}[!tbhp]
\centerline{
\includegraphics[width=3.5in]{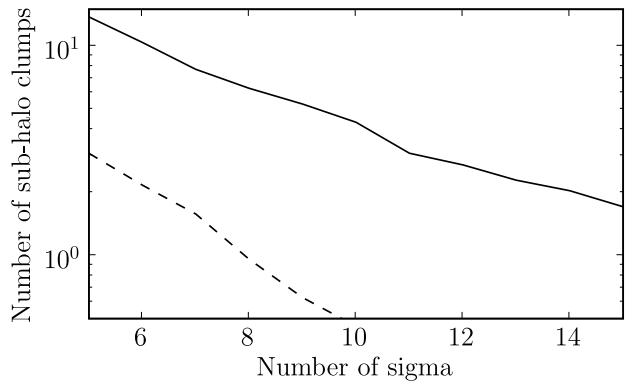}
}
\caption{Number of clumps
observed by \Fermi/LAT vs.\ number of $\sigma$ significance in 5 years of LAT observations 
in all-sky scanning mode (solid line); 1 year of observation (dashed line).
A generic WIMP of mass 100 GeV and $\langle\sigma v\rangle = 2.3\times10^{-26}$ 
cm$^3$ s$^{-1}$, a halo clump distribution from \citeauthor{Taylor2005a}
(\citeyear{Taylor2005a,Taylor2005b})
and diffuse \gray{} backgrounds according to \citet{Sreekumar1998} 
and  \citet{Strong2000} have been assumed.
}
\label{f3.10}
\end{figure}

Detailed calculations of LAT sensitivities to Dark Matter are
described in a separate paper \citep{Baltz2008}.  Generally,
sensitivities are in the cosmologically interesting region of $\langle\sigma v\rangle
\sim10^{-26}-10^{-25}$ cm$^3$ s$^{-1}$, in the mass range between 40 and 200 GeV.
Figure~\ref{f3.10} shows the expected number of halo clumps vs.\ detection
significance for a generic WIMP of mass 100 GeV and $\langle\sigma v\rangle
 = 2.3\times10^{-26}$ cm$^3$ s$^{-1}$
assuming the distribution of halo clumps given by \citeauthor{Taylor2005a}
(\citeyear{Taylor2005a,Taylor2005b})
in which about 30\% of the halo mass is concentrated in
halo clumps.  The diffuse background was assumed to consist of an
isotropic extragalactic component \citep{Sreekumar1998} and a
Galactic component \citep{Strong2000}.  

The intensity needed to
detect a \gray{} line with $5\sigma$ significance is in the vicinity
of $10^{-9}$ ph cm$^{-2}$ s$^{-2}$ sr$^{-1}$ for an annulus around the
Galactic Center (masking the galactic plane to $\pm15^\circ$).

In \citet{Baltz2006}, information obtainable with \Fermi{} is compared
with what may be learned at upcoming accelerator-based experiments,
for a range of particle Dark Matter models.  Over sizable ranges of
particle model parameter space, \Fermi{} has significant sensitivity and
will provide key pieces of the puzzle.  The challenge will be to
untangle the annihilation signals from the astrophysical backgrounds
due to other processes.

\placefigure{f3.10}

\subsection{Use high-energy \gray{s} to probe the early universe}\label{s3.5}

Photons above 10 GeV can probe the era of galaxy formation through
absorption by near UV, optical, and near IR extragalactic background
light (EBL).  The EBL at IR to UV wavelengths is accumulated radiation
from structure and star formation and its subsequent evolution in the
universe with the main contributors being the starlight in the optical
to UV band, and IR radiation from dust reprocessed starlight \citep[see
e.g.,][]{Madau1996,MacMinn1996,Primack2001,Hauser2001}.

Since direct measurements of EBL suffer from large systematic
uncertainties due to contamination by the bright foreground
(e.g., interplanetary dust, stars and gas in the Milky Way, etc.), the
indirect probe provided by absorption of high-energy \gray{s} via
pair production ($\gamma + \gamma \to e^+ + e^-$), emitted from blazars, during their
propagation in the EBL fields, can be a powerful tool for probing the
EBL density.  For example, observations of relatively nearby TeV
blazars by the HESS atmospheric Cherenkov telescope \citep{Aharonian2006c} 
have placed significant limits on the EBL at IR energies in
the local universe.  The photon-photon pair production cross section
has a pronounced maximum at $E_\gamma \approx 0.8$ TeV (1 eV/EEBL) (interaction
angle averaged), close to the pair production threshold.  Hence the
LAT energy range extending to greater than 300 GeV is ideal for
probing the EBL in the largely unexplored optical-UV band.  According
to current EBL models \citep[e.g.,][]{Primack1999,Stecker2006,Kneiske2004}, 
absorption breaks in the LAT energy range are
expected for sources located at $z \ge 0.5$.  This offers for the first
time the opportunity to constrain the \emph{evolution} of the EBL.  For this
purpose, at least two methods have been developed:  probing the
horizon of extragalactic \gray{s} through measurements of either the
ratio of absorbed to unabsorbed flux versus redshift \citep{Chen2004}, 
or detection of the e-folding cutoff energy $E(\tau_{\gamma\gamma} = 1)$ as a
function of redshift \citep{Fazio1970,Kneiske2004} in a
large number of suitable sources.  With the expected LAT flux
sensitivity the number of detected \gray{} loud blazars will increase
to potentially several thousand sources (see \S\ref{s3.2.1}) with redshifts
up to $z \sim 5-6$.  Such a large number of sources will be required for a
statistically meaningful search for evolutionary behavior of spectral
absorption features in bright and hard-spectrum AGNs.  Any of the
analysis methods employed requires disentangling source intrinsic
opacity effects, particularly if they are evolutionary with redshift,
from the absorption due to EBL.  Absorption in the local environment
of AGN but external to the jet radiation fields has been shown to
mimic an absorption pattern similar to what is expected from EBL
attenuation of \gray{s} \citep{Reimer2007}, i.e. higher \gray{} opacities
from higher redshift sources.  Careful source selection and a
statistical assessment of the radiation field density at the \gray{}
source site will be an integral part of the analysis.  Monitoring of
external photon fields in AGN (e.g., broad-line region lines) and
correlating with the observed \gray{} cutoff energy may offer
verification, and possibly quantification, of this effect.

\section{Summary}\label{s4}

The Large Area Telescope, the primary instrument on the \FGST,
is a state-of-the-art, high-energy \gray{} telescope.  The
LAT's combination of wide field-of-view, large effective area,
excellent single photon angular resolution (particularly at high
energies), good energy resolution, excellent time resolution and low
instrumental dead time, will push back several frontiers in
high-energy astrophysics.  Data from the LAT and software analysis
tools will be available to the entire scientific community.

\acknowledgements

The \Fermi/LAT Collaboration acknowledges the generous ongoing support of a
number of agencies and institutes that have supported both the development
and the operation of the LAT as well as scientific data analysis.  
These include the National Aeronautics and Space
Administration and the Department of Energy in the United States, the
Commissariat \`a l'Energie Atomique and the Centre National de la
Recherche Scientifique / Institut National de Physique Nucl\'eaire et de
Physique des Particules in France, the Agenzia Spaziale Italiana and the
Instituto Nazionale di Fisica Nucleare
in Italy, the Ministry of Education, Culture, Sports,
Science and Technology (MEXT), High Energy Accelerator Research
Organization (KEK) and Japan Aerospace Exploration Agency (JAXA) in
Japan, and the K. A. Wallenberg Foundation, the Swedish Research Council, 
and the Swedish National Space Board in Sweden. 
Additional support from the following agencies is also gratefully acknowledged:  
the Istituto Nazionale di Astrofisica in Italy and the K. A. Wallenberg Foundation 
for providing a grant in support of a Royal Swedish Academy of Sciences 
Research fellowship for JC.

\clearpage


\begin{deluxetable}{lc}
\tablecaption{\label{t1.1} Summary of Large Area
Telescope Instrument parameters and estimated performance}
\tablecolumns{2}
\tablewidth{0pt}
\tablehead{
\colhead{Parameter} & \colhead{Value or Range}
}
\startdata
Energy range                                        & 20 MeV -- 300 GeV\\
Effective area at normal incidence\tablenotemark{a} & 9,500 cm$^2$\\
Energy resolution (equivalent Gaussian $1\sigma$):\\
$\quad$ 100 MeV -- 1 GeV (on axis)                 & 9\%--15\%\\
$\quad$ 1 GeV -- 10 GeV (on axis)                  & 8\%--9\%\\
%
$\quad$ 10 GeV -- 300 GeV (on-axis)                 & 8.5\%--18\%\\
$\quad$ $>$10 GeV ($>$$60^\circ$ incidence)         & $\le$6\%\\
Single photon angular resolution (space angle)\\
on-axis, 68\% containment radius: \\
$\quad$ $>$10 GeV                                   & $\le$$0.15^\circ$\\
$\quad$ 1 GeV                                       & $0.6^\circ$\\
$\quad$ 100 MeV                                     & $3.5^\circ$\\
$\quad$ on-axis, 95\% containment radius & $<3\times\theta_{68\%}$\\ 
$\quad$ off-axis containment radius at $55^\circ$ & $<1.7\times$ on-axis value\\
Field of View (FoV) & 2.4 sr\\
Timing accuracy & $<10$ $\mu$sec\\
Event readout time (dead time) & $26.5$ $\mu$sec\\
\hline
\noindent{\smallskip}
GRB location accuracy on-board\tablenotemark{b} & $<10\arcmin$\\
GRB notification time to spacecraft\tablenotemark{c} & $<$5 sec\\
Point source location determination\tablenotemark{d} & $<0.5\arcmin$\\
Point source sensitivity ($>$100 MeV)\tablenotemark{e} & $3\times10^{-9}$ ph cm$^{-2}$ s$^{-1}$
\smallskip
\enddata

\tablenotetext{a}{Maximum (as function of energy) effective area at normal
incidence.  Includes inefficiencies necessary to achieve required
background rejection.  Effective area peak is typically in the 1 to 10
GeV range.}

\tablenotetext{b}{For burst ($<$20 sec duration) with $>$100 photons above
1 GeV.  This corresponds to a burst of $\sim$5 cm$^{-2}$ s$^{-1}$ peak rate in the 50
-- 300 keV band assuming a spectrum of broken power law at 200 keV from
photon index of --0.9 to --2.0.  Such bursts are estimated to occur in
the LAT FoV $\sim$10 times per year.}

\tablenotetext{c}{Time relative to detection of GRB.}

\tablenotetext{d}{High latitude source of $10^{-7}$ cm$^{-2}$ s$^{-1}$ 
flux at $>$100 MeV with a photon
spectral index of --2.0 above a flat background and assuming no
spectral cut-off at high energy; $1\sigma$ radius; 1-year survey.}

\tablenotetext{e}{For a steady source after 1 year sky survey, assuming a
high-latitude diffuse flux of $1.5\times10^{-5}$ cm$^{-2}$ s$^{-1}$ sr$^{-1}$  ($>$100 MeV) and a
photon spectral index of --2.1, with no spectral cut-off.}
\end{deluxetable}

\newcommand{\cola}{5cm}
\newcommand{\colb}{8cm}
\newcommand{\colc}{3cm}

\begin{deluxetable}{lcl}
\tablecaption{\label{t2.1} Key LAT tracker parameters}
\tablecolumns{3}
\tablewidth{0pt}
\tablehead{
\colhead{Parameter} & \colhead{Value} & \colhead{Performance Drivers and Constraints}
}
\startdata

\begin{minipage}[l]{\cola}
Noise occupancy (fraction of channels with noise hits per trigger)
\end{minipage}
 & $10^{-6}$ &
\begin{minipage}[l]{\colb}
Trigger rate, data volume, track reconstruction.  
The requirement, driven by the trigger rate, is $<10^{-4}$
\end{minipage}
\noindent{\medskip}\\

\begin{minipage}[l]{\cola}
Single channel efficiency for minimum ionizing particle (MIP), 
within fiducial volume
\end{minipage}
 & $>$99\% &
\begin{minipage}[l]{\colb}
PSF, especially at low energy.  
It is important to measure the tracks in the first 2 planes following the conversion point.
\end{minipage}
\noindent{\medskip}\\

\begin{minipage}[l]{\cola}
Ratio of strip pitch to vertical spacing between tracker planes
\end{minipage}
 & 0.0071 &
\begin{minipage}[l]{\colb}
High-energy ($>$1 GeV) PSF
\end{minipage}
\noindent{\medskip}\\

\begin{minipage}[l]{\cola}
Silicon-strip detector pitch (center-to-center distance between strips)
\end{minipage}
 & 228 $\mu$m &
\begin{minipage}[l]{\colb}
Small value needed to maintain a small pitch-to-plane-spacing ratio without destroying the FoV.
\end{minipage}
\noindent{\medskip}\\

\begin{minipage}[l]{\cola}
Aspect ratio (height/width)
\end{minipage}
 & 0.4 &
\begin{minipage}[l]{\colb}
Large FoV for photons with energy determination
\end{minipage}
\noindent{\medskip}\\

\begin{minipage}[l]{\cola}
Front converter foil thickness in radiation lengths (100\% W)
\end{minipage}
 & 
\begin{minipage}[c]{\colc}
$12\times0.03$\\ (0.010 cm/foil)
\end{minipage}
 &
\begin{minipage}[l]{\colb}
Minimize thickness per plane for low-energy PSF, but not so much that support 
material dominates. Maximize total thickness to maximize effective area.
\end{minipage}
\noindent{\medskip}\\

\begin{minipage}[l]{\cola}
Back converter foil thickness in radiation lengths (93\% W)
\end{minipage}
 & 
\begin{minipage}[r]{\colc}
$4\times0.18$\\ (0.072 cm/foil)
\end{minipage}
 &
\begin{minipage}[l]{\colb}
Effective area and FoV at high energies
\end{minipage}
\noindent{\medskip}\\

\begin{minipage}[l]{\cola}
Support material and detector material per $x-y$ plane (radiation lengths)
\end{minipage}
 & 0.014 &
\begin{minipage}[l]{\colb}
Stable mechanical support is needed, but much of this material is in a non-optimal 
location for the PSF.  Minimize to limit PSF tails from conversions occurring in 
support material.
\end{minipage}
\noindent{\smallskip}

\enddata
\end{deluxetable}

\renewcommand{\cola}{5cm}
\renewcommand{\colb}{8cm}

\begin{deluxetable}{lcl}
\ifthenelse{\equal{\ms}{preprint}}
{}
{
\rotate
}
\tablecaption{\label{t2.2} Key LAT calorimeter parameters}
\tablecolumns{3}
\tablewidth{0pt}
\tablehead{
\colhead{Parameter} & \colhead{Value} & \colhead{Performance Drivers and Constraints}
}
\startdata

Depth, including tracker (radiation lengths)
 & 10.1 &
\begin{minipage}[l]{\colb}
Calorimeter depth is a compromise in shower containment against maximum permitted mass.  
Use segmentation and shower profile analysis to improve energy measurement at high energies
\end{minipage}
\noindent{\medskip}\\

Sampling (angle dependent)
 & $>$90\% active &
\begin{minipage}[l]{\colb}
Energy loss in passive material causes low-energy tails on measured energy and affects 

energy resolution.
\end{minipage}
\noindent{\medskip}\\

Longitudinal segmentation
 & 8 segments &
\begin{minipage}[l]{\colb}
Shower profile analysis permits estimation of and correction for energy leakage
\end{minipage}
\noindent{\medskip}\\

Lateral segmentation
 & $\sim$1 Moli\`ere radius &
\begin{minipage}[l]{\colb}
Correlation of energy deposition in calorimeter with extrapolated tracks in tracker 
is critical part of background rejection.
\end{minipage}
\noindent{\smallskip}

\enddata
\end{deluxetable}

\renewcommand{\cola}{5cm}
\renewcommand{\colb}{8cm}

\begin{deluxetable}{lcl}
\ifthenelse{\equal{\ms}{preprint}}
{}
{
\rotate
}
\tablecaption{\label{t2.3} Key LAT anticoincidence detector parameters}
\tablecolumns{3}
\tablewidth{0pt}
\tablehead{
\colhead{Parameter} & \colhead{Value} & \colhead{Performance Drivers and Constraints}
}
\startdata

Segmentation into tiles
 & $<$1000 cm$^2$ each &
\begin{minipage}[l]{\colb}
Minimize self-veto, especially at high energy.  This value is for the top. 
Side tiles are smaller, to achieve a similar solid angle, as seen from the calorimeter.
\end{minipage}
\noindent{\medskip}\\

Efficiency of a tile for detecting a MIP
 & $>$0.9997 &
\begin{minipage}[l]{\colb}
Cosmic ray rejection, to meet a requirement of 0.99999 when combined with the other subsystems.
\end{minipage}
\noindent{\medskip}\\

Number of layers
 & 1 &
\begin{minipage}[l]{\colb}
Minimize material, mass, and power.  Dual readout on each tile for redundancy.
\end{minipage}
\noindent{\medskip}\\

Micrometeoroid / thermal blanket thickness 
 & 0.39 g cm$^{-2}$ &
\begin{minipage}[l]{\colb}
Small value needed to minimize \gray{} production in this passive material from 
cosmic-ray interactions.
\end{minipage}
\noindent{\medskip}\\

Total thickness (radiation lengths) 
 & 10.0 mm (0.06) 
 & Minimize absorption of incoming gamma radiation
\begin{minipage}[l]{\colb}
\end{minipage}
\noindent{\smallskip}

\enddata
\end{deluxetable}

\begin{deluxetable}{lccc}
\ifthenelse{\equal{\ms}{preprint}}
{}
{
\rotate
}
\tablecaption{\label{t2.4} Data sources for background model}
\tablecolumns{4}
\tablewidth{0pt}
\tablehead{
 & \multicolumn{3}{c}{Energy range}\\
 & \colhead{$>$ local geomagnetic cutoff}
 & \colhead{150 MeV to geomagnetic cutoff}
 & \colhead{10 MeV -- 150 MeV}
}
\startdata

Galactic Cosmic Rays \\ 
  $\quad$ protons + antiprotons & AMS\\
  $\quad$ electrons & AMS\\
  $\quad$ positrons & AMS\\
  $\quad$ He & AMS\\
  $\quad$ $Z > 2$ nuclei & HEAO--3 \medskip\\ 

Splash Albedo\\
  $\quad$ protons & & AMS & Nina\\
  $\quad$ electrons & & AMS & Mariya\\
  $\quad$ positrons & & AMS & Mariya \medskip\\

Re-entrant Albedo\\
  $\quad$ protons & & Nina\\
  $\quad$ electrons & & Mariya\\
  $\quad$ positrons & & Mariya \medskip\\

Earth albedo \gray{s} &
\multicolumn{3}{c}{10 MeV -- 100 GeV, EGRET} \medskip\\

Neutrons &
\multicolumn{3}{c}{10 MeV -- 1 TeV, various sources} \smallskip

\enddata
\tablecomments{
Data Sources:   AMS: \citet{Aguilar2002};
                    Nina: \citet{Bidoli2002};
                    Mariya: \citet{Voronov1991}, \citet{Mikhailov2002};
                    EGRET: \citet{Petry2005};
		    HEAO--3: \citet{Engelmann1990};
                    neutrons: \citet{Selesnik2007} 
}
\end{deluxetable}

\renewcommand{\colb}{8cm}
\begin{deluxetable}{lcl}
\tablecaption{\label{t2.5} LAT analysis classes}
\tablecolumns{3}
\tablewidth{0pt}
\tablehead{
\colhead{
Analysis class} & \colhead{Residual background} & \colhead{Characteristics}\\
\colhead{} & \colhead{rate (Hz)} & \colhead{}
}
\startdata

Transient & 2 &
\begin{minipage}[l]{\colb}
Maximize effective area, particularly at low energy, at the expense of higher 
residual background rate; suitable for study of localized, transient sources
\end{minipage}
\noindent{\medskip}\\

Source & 0.4 &
\begin{minipage}[l]{\colb}
Residual background rate comparable to extragalactic diffuse rate estimated 
from EGRET; suitable for study of  localized sources
\end{minipage}
\noindent{\medskip}\\

Diffuse & 0.1 &
\begin{minipage}[l]{\colb}
Residual background rate comparable to irreducible limit and tails of PSF at 
high-energy minimized; suitable for study of the weakest diffuse sources expected
\end{minipage}
\noindent{\smallskip}

\enddata
\end{deluxetable}

\renewcommand{\cola}{4cm}
\renewcommand{\colb}{6cm}
\begin{deluxetable}{lll}
\tablecaption{\label{t3.1} LAT searches for dark matter}
\tablecolumns{3}
\tablewidth{0pt}
\tablehead{
\colhead{Astrophysical source}\\
\colhead{or search technique} & \colhead{Advantages} & \colhead{Disadvantages}
}
\startdata

Galactic center &
\begin{minipage}[l]{\cola}
Large number of photons
\end{minipage} &
\begin{minipage}[l]{\colb}
Disturbance by many point sources, uncertainty in diffuse background prediction.
\end{minipage}
\noindent{\medskip}\\

Satellites, sub-halos &
\begin{minipage}[l]{\cola}
Low celestial diffuse background, good identification of source 
\end{minipage} &
\begin{minipage}[l]{\colb}
Low number of photons.
\end{minipage}
\noindent{\medskip}\\

Milky way halo &
\begin{minipage}[l]{\cola}
Large number of photons
\end{minipage} &
\begin{minipage}[l]{\colb}
Uncertainty in Galactic diffuse background prediction
\end{minipage}
\noindent{\medskip}\\

Extragalactic &
\begin{minipage}[l]{\cola}
Large number of photons
\end{minipage} &
\begin{minipage}[l]{\colb}
Astrophysical uncertainties, uncertainty in Galactic diffuse contribution.
\end{minipage}
\noindent{\medskip}\\

Spectral lines &
\begin{minipage}[l]{\cola}
No astrophysical uncertainties, smoking gun signal
\end{minipage} &
\begin{minipage}[l]{\colb}
Very low number of photons.
\end{minipage}
\noindent{\smallskip}

\enddata
\end{deluxetable}

\clearpage


\clearpage

\begin{thebibliography}{}
\bibitem[Abdo et al.(2008)]{Abdo2008} 
Abdo, A., et al. 2008, Science, 322, 1218

\bibitem[Agostinelli et al.(2003)]{Agostinelli2003} 
Agostinelli, S., et al. 2003, Nucl.\ Instr.\ Meth.\ A, 506, 250

\bibitem[Aguilar et al.(2002)]{Aguilar2002} 
Aguilar, M., et al. 2002, \physrep, 366, 331

\bibitem[Aharonian et al.(2005a)]{Aharonian2005a} 
Aharonian, F. A., et al. 2005a, \aap, 432, L25

\bibitem[Aharonian et al.(2005b)]{Aharonian2005b} 
Aharonian, F. A., et al. 2005b, \aap, 437, L7

\bibitem[Aharonian et al.(2006a)]{Aharonian2006a} 
Aharonian, F. A., et al. 2006a, \aap, 456, 245

\bibitem[Aharonian et al.(2006b)]{Aharonian2006b} 
Aharonian, F. A., et al. 2006b, \aap, 449, 223

\bibitem[Aharonian et al.(2006c)]{Aharonian2006c} 
Aharonian, F. A., et al. 2006c, \nat, 440, 1018

\bibitem[Albert et al.(2007)]{Albert2007} 
Albert, J., et al. 2007, \apj, 665, L51

\bibitem[Alfaro et al.(2002)]{Alfaro2002} 
Alfaro, J., Morales-Tecotl, H. A., \& Urrutia, L. F. 2002, \prd, 65, 103509

\bibitem[Allen et al.(1997)]{Allen1997} 
Allen, G. E., et al. 1997, \apj, 487, L97

\bibitem[Allison et al.(2006)]{Allison2006} 
Allison, J., et al. 2006, \ieeenuc, 53, 270

\bibitem[Amelino-Camelia et al.(1998)]{Amelino-Camelia1998} 
Amelino-Camelia, G., Ellis, J., Mavromatos, N.E., Nanopoulos, D.V., \& Sarkar, S. 1998, \nat, 395, 525


\bibitem[Atwood et al.(1994)]{Atwood1994}
Atwood, W. B., et al. 1994, Nucl.\ Instr.\ Meth.\ A, 342, 302

\bibitem[Atwood et al.(2000)]{Atwood2000}
Atwood, W. B., et al. 2000, Nucl.\ Instr.\ Meth.\ A, 446, 444

\bibitem[Atwood et al.(2006)]{Atwood2006}
Atwood, W. B., Ziegler, M., Johnson, R. P., \& Baughman, B. M. 2006, \apj, 652, L49

\bibitem[Atwood et al.(2007)]{Atwood2007}
Atwood, W. B., et al. 2007, \app, 28, 422

\bibitem[Baldini et al.(2006)]{Baldini2006}
Baldini, L., et al.  2006, \ieeenuc, 53, 466

\bibitem[Baldini et al.(2007)]{Baldini2007}
Baldini, L., et al.  2007, in Proc.\ of $1^{\rm st}$ GLAST Symposium, eds.\
S. Ritz, P. Michelson, \& C. Meegan, (NY: AIP), AIP Conf.\ Proc.\ 921, 190

\bibitem[Baltz et al.(2006)]{Baltz2006}
Baltz, E. A., Battaglia, M., Peskin, M. E., \& Wizansky, T. 2006, \prd, 74, 103251

\bibitem[Baltz et al.(2008)]{Baltz2008}
Baltz, E. A., et al. 2008, \jcap, 07, 013

\bibitem[Band(1993)]{Band1993}
Band, D., et al. 1993, \apj, 413, 281

\bibitem[Baring \& Harding(1997)]{Baring1997}
Baring, M. G., \& Harding, A. K. 1997, \apj, 491, 663

\bibitem[Baring(2006)]{Baring2006}
Baring, M. G. 2006, \apj, 650, 1004

\bibitem[Barrand et al.(2001)]{Barrand2001}
Barrand, G., et al. 2001, Computer Phys.\ Comm., 140, 45

\bibitem[Baughman et al.(2007)]{Baughman2007}
Baughman, B. M., Atwood, W. B., Johnson, R. P., Porter, T. A., \&  Ziegler M.,
2007, in Proc.\ 30th Int.\ Cosmic Ray Conf., in press (arXiv:0706.0503)
 
\bibitem[Berezinsky et al.(1997)]{Berezinsky1997}
Berezinsky, V. S., Blasi, P., \& Ptuskin, V. S. 1997, \apj, 487, 529

\bibitem[Berezinsky et al.(2001)]{Berezinsky2001}
Berezinsky, V., Hnatyk, B., \& Vilenkin, V. 2001, \prd, 64, 043004

\bibitem[Berger et al.(2007)]{Berger2007}
Berger, E., et al. 2007, \apj, 660, 496

\bibitem[Bergstr\"om(2000)]{Bergstrom2000}
Bergstr\"om, L. 2000, Rept.\ Prog.\ Phys., 63, 793

\bibitem[Bergstr\"om et al.(2001)]{Bergstrom2001}
Bergstr\"om, L., Edsj\"o, J., \& Ullio, P. 2001, \prl, 87, 251301

\bibitem[Bidoli et al.(2002)]{Bidoli2002}
Bidoli, V., et al. 2002, Annales Geophysicae, 20, 1693

\bibitem[Bignami et al.(1975)]{Bignami1975}
Bignami, G. F. et al. 1975, Space Sci.\ Instr., 1, 245

\bibitem[Bignami et al.(1979)]{Bignami1979}
Bignami, G. F., Fichtel, C. E., Hartman, R. C., \& Thompson D. J. 1979, \apj, 323, 649

\bibitem[Bignami \& Caraveo(1996)]{Bignami1996}
Bignami, G. F., \& Caraveo, P. A. 1996, \araa, 34,331

\bibitem[Blandford \& Znajek(1977)]{Blandford1977}
Blandford, R. D., \& Znajek, R. L. 1977, \mnras, 179, 433

\bibitem[Bloom et al.(2002)]{Bloom2002}
Bloom, J. S., Kulkarni, S. R., \& Djorgovski, S. G. 2002, \aj, 123, 1111

\bibitem[Bloom et al.(2006)]{Bloom2006}
Bloom, J. S., et al. 2006, \apj, 638, 354

\bibitem[B\"ottcher \& Dermer(1998)]{Bottcher1998}
B\"ottcher, M., \& Dermer, C. D. 1998, \apj, 499, L131

\bibitem[Bouvier at al.(2008)]{Bouvier2008}
Bouvier, A., et al. 2008, GCN Circular 8183

\bibitem[Breiman et al.(1984)]{Breiman1984}
Breiman, L., Friedman, J., Stone, C. J., \& Olshen, R. A. 1984, Clasification and Regression Trees, 
Wadsworth International Group, Belmont, CA

\bibitem[Calc\'aneo-Rold\'an \& Moore(2000)]{Calcaneo-Roldan2000}
Calc\'aneo-Rold\'an, C., \& Moore, B. 2000, \prd, 62, 123005

\bibitem[Caraveo et al.(1996)]{Caraveo1996}
Caraveo, P. A., Bignami, G. F., \&  Tr\"umper, J., 1996, \aapr, 7, 209

\bibitem[Caraveo \&  Reimer(2007)]{Caraveo2007}
Caraveo, P. A., \&  Reimer, O. 2007, in Proc. of 1st GLAST Symposium, 
eds. S. Ritz, P. Michelson \& C. Meegan,  (NY:AIP), AIP Conf. Proc. 921, 289 

\bibitem[Carson et al.(1996)]{carson1996}
Carlson, P., et al. 1996, Nucl.\ Instr.\ Meth.\ A, 376, 271  

\bibitem[Casanova et al.(2007)]{Casanova2007}
Casanova, S., Dingus, B. L., \& Zhang, B. 2007, \apj, 656, 306

\bibitem[Chandler et al.(2001)]{Chandler2001}
Chandler, A. M., et al. 2001 \apj, 556, 59

\bibitem[Chen et al.(2004)]{Chen2004}
Chen, A., Reyes, L.C., \& Ritz, S. 2004, \apj 608, 686

\bibitem[Cheng et al.(1986)]{Cheng1986}
Cheng, K. S., Ho, C., \& Ruderman, M. 1986, \apj, 300, 500

\bibitem[Chiang \& Mukherjee(1998)]{Chiang1998}
Chiang, J., \& Mukherjee, R. 1998, \apj, 496, 752

\bibitem[Colafrancesco(2002)]{Colafrancesco2002}
Colafrancesco, S., 2002, \aap, 396, 31

\bibitem[Colafrancesco \& Blasi(1998)]{Colafrancesco1998}
Colafrancesco, S., \& Blasi, P. 1998, \app, 9, 227

\bibitem[Couto \'e Silva et al.(2001)]{Couto_e_Silva2001}
Couto \'e Silva, E. do, et al. 2001, Nucl.\ Instr.\ Meth.\ A, 474, 19

\bibitem[Dar \& Shaviv(1995)]{Dar1995}
Dar, A., \& Shaviv, N. 1995, \prl, 75, 3052

\bibitem[Daugherty \& Harding(1996)]{Daugherty1996}
Daugherty, J. K., \& Harding, A. K. 1996, \apj, 458, 278

\bibitem[de Boer et al.(2005)]{de_Boer2005}
de Boer, W., et al. 2005, \prl, 95, 209001

\bibitem[de Jager et al.(2006)]{de_Jager2006}
de Jager, O. C., et al. 2006, \aaps, 120, 441

\bibitem[Dermer et al.(2000)]{Dermer2000}
Dermer, C. D., Chiang, J., \& Mitman, K. E. 2000, \apj, 537, 785

\bibitem[Dermer(2007)]{Dermer2007}
Dermer, C. D. 2007, \apj, 659, 958

\bibitem[Dixon et al.(1998)]{Dixon1998}
Dixon, D. D., Hartmann, D. H., Kolaczyk, E. D., Samimi, J., Diehl, R., Kanbach, G., 
Mayer-Hasselwander, H., \& Strong, A. W. 1998, \na, 3, 539
 
\bibitem[Dolgov \& Silk(1993)]{Dolgov1993}
Dolgov, A., \& Silk, J. 1993, \prd, 47, 4244 

\bibitem[Dyks \& Rudak(2003)]{Dyks2003}
Dyks, J., \& Rudak, B. 2003, \apj, 598, 1201

\bibitem[Els\"asser \& Mannheim(2005)]{Elsasser2005}
Els\"asser, D., \& Mannheim, K. 2005, \prl, 94, 171302

\bibitem[Engelmann et al.(1990)]{Engelmann1990}
Engelmann, J. J., et al. 1990, \aap, 233, 96

\bibitem[Ensslin et al.(1997)]{Ensslin1997}
Ensslin, T. A., Biermann, P. L., Kronberg, P. P., \& Wu, X.-P., 1997, \apj, 477, 560

\bibitem[Esposito et al.(1996)]{Esposito1996}
Esposito, J. A., et al. 1996, \apj, 461, 820

\bibitem[Fazio \& Stecker(1970)]{Fazio1970}
Fazio, G. G., \& Stecker, F. W. 1970, \nat, 226, 135

\bibitem[Fenimore et al.(1999)]{Fenimore1999}
Fenimore, E. E., Ramirez-Ruiz, E., \& Wu, B. 1999, \apj, 518, 73

\bibitem[Feretti et al.(2004)]{Feretti2004}
Feretti, L., Burigana, C., \& Ensslin, T. 2004, \na, 48, 1134

\bibitem[Ferreira et al.(2004)]{Ferreira2004}
Ferreira, O., et al. 2004, \nimA, 530, 323

\bibitem[Fichtel et al.(1975)]{Fichtel1975}
Fichtel, C. E., et al. 1975, \apj, 198, 163

\bibitem[Fichtel et al.(1994)]{Fichtel1994}
Fichtel, C. E., et al. 1994, \apj, 434, 557

\bibitem[Fossati et al.(1998)]{Fossati1998}
Fossati G., et al. 1998, \mnras, 299, 433

\bibitem[Fr\"uhwirth et al.(2000)]{Fruhwirth2000}
Fr\"uhwirth, R., Regler, M., Bock, R. K., Grote, H. \& Notz, D. 2000, 
Data Analysis Techniques for High-Energy Physics, 2nd edition, Cambridge University Press

\bibitem[Fryer \& Heger(2005)]{Fryer2005}
Fryer, C. L., \& Heger, A. 2005, \apj, 623, 302

\bibitem[Funk(2008)]{Funk2008}
Funk, S., et al. 2008, \apj, 679, 1299

\bibitem[Gabici \& Blasi(2003)]{Gabici2003}
Gabici, S., \& Blasi, P. 2003, \app, 19, 679

\bibitem[Gao et al.(1990)]{Gao1990}
Gao, Y.-T., Stecker, F. W., Gleiser, M., \& Cline, D. B. 1990, \apj, 361, 37

\bibitem[Gehrels et al.(2000)]{Gehrels2000}
Gehrels, N., et al. 2000, \nat, 404, 363

\bibitem[Giommi et al.(2006)]{Giommi2006}
Giommi, P., et al. 2006, \aap, 456, 911

\bibitem[Giovannini et al.(1999)]{Giovannini1999}
Giovannini, G., Tordi, M., \& Feretti, L. 1999, \na, 4, 141

\bibitem[Gonthier et al.(2007)]{Gonthier2007}
Gonthier, P. L., et al. 2007, \apss, 309, 245

\bibitem[Gonzalez et al.(2003)]{Gonzalez2003}
Gonzalez, M. M., et al. 2003, \nat, 424, 74

\bibitem[Grechnev et al.(2008)]{Grechnev2008}
Grechnev, V. V. 2008, \solphys, 252, 149

\bibitem[Grenier(2000)]{Grenier2000}
Grenier, I. A. 2000, \aap, 364, L93

\bibitem[Grenier et al.(2000)]{Grenier2000a}
Grenier, I. A., Casandjian, J.-M., \& Terrier, R. 2005, \sci, 307, 1292

\bibitem[Grove et al.(2008)]{Grove2008}
Grove, E., et al. 2008, in preparation

\bibitem[Haino et al.(2004)]{Haino2004}
Haino, S., et al. 2004, \plb, 594, 35

\bibitem[Halpern et al.(2001)]{Halpern2001}
Halpern, J. P., et al. 2001, \apj, 552, L125

\bibitem[Halpern et al.(2002)]{Halpern2002}
Halpern, J. P., et al. 2002, \apj, 573, L41

\bibitem[Harding et al.(2007)]{Harding2007}
Harding, A. K., Grenier, I. A., \& Gonthier, P. L. 2007, \apss, 309, 221

\bibitem[Hartman et al.(1999)]{Hartman1999}
Hartman, R. C., et al. 1999, \apj{S}, 123, 79

\bibitem[Hauser \& Dwek(2001)]{Hauser2001}
Hauser, M. G., \& Dwek, E. 2001, \araa, 39, 249

\bibitem[Hawking(1974)]{Hawking1974}
Hawking, S. W., 1974, \nat, 248, 30

\bibitem[Hunter et al.(1997)]{Hunter1997}
Hunter, S. D., et al. 1997, \apj, 481, 205

\bibitem[Hurley(1994)]{Hurley1994}
Hurley, K.  1994, \nat, 372, 652

\bibitem[Jackson et al.(2005)]{Jackson2005}
Jackson, B., et al. 2005, IEEE Signal Processing Letters, 12, 105

\bibitem[Jakobsson et al.(2006)]{Jakobsson2006}
Jakobsson, P., et al. 2006, \aap, 447, 897

\bibitem[Johnson et al.(2001)]{Johnson2001}
Johnson, W. N., et al. 2001, \ieeenuc, 48, 1182

\bibitem[Jungman et al.(1996)]{Jungman1996}
Jungman, G., Kamionkowski, M., \& Griest, K. 1996, \prd, 267, 195

\bibitem[Kaaret \& Cottam(1996)]{Kaaret1996}
Kaaret, P., \& Cottam, J. 1996, \apj, 462, L35

\bibitem[Kaneko et al.(2006)]{Kaneko2006}
Kaneko, Y., et al. 2006, \apj{S}, 166, 298

\bibitem[Kanbach et al.(1993)]{Kanbach1993}
Kanbach, G., et al. 1993, \aaps, 97, 349

\bibitem[Keshet et al.(2003)]{Keshet2003}
Keshet, U., et al. 2003, \apj, 585, 128

\bibitem[Kneiske et al.(2004)]{Kneiske2004}
Kneiske, T. et al. 2004, \aap, 413, 807 

\bibitem[Kniffen et al.(1993)]{Kniffen1993}
Kniffen, D. A., et al. 1993, \apj, 411, 133

\bibitem[Kramer et al.(2003)]{Kramer2003}
Kramer, M., et al. 2003, \mnras, 342, 1299


\bibitem[Lake(1990)]{Lake1990}
Lake, G. 1990, \nat, 346, 39

\bibitem[Lithwick \& Sari(2001)]{Lithwick2001}
Lithwick, Y., \& Sari, R. 2001, \apj, 555, 540

\bibitem[Loeb \& Waxman(2000)]{Loeb2000}
Loeb, A., \& Waxman, E. 2000, \nat, 405, 156

\bibitem[MacMinn \& Primack(1996)]{MacMinn1996}
MacMinn, D., \& Primack, J. R. 1996, \ssr, 75, 413

\bibitem[Madau \& Phinney(1996)]{Madau1996}
Madau, P., \& Phinney, E. S. 1996, \apj, 456, 124

\bibitem[Maki et al.(1996)]{Maki1996}
Maki, K., Mitsui, T., \& Orito, S. 1996, \prl, 76, 3474

\bibitem[Mattox et al.(1996)]{Mattox1996}
Mattox, J., et al. 1996, \aaps, 120, 95

\bibitem[Mattox et al.(2001)]{Mattox2001}
Mattox, J., Hartman, R. C., \& Reimer, O. 2001, \apj{S}, 135, 155

\bibitem[M\'esz\'aros et al.(1994)]{Meszaros1994}
M\'esz\'aros, P., Rees, M. J., \& Papathanassion, H. 2004, \apj, 432, 181
 
\bibitem[McEnery et al.(2008)]{McEnery2008}
McEnery, J., et al. 2008, GCN Circular 8407

\bibitem[Mikhailov et al.(2002)]{Mikhailov2002}
Mikhailov, V., et al. 2002, J.\ Mod.\ Phys.\ A, 17, 1695

\bibitem[Miniati(2002)]{Miniati2002}
Miniati, F. 2002, \mnras, 337, 199

\bibitem[Mirabal \& Halpern(2001)]{Mirabal2001}
Mirabal, N., \& Halpern, J. P. 2001, \apj, 547, L137

\bibitem[Mizuno et al.(2004)]{Mizuno2004}
Mizuno, T., et al. 2004, \apj, 614, 1113

\bibitem[Moderski et al.(2004)]{Moderski2004}
Moderski, R., et al. 2004, \apj, 611, 77

\bibitem[Moiseev et al.(2004)]{Moiseev2004}
Moiseev, A. A., et al. 2004, \app, 22, 275

\bibitem[Moiseev et al.(2007)]{Moiseev2007}
Moiseev, A. A., et al. 2007, \app, 27, 339

\bibitem[Montmerle(1979)]{Montmerle1979}
Montmerle, T. 1979, \apj, 231, 95

\bibitem[Moskalenko \& Strong(2000)]{Moskalenko2000}
Moskalenko, I. V., \& Strong, A. W. 2000, \apj, 528, 357

\bibitem[Moskalenko \& Strong(2005)]{Moskalenko2005}
Moskalenko, I. V., \& Strong, A. W. 2005, in Proc.\ Int.\ Conf.\ on Astrophysical 
Sources of High Energy Particles and Radiation, eds. T. Bulik, et al. (NY: AIP), AIP Conf.\ Proc.\ 801, 57

\bibitem[Moskalenko et al.(2006)]{Moskalenko2006}
Moskalenko, I. V., Porter, T. A., \& Digel, S. W. 2006, \apj, 652, L65

\bibitem[Moskalenko \& Porter(2007)]{Moskalenko2007a}
Moskalenko, I. V., \& Porter, T. A. 2007, \apj, 670, 1467

\bibitem[Moskalenko et al.(2007)]{Moskalenko2007}
Moskalenko, I. V., Digel, S. W., Porter, T. A., Reimer, O., \& Strong A. W. 2007, 
Nucl.\ Phys.\ B (Proc.\ Suppl.), 173, 44

\bibitem[Moskalenko et al.(2008)]{Moskalenko2008}
Moskalenko, I. V., Porter, T. A., Digel, S. W., Michelson, P. F., \& Ormes, J. F. 2007, 
\apj, 681, 1708

\bibitem[M\"ucke \& Pohl(2000)]{Mucke2000}
M\"ucke, A., \& Pohl, M. 2000, \mnras, 312, 177

\bibitem[Mukherjee \& Chiang(1999)]{Mukherjee1999}
Mukherjee, R., \& Chiang, J. 1999, \app, 11, 213


\bibitem[Muslimov \& Harding(2003)]{Muslimov2003}
Muslimov, A., \& Harding, A. K. 2003, \apj, 588, 430

\bibitem[Nakar(2007)]{Nakar2007}
Nakar, E. 2007, \physrep, 442, 166 

\bibitem[Narumoto \& Totani(2006)]{Narumoto2006}
Narumoto T., \& Totani, T. 2006, \apj, 643, 81

\bibitem[Ng et al.(2005)]{Ng2005}
Ng, C.-Y., Roberts, M. S. E., \& Romani, R. W. 2005, \apj, 627, 904

\bibitem[Nolan et al.(2003)]{Nolan2003}
Nolan, P. L., et al. 2003, \apj, 597, 615

\bibitem[Norris et al.(1996)]{Norris1996}
Norris, J. P., et al. 1996, \apj, 459, 393

\bibitem[Omodei et al.(2008)]{Omodei2008}
Omodei, N., et al. 2008, GCN Circular 8407

\bibitem[Orlando \& Strong(2007)]{Orlando2007}
Orlando, E., \& Strong, A. W. 2007, \apss, 309, 359

\bibitem[Orlando \& Strong(2008)]{Orlando2008}
Orlando, E., \& Strong, A. W. 2008, \aap, 480, 847

\bibitem[Page \& Hawking(1976)]{Page1976}
Page, D. N., \& Hawking, S. W. 1976, \apj, 206, 1

\bibitem[Paredes et al.(2000)]{Paredes2000}
Paredes, J. M., Mart\'i, J., Rib\'o, M., \& Massi, M. 2000, \sci, 288, 2340

\bibitem[Pavlidou \& Fields(2001)]{Pavlidou2001}
Pavlidou, V., \& Fields, B. D. 2001, \apj, 558, 63

\bibitem[Pavlidou \& Fields(2002)]{Pavlidou2002}
Pavlidou, V., \& Fields, B. D. 2002, \apj, 575, L5

\bibitem[Picozza et al.(2007)]{Picozza2007}
Picozza, P., et al. 2007, \app, 27, 296

\bibitem[Petry(2005)]{Petry2005}
Petry, D. 2005, in High Energy Gamma Ray Astronomy: 2nd International Symposium, 
Heidelberg, July, 2004, eds.\ F. A. Aharonian, H. J. Volk, \& D. Horns, (NY:AIP), AIP Conf.\ Proc.\ 745, 709

\bibitem[Porciani \& Madau(1999)]{Porciani1999}
Porciani, C., \& Madau, P. 1999, \apj, 526, L522

\bibitem[Porter et al.(2006)]{Porter2006}
Porter, T. A., Moskalenko, I. V., \& Strong A. W. 2006, \apj, 648, L29

\bibitem[Porter et al.(2008)]{Porter2008}
Porter, T. A., Moskalenko, I. V., Strong A. W., Orlando, E., \& Bouchet, L. 2008, 
\apj, 682, 400

\bibitem[Preece et al.(2000)]{Preece2000}
Preece, R. D., et al. 2000, \apj{S}, 126, 19


\bibitem[Primack et al.(1999)]{Primack1999}
Primack, J. R., Bullock, J. S., Sommerville, R. S. \& MacMinn, D. 1999, \app, 11, 93

\bibitem[Primack et al.(2001)]{Primack2001}
Primack, J. R., Somerville, R. S., Bullock, J. S. \& Devriendt, J. E. G. 2001, AIP, 558, 463

\bibitem[Primack et al.(2005)]{Primack2005}
Primack, J. R., Bullock, J. S., \& Sommerville, R. S. 2005, in Proc.\ of the Gamma 2004 
Symposium on High Energy Gamma Ray Astronomy (NY:AIP), AIP Conf.\ Proc.\ 745, 23

\bibitem[Reimer(2007)]{Reimer2007}
Reimer, A. 2007, \apj, 665, 1023

\bibitem[Reimer et al.(2001)]{Reimer2001}
Reimer, O., et al. 2001, \mnras, 324, 772

\bibitem[Reimer et al.(2003)]{Reimer2003}
Reimer, O., Pohl, M., Sreekumar, P., \& Mattox, J. R. 2003, \apj, 558, 155

\bibitem[Rephaeli et al.(2008)]{Rephaeli2008}
Rephaeli, Y., Nevalainen, J., Ohashi, T., \& Bykov, A. M. 2008, \ssr, 134, 71

\bibitem[Roberts et al.(2001)]{Roberts2001}
Roberts, M. S. E., Romani, R. W., \& Kawai, N. 2001, \apj{S}, 133, 451 

\bibitem[Romani(1996)]{Romani1996}
Romani, R. W. 1996, \apj, 470, 469

\bibitem[Romero et al.(1999)]{Romero1999}
Romero, G. E., et al. 1999, \aap, 348, 868R

\bibitem[Rudaz \& Stecker(1991)]{Rudaz1991}
Rudaz, S., \& Stecker, F. W. 1991, \apj 368, 406

\bibitem[Scargle(1998)]{Scargle1998}
Scargle, J. D. 1998, \apj, 504, 405  

\bibitem[Scargle et al.(2008)]{Scargle2008}
Scargle, J. D., Norris, J. P., \& Bonnell, J. T. 2008, \apj, 673, 972  

\bibitem[Seckel et al.(1991)]{Seckel1991}
Seckel, D., Stanev, T., \& Gaisser, T. K. 1991, \apj, 382, 652

\bibitem[Selesnik et al.(2007)]{Selesnik2007}
Selesnik, R. S., Looper, M. D., \& Mewaldt, R. A. 2007, Space Weather, 5, S04003

\bibitem[Share et al.(2006)]{Share2006}
Share, G.H., et al. 2006, \baas, 38, 255

\bibitem[Sikora \& Madejski(2000)]{Sikora2000}
Sikora, M., \& Madejski, G. 2000, \apj, 534, 109

\bibitem[Silk \& Srednicki(1984)]{Silk1984}
Silk, J., \& Srednicki, M., 1984, \prl, 53, 624

\bibitem[Smith et al.(2008)]{Smith2008}
Smith, D. A., et al. 2008, \aap, to be published

\bibitem[Sommer et al.(1994)]{Sommer1994}
Sommer, M., et al. 1994, \apj, 422, L63

\bibitem[Sowards-Emmerd et al.(2003)]{Sowards-Emmerd2003}
Sowards-Emmerd, D., Romani, R. W., \& Michelson, P. F. 2003, \apj, 590, 109

\bibitem[Sowards-Emmerd et al.(2004)]{Sowards-Emmerd2004}
Sowards-Emmerd, D., et al. 2004, \apj, 609, 564

\bibitem[Sowards-Emmerd et al.(2005)]{Sowards-Emmerd2005}
Sowards-Emmerd, D., et al. 2005, \apj, 626, 95 

\bibitem[Spergel et al.(2007)]{Spergel2007}
Spergel, D. N., et al. 2007, \apj{S}, 170, 377

\bibitem[Sreekumar et al.(1998)]{Sreekumar1998}
Sreekumar, P., et al. 1998, \apj, 494, 523

\bibitem[Stecker et al.(1971)]{Stecker1971}
Stecker, F. W., Vette, J. I., \& Trombka, J. I. 1971, \nat, 231, 122

\bibitem[Stecker \& Salamon(1996)]{Stecker1996}
Stecker, F., \& Salamon, M. 1996, \apj, 464, 600

\bibitem[Stecker et al.(2006)]{Stecker2006}
Stecker, F. W., Malkan, A., \& Scully, S. T. 2006, \apj, 648, 774; and erratum: \apj, 658, 1392 (2007)

\bibitem[Stecker et al.(2008)]{Stecker2008}
Stecker, F., Hunter, S. D., \& Kniffen, D. A. 2008, \app, 29, 25 

\bibitem[Strong et al.(2000)]{Strong2000}
Strong, A. W., Moskalenko, I. V., \& Reimer, O. 2000, \apj, 537, 763

\bibitem[Strong et al.(2004a)]{Strong2004a}
Strong, A. W., Moskalenko, I. V., \& Reimer, O. 2004a, \apj, 613, 962 

\bibitem[Strong et al.(2004b)]{Strong2004b}
Strong, A. W., Moskalenko, I. V., \& Reimer, O. 2004b, \apj, 613, 956 

\bibitem[Strong et al.(2004c)]{Strong2004c}
Strong, A. W., Moskalenko, I. V., Reimer O., Digel S., \& Diehl R. 2004c, \aap, 422, L47

\bibitem[Sturner \& Dermer(1995)]{Sturner1995}
Sturner, S. J., \& Dermer, C. D. 1995, \aap, 291, L17

\bibitem[Tajima et al.(2008)]{Tajima2008}
Tajima, H., et al. 2008, GCN Circular 8246

\bibitem[Takata et al.(2006)]{Takata2006}
Takata, J., et al. 2006, \mnras, 366, 1310

\bibitem[Tavani et al.(1998)]{Tavani1998}
Tavani, M., et al. 1998, \apj, 497, L89

\bibitem[Tavani et al.(2008)]{Tavani2008}
Tavani, M., et al. 2008, \nimA, 588, 52

\bibitem[Taylor \& Babul(2005a)]{Taylor2005a}
Taylor, J. E., \& Babul, A. 2005a, \mnras, 364, 515

\bibitem[Taylor \& Babul(2005b)]{Taylor2005b}
Taylor, J. E., \& Babul, A. 2005b, \mnras, 364, 535

\bibitem[Thompson et al.(1993)]{Thompson1993}
Thompson, D. J., et al. 1993, \apj{S}, 86, 629 

\bibitem[Thompson et al.(1997a)]{Thompson1997a}
Thompson, D. J., et al. 1997a, Proc. 4th Compton Symp., ed. C.D. Dermer, et al. (New York, AIP), pp.39-51

\bibitem[Thompson et al.(1997b)]{Thompson1997b}
Thompson, D. J., Bertsch, D. L., Morris, D. J., \& Mukherjee, R. 1997b, \jgr, 102, 14735

\bibitem[Thompson et al.(2002)]{Thompson2002}
Thompson, D. J., et al. 2002, \ieeenuc, 49, 1898 

\bibitem[Thompson(2001)]{Thompson2001}
Thompson, D. J. 2001, Proc.\ of High-Energy Gamma-ray Astronomy Symposium, eds. F.A. Aharonian
\& H.J. V\"olk (New York, AIP), AIP Conf.\ Proc., 558, 103

\bibitem[Thompson(2004)]{Thompson2004}
Thompson, D. J. 2004, in Cosmic Gamma-Ray Sources, eds.\ K. S. Cheng,
\& G. E. Romero (Dordrecht Boston London: Kluwer), 149

\bibitem[Torres et al.(2004)]{Torres2004}
Torres, D. F., et al. 2004, \apj, 607, L99

\bibitem[Torres \& Reimer(2005)]{Torres2005}
Torres, D. F., \& Reimer, O. 2005, \apj, 629, L141

\bibitem[Totani \& Kitayama(2000)]{Totani2000}
Totani, T., \& Kitayama, T. 2000, \apj, 545, 572

\bibitem[Ullio et al.(2002)]{Ullio2002}
Ullio, P., et al. 2002, \prd, 66, 123502

\bibitem[Urry \& Padovani(1995)]{Urry1995}
Urry, M., \& Padovani, P. 1995, \pasp, 107, 803

\bibitem[Vercellone et al.(2008)]{Vercellone2008}
Vercellone, S., et al. 2008, \apj, 676, L13

\bibitem[Voronov et al.(1991)]{Voronov1991}
Voronov, S. A., et al. 1991, Cosmic Research (English Translation) 29(4), 567

\bibitem[Willis(1996)]{Willis1996}
Willis, T. D. 1996, PhD thesis, Stanford University

\bibitem[Yadigaroglu \& Romani(1997)]{Yadigaroglu1997}
Yadigaroglu, I. A., \& Romani, R. W. 1997, \apj, 476, 347

\bibitem[Zhang \& M\'esz\'aros(2004)]{Zhang2004}
Zhang, B., \& M\'esz\'aros, P.  2004, Intl.\ J.\ Mod.\ Phys.\ A, 19, 238

\bibitem[Zhang et al.(2004)]{Zhang2004a}
Zhang, W., Woosley, S. E., \& Heger, A. 2004, \apj, 608, 365

\bibitem[Ziegler et al.(2008)]{Ziegler2008}
Ziegler, M., Baughman, B. M., Johnson, R. P., \& Atwood, W. B. 2008, \apj, 680, 620

\bibitem[Zuccon et al.(2003)]{Zuccon2003}
Zuccon, P., et al. 2003, \app, 20, 221


\end{thebibliography}
\end{document}